\documentclass[graybox, secnum]{svmult}

\usepackage{mathptmx}       % selects Times Roman as basic font
\usepackage{helvet}         % selects Helvetica as sans-serif font
\usepackage{courier}        % selects Courier as typewriter font
\usepackage{type1cm}        % activate if the above 3 fonts are
\usepackage{makeidx}         % allows index generation
\usepackage{graphicx}        % standard LaTeX graphics tool
\usepackage{multicol}        % used for the two-column index
\usepackage[bottom]{footmisc}% places footnotes at page bottom
\usepackage{hyperref}        %for hyperlinks
\usepackage{soul}            % for high-lighting of text

 \usepackage[export]{adjustbox}

\hypersetup{colorlinks=true,urlcolor=blue}
\usepackage[square,numbers]{natbib}
\makeindex             % used for the subject index
\def\ie{{\it i.e.}}
\def\eg{{\it e.g.}}
\def\etc{{\it etc.}}
\def\etal{{\it et al.}}
                       
\def\lsim{\mathrel{\raise.3ex\hbox{$<$\kern-.75em\lower1ex\hbox{$\sim$}}}}
\def\gsim{\mathrel{\raise.3ex\hbox{$>$\kern-.75em\lower1ex\hbox{$\sim$}}}}
\def\Pl{{\mbox{\scriptsize Pl}}}

\def\CM{{\mbox{\scriptsize CM}}}
\def\GeV{{\rm GeV}}

\newcommand\zosoir{\includegraphics[width=0.09\hsize,valign=c]{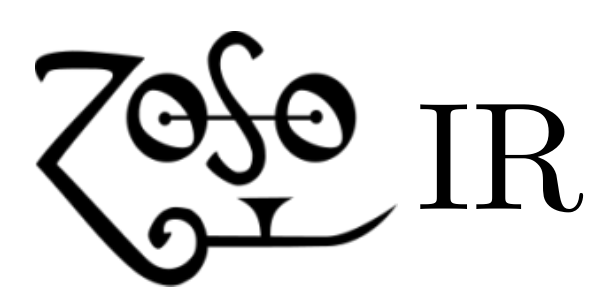}}
\newcommand\zosouv{\includegraphics[width=0.09\hsize,valign=c]{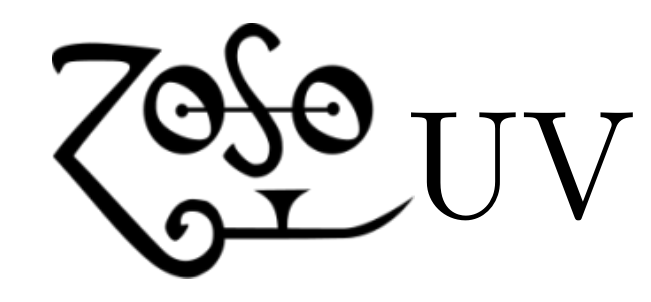}}

\begin{document}
%\tableofcontents{}
\title*{Effective Field Theory for Compact Binary Dynamics}
\author{Walter D. Goldberger} %\thanks{corresponding author}}
\institute{ \at  Sloane Physics Laboratory, Yale University, New Haven CT 06520, USA, \email{walter.goldberger@yale.edu}
%\and Second Author \at Institute 2, Address of Institute 2 \email{name2@email.address}
}
\maketitle
\abstract{I review the effective field theory (EFT) description of gravitating compact objects.   The focus is on kinematic regimes where gravity is perturbative, in particular the adiabatic inspiral phase relevant to gravitational wave detection.   For such configurations, there is a  hierarchy of length scales which all play a role in the dynamics, ranging from the gravitational radius, to the size of the objects, to their typical orbital separation, and finally the wavelength of the radiation emitted by the system.  To disentangle these scales, and to achieve manifest power counting in the expansion parameter, it is necessary to construct a tower of  EFTs of gravity, each coupled to distinct line defect localized degrees of freedom.   I describe the relevant effective theories at each scale as well as the matching between these theories across each physical threshold.   While the main applications of these methods are to classical dynamics, quantum gravity effects, \eg~Hawking graviton exchange, can be systematically incorporated if the momentum transfers are small compared to the Planck mass.   } 

\section*{Keywords} 
Effective field theory of gravity, black holes, gravitational waves, radiation, Post-Newtonian expansion, Post-Minkowskian expansion, compact binary system.

\section{Introduction}

The observation by LIGO~\cite{GW} in 2015 of gravitational waves sourced by the merger of two extra-galactic black holes~\cite{LIGOScientific:2016aoc} has brought renewed focus on understanding the gravitational dynamics of bound compact (black hole or neutron star) binary systems.   When the objects are separated by distances close to the typical Schwarzschild radius $r_s=2G_N M$, they are in the regime of large spacetime curvature and strong gravitational fields, quantitatively tractable only by the methodology of numerical general relativity~\cite{Lehner:2014asa}.  However, in the early \emph{adiabatic inspiral} stage, when the orbital separation is large, $r\gg r_s$, the evolution of the system is slow and admits a systematic expansion in powers of $r_s/r\ll 1$.    By the virial theorem of Newtonian gravity, the adiabatic inspiral is necessarily characterized by non-relativistic orbits, with typical relative velocities\footnote{\emph{Conventions:}   We adopt units where $c=\hbar =1$,  and define $\eta_{\mu\nu}=\mbox{diag}(1,-1,-,1-1)$,  $R^\mu{}_{\nu\rho\sigma}=\partial_\rho\Gamma^\mu{}_{\sigma\nu}+\Gamma^\mu{}_{\rho\lambda}\Gamma^\lambda_{\sigma\nu}-(\rho\leftrightarrow\sigma)$.} of order $$v^2 \sim {G_N M\over r}\sim {r_s\over r} \ll 1.$$

In order to optimize the detection of merger signals buried in the noisy gravitational wave data, for the purposes of parameter extraction (binary masses, spins, \etc), and to interface with numerical relativity simulations, it is crucial to carry out the analytical expansion of Einstein's equations to rather high order in powers of the velocity $v$.   Accurate theoretical wave templates are needed to at least order $v^{10}$~\cite{zimmerman} relative to the zeroth order solution, consisting of predominantly quadrupolar gravitational radiation sourced by nearly Newtonian orbits.   The traditional approach to the ``post-Newtonian'' (PN) expansion of the solution to Einstein equations as a perturbative series in powers $v$ has a long history see ~\cite{blanchet,Schafer:2018kuf} for reviews and complete list of references.

A more modern approach to perturbative gravitational dynamics, first proposed in~\cite{Goldberger:2004jt}, recasts the PN expansion in a language more familiar to particle physicists, as an effective field theory (EFT) of self-interacting gravitons coupled to classical worldline sources.   In this formulation, the graviton fluctuations about a fixed yet arbitrary configuration of point-like defects are integrated out, resulting in an effective action functional whose extrema yield the classical dynamics of the binary system.  In this way, the theory systematically captures both the effects of conservative gravitational forces (``potentials'') as well as radiation reaction on the evolution of the orbits.  Once the classical extrema have been determined, one uses them to calculate the one-point function of the graviton $\langle h_{\mu\nu}\rangle$ sourced by the binary, which encodes the waveform seen by detectors placed at asymptotic future null infinity ${\cal I}^+$.

The EFT approach to gravitationally bound systems relies on the observation that binary dynamics in the PN regime involves a hierarchy of well separated length scales:
\begin{eqnarray*}
r_s &=&\mbox{gravitational radius}\sim 2 G_N M, \\
{\cal R} &=& \mbox{typ. size of binary constituents},\\
r &=&\mbox{typ. orbital radius}\sim v^{-2} r_s,\\
\lambda &=&\mbox{typ. radiation wavelength}\sim v^{-1} r,
\end{eqnarray*}
with 
$$
r_s\lsim {\cal R} \ll r  \ll \lambda. 
$$

First, we assume that at distances larger than the Planck length, $\ell_{\Pl}=m_{\Pl}^{-1}\equiv\sqrt{32\pi G_N}\sim 10^{-19}\GeV^{-1}$, gravity is described by the Einstein-Hilbert Lagrangian~\cite{Einstein:1916vd} 
\begin{equation}
\label{eq:EH}
S_{EH}=-2 m_{\Pl}^2\int d^4 x \sqrt{g} R + \cdots,
\end{equation}
expanded around a fixed background, e.g. the Minkowski vacuum, $g_{\mu\nu}=\eta_{\mu\nu}+h_{\mu\nu}/m_{\Pl}$.    We take the point of view~\cite{Donoghue:1994dn} that Eq.~(\ref{eq:EH}) represents the unique~\cite{Weinberg:1964ew,Weinberg:1965rz} low energy effective theory of quantum gravity, with well defined Feynman rules~\cite{Gupta:1952zz,Feynman:1963ax,DeWitt:1967yk,DeWitt:1967ub,DeWitt:1967uc},  with sensible infrared (IR) behavior~\cite{Weinberg:1965nx},  and ultraviolet (UV) renormalization properties~\cite{tv,Goroff:1985th}.   We have suppressed in Eq.~(\ref{eq:EH}) an infinite tower of local higher curvature terms, generated by loop corrections, which are assumed to be kinematically suppressed by powers of $(E_{\CM}/m_{\Pl})^2\ll 1$ in a typical process.    More precisely, for the applications of interest in this review, we consider the classical limit with $\ell_{Pl}\rightarrow 0$  and gravitational radius $r_s\sim G_N E_{\CM}$ held fixed to be somewhat smaller than the physical radius ${\cal R}\gg\ell_{\Pl}$ of the compact objects, in which case the higher order terms in Eq.~(\ref{eq:EH}) give rise to corrections suppressed by powers of $(\ell_{\Pl}/r_s)^2\ll 1$ .

For a compact object of a given mass, the scale ${\cal R}$ depends on the detailed internal structure via a thermodynamic equation of state.   By definition, a compact object is one with $\kappa={\cal R}/2 G_N M\gsim {\cal O}(1)$ (\eg~$\kappa_{BH}=1$ for a Schwarzschild black hole, while $\kappa_{NS}\sim {\cal O}(10)$ for neutron stars).   Even though the typical energy scale $E_{\CM}$ is super-Planckian, the orbital separation $\sim r$ in a binary encounter is taken to be large, $r\gg {\cal R}\gg r_s$, so that observables at scales $r$ can be computed by treating the compact constituents as point defects whose worldlines deflect via graviton exchange and which source and absorb radiation.  This theory of classical worldlines coupled to gravitons is suitable for computing gravitational radiation in compact binaries, as an expansion in powers of $r_s/r$ (formally, this is an expansion in powers of $G_N$, the so-called ``post-Minkowskian'' (PM) expansion of general relativity).   In such a kinematic regime, the relevant graviton modes have typical four-momenta $k^\mu\sim 1/r$.

 For applications to gravitational wave astronomy, one is in addition interested in adiabatic binary inspirals, with bound non-relativistic orbits, $v\ll 1$.  Consequently, there is now an additional separation of scales between orbital dynamics at the scale $r$ and radiation at a characteristic wavelength set by the multipole expansion, $\lambda\sim r/v\gg r$.   Thus the various scales in the problem become \emph{correlated}, in the sense that the single expansion parameter $v$ controls the relative contribution of physics arising at widely separated scales, $r_s\sim v^2 r \sim v^3 \lambda$.
 
 %%%%%%%%%%%%%%%%%%%%%%%%%%%%%%%%%%%%%%%%%%%%%%%%%%%%%%%%%%%%%%%%%%%%%%%%%
\begin{figure}
\begin{center}
\includegraphics[width=0.90\hsize]{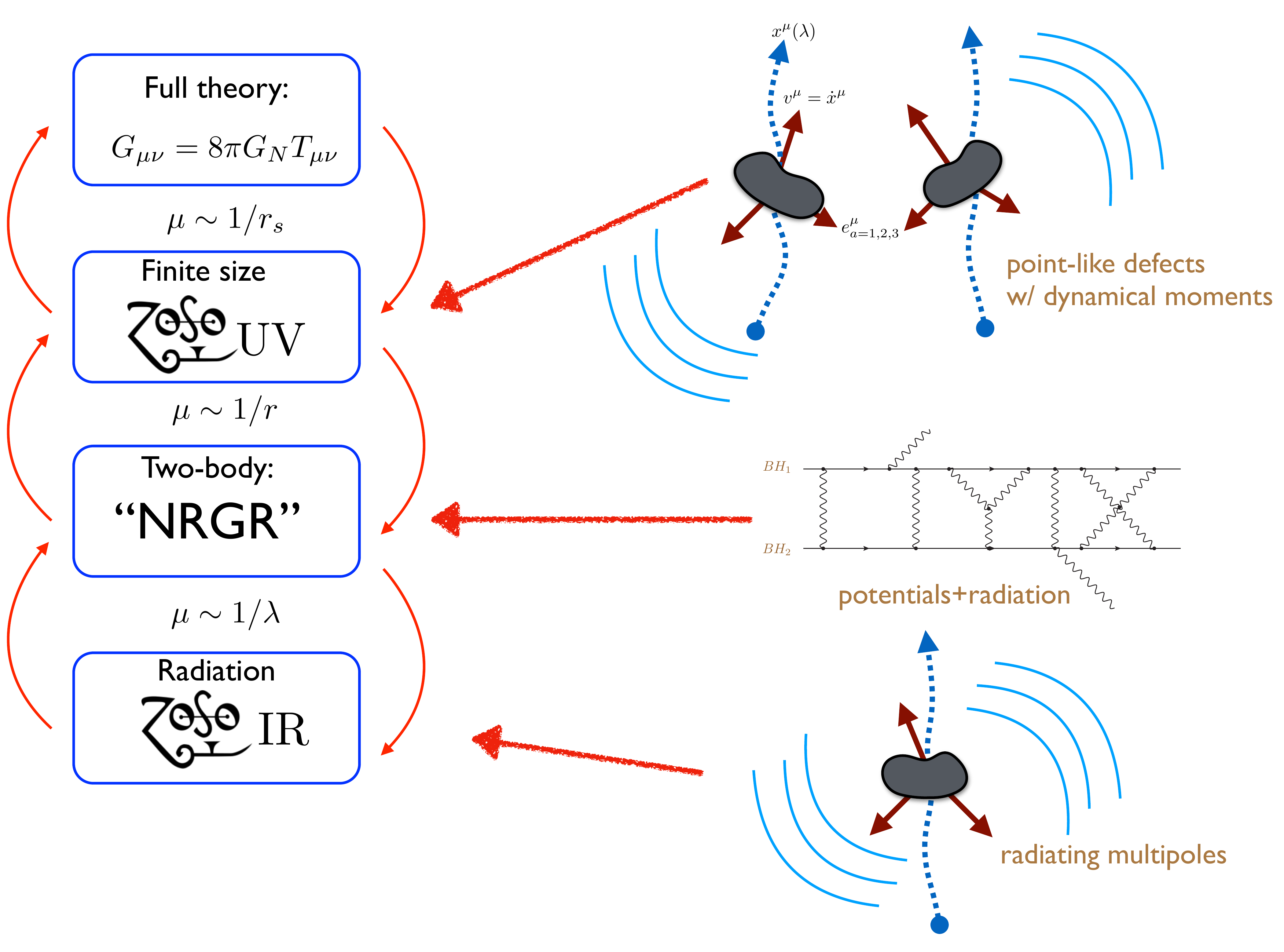}
\end{center}
\caption{Tower of gravity EFTs for non-relativistic compact bound states.}
\label{fig:tower}
\end{figure}
%%%%%%%%%%%%%%%%%%%%%%%%%%%%%%%%%%%%%%%%%%%%%%%%%%%%%%%%%%%%%%%%%%%%%%%%%%%

In order to disentangle the various effects at a given order in $v$, it  is natural to organize the physics in terms of a tower of Wilsonian EFTs of gravity~\cite{Goldberger:2004jt,Goldberger:2006bd}, as depicted in fig.~\ref{fig:tower}.      Reformulating binary dynamics within the framework of EFT then leads to conceptual and technical simplifications, for exactly the same reasons as in the well-established applications of EFT to systems that do not contain gravity (\eg  in high energy physics or condensed matter):
\begin{itemize}
\item \emph{Power counting}:   The Wilson coefficients of local operators in the effective Lagrangian depend only on the UV energy scale $\Lambda_{UV}$ corresponding to modes that propagate over short distances which have been integrated out from the Lagrangian.    Power counting of corrections to observables defined at an IR energy scale $\Lambda_{IR}$, in the expansion parameter $\Lambda_{IR}/\Lambda_{UV}\ll 1$, is manifest.
\item \emph{Analyticity of short distance contributions}:   UV effects are in one-to-one correspondence with local operators in the effective Lagrangian.   Thus at any given order in $\Lambda_{IR}/\Lambda_{UV},$ the most general Lagrangian that is consistent with the symmetries of the relevant degrees of freedom accessible to experiments at energies $\sim\Lambda_{IR}$ necessarily describes the UV physics in a model-independent way.   For suitably defined observables, these short distance contributions depend analytically on the kinematics.
\item  \emph{Renormalization group (RG) evolution}:   Non-analytic contributions, in the form of (potentially) large logarithms $\ln \Lambda_{UV}/\Lambda_{IR}\gg 1$, can be understood as the RG evolution of the EFT Wilson coefficients from a matching scale $\mu\sim \Lambda_{UV}$ where the EFT is defined (by \emph{matching} to a more complete microscopic theory) down to the IR at a scale $\mu\sim\Lambda_{IR}$.  The scaling dimensions of the Wilson coefficients are calculable in the EFT and non-analytic effects in $\Lambda_{IR}/\Lambda_{UV}$ are universal, independent of the detailed microscopic physics that might not yet be experimentally resolvable.
\end{itemize}

The goal of this chapter is to present a detailed overview of the EFT interpretation of binary dynamics, in the regime $r_s/r\ll 1$ where perturbation theory applies.   In sec.~\ref{sec:1B} we integrate out the internal structure of an isolated compact object.  We show how, at distance scales larger than the radius ${\cal R}$, the response to external gravitational fields is systematically encoded in the Wilson coefficients of a local worldline action, suppressed by powers of  ${\cal R}$.   For simplicity we assume in sec.~\ref{sec:1B} that the internal dynamics is gapped, \ie~that there is no absorption or emission of bulk gravitons.

In sec.~\ref{sec:2B} we formulate the two-body problem as a theory of gravitons coupled to the compact object worldlines.   Because of the hierarchy between conservative orbital dynamics at scales $\sim r$, and radiative effects at wavelengths $\lambda\sim r/v\gg r$  in the non-relativistic limit, ensuring manifest scaling in powers of velocity requires the construction of two a priori independent EFTs of gravity, whose structure is discussed in secs.~\ref{sec:NRGR},~\ref{sec:zosoir}  respectively.  One EFT, NRGR (sec.~\ref{sec:NRGR}), is a theory of potential and radiation graviton modes which is operative at distance scales between ${\cal R}$ and $r$, while at distances $\gsim r$ , the system is described by an EFT of non-linear radiation coupled to a set of multipole moments localized on a ``defect'' worldline (sec~\ref{sec:zosoir}).

Sec.~\ref{sec:nonlocal} provides a survey of various types of non-local in time phenomena associated with binary dynamics.   First, in sec.~\ref{sec:rr} we explain how the EFT consistently predicts the backreaction of the emitted gravitational waves (radiation reaction) on the evolution of the non-relativistic orbits.   In sec.~\ref{sec:horizon}, we account for the non-trivial effects associated with the event horizon of a black hole in a binary bound state.    Because the excitations associated with perturbations of the black hole horizon propagate over distance scales of order the Schwarzschild radius $r_s$, finite size effects are no longer gapped, and the local worldline description presented in sec.~\ref{sec:1B} does not correctly capture the IR physics.   Nevertheless, we show in sec.~\ref{sec:horizon} that finite size dissipative effects on the binary system, such as the absorption of energy and momentum by the horizon, and even super-radiant amplification of radiation, can still be described model-independently within an EFT that contains additional  worldline localized degrees of freedom whose coupling to gravitons is constrained by diffeomorphism invariance.   In this section, we also provide a gauge invariant definition of the ``Love numbers'' that characterize the tidal response of a black hole, using the language of (linear) response theory.

While not relevant to phenomenology, the EFTs presented in this review are also capable of capturing \emph{quantum corrections} to gravitationally bound states of black holes.   To illustrate this point, in sec.~\ref{sec:hawkrad} we extend the formalism of sec.~\ref{sec:horizon} to incorporate the effects of Hawking radiation~\cite{Hawking:1974sw} on black hole two-body interactions.   As a simple example, we analyze how the exchange of \emph{virtual} Hawking radiation leads to new calculable features in the inelastic scattering of an elementary particle of mass $\ll m_{\Pl}$ by a Schwarzschild black hole, giving rise to effects at the same order in $(E_{\CM}/m_{\Pl})^2$ as the leading quantum corrections to scattering in quantum gravity due to graviton loops, of the sort first studied in~\cite{Donoghue:1993eb,Donoghue:1994dn,Bjerrum-Bohr:2002gqz}.

My hope is that this review will give the reader a sense of how the tower of EFTs depicted in fig.~\ref{fig:tower} gives a complete description of compact binaries in the perturbative regime, including all effects between the size of the compact objects themselves up to the scale of the radiation.    It is beyond the scope here to provide full technical details of the calculations that have been performed using the EFT.   There exist already several review articles that treat the various technical aspects in more details, see~\cite{Goldberger:2006bd,Goldberger:2007hy,Foffa:2013qca,Porto:2016pyg,Levi:2018nxp}.   Finally, I apologize in advance that space limitations prevent me from giving here a complete guide to the vast literature on the subject.  For example, not discussed here at all are the very recent applications of the EFT to the calculation of PM scattering observables, a topic at that has drawn together the scattering amplitude, effective field theory, and traditional general relativity communities.  A review of this rapidly developing subject can be found in~\cite{Buonanno:2022pgc}.   Similarly I refer to the Snowmass 2021 article~\cite{Goldberger:2022ebt}, which provides an up-to-date and exhaustive compilation of references on EFTs of gravity and their application to gravitational wave sources.  

\emph{Cross-references}:   
A general review on the description of low energy quantum gravity as an effective quantum field theory can be found in the contribution of J. Donoghue to this volume.    See also the review articles~\cite{Burgess:2003jk,Donoghue:1995cz}.  The chapter by E. Bjerrum-Bohr \etal ~provides a detailed description of the Feynman rules for perturbative gravity expanded about flat spacetime (see also the section \emph{Perturbative Quantum Gravity} edited by I. Shapiro).   The contribution by C. P. Burgess~\etal~has intellectual overlap with the discussion of dissipation and black hole horizons in sec.~\ref{sec:horizon}

\section{The one-body sector}
\label{sec:1B}

We begin by constructing an EFT that characterizes the low-frequency response of an isolated compact astrophysical object to external gravitational perturbations.    To that end, we imagine that we start first with the system in isolation. The details of its internal shape or composition depend sensitively on the microscopic theory.  We assume in this review that the microscopic theory consists of GR coupled to the Standard Model (SM) of strong plus electroweak interactions.  In this case, the compact object is, by definition, just some complicated many-body equilibrium state ${\hat\rho}$ with average total (ADM~\cite{Arnowitt:1962hi}) energy $M$, angular momentum $J$ and either zero (a black hole\footnote{We assume black holes with $Q_{em}=0$ in this review.}) or very large baryon number.   Although the focus in on the SM, the methods that we introduce in this review have been generalized (see \cite{Goldberger:2022ebt}  for a complete set of references) to include possible extensions of the SM that carry additional (e.g dark matter) fields and therefore a richer zoology of compact stellar objects.

Regardless of the details of the internal structure, any type of self-gravitating distribution of matter will appear at long distances much larger than its radius ${\cal R}$ to be approximately point-like, with a well-defined  ``center-of-mass" worldline $x^\mu(s)$.  For example, the long distance gravitational field of the isolated object has the same universal form $\sim G_N M/|{\vec x}|$ at spatial infinity, indistinguishable from a static point particle at the origin of the coordinate system. By going to large but finite distance, the gravitational field also encodes the angular momentum $J$ as well as other multipole moments which do depend on the precise microscopic state ${\hat\rho}$.  From the point of view of distant observers, these can also be described in terms of degrees of freedom localized on the defect worldline $x^\mu(s)$ (see secs.~\ref{sec:zosoir}  and~\ref{sec:horizon}).

Next, we consider how the equilibrium state of the compact object responds to external gravitational perturbations.  Physically, we might imagine these perturbations to correspond to, \eg on-shell gravitons coming in from past null infinity ${\cal I}^-$ and scattering off the object out to future null infinity ${\cal I}^+$, or perhaps to massive particles incoming from past timelike infinity $i^-$that get caught in the object's gravitational field and generate, via off-shell exchange, tidal deformations of its shape.   Within the point particle description, we can think of such probes generically as if we were turning on some gravitational field $g_{\mu\nu}$ with the appropriate boundary conditions at infinity, which interacts with the compact object.    As long as the curvature length scale ${\cal L}$ associated with $g_{\mu\nu}$ is large compared to the size of the object $\sim {\cal R}$, we can continue to describe the response of the object systematically within a worldline EFT, as an expansion in powers in $G_N M/{\cal R}\ll 1$ and $L/{\cal R}\ll 1$.

In this section, we assume for simplicity that the internal dynamics of the compact object is `gapped' at some frequency scale much larger than the scale $1/{\cal L}$ set by the curvature.  In this limit, dissipation, \eg ~the possibility of absorption of gravitational energy-momentum by the object, is suppressed.   We will come back to the inclusion of such non-conservative effects later on in sec.~\ref{sec:horizon}.  Thus, by assumption, the relevant degrees of freedom in the IR consist of:
\begin{itemize}
\item The spacetime metric $g_{\mu\nu}(x)$.
\item A worldline $x^\mu(s)$, describing the center-of-mass motion of the object.
\item A local Lorentz frame frame $e^\mu{}_a(s)$,
\begin{eqnarray*}
g_{\mu\nu}(x(s)) e^\mu{}_a e^\nu{}_b=\eta_{ab} , & \eta^{ab}  e^\mu{}_a e^\nu{}_b=g_{\mu\nu}(x(s)),
\end{eqnarray*}
at each point $x^\mu(s)$ along the worldline.   This frame describes the orientation of the compact object relative to distant observers, in particular the object's rotational velocity is
\begin{equation}
\label{eq:om}
\Omega^{ab} = g^{\mu\nu} e^a{}_\mu \left({\dot x}^\rho \nabla_\rho\right) e^b{}_\nu = -\Omega^{ba}.
\end{equation}
\end{itemize}

The local frame $e^\mu{}_a(s)$ is necessary to describe objects with non-zero spin.    In the absence of gravity, it was introduced by Regge and Hanson~\cite{Hanson:1974qy} to treat the classical motion of relativistic spinning particles coupled to electromagnetic fields.   In flat spacetime, the worldline degrees of freedom $(x^\mu,e^\mu{}_a)$ parameterize points on the Poincare group of isometries, which are in general \emph{spontaneously broken} by the presence of the compact object.   The extension of the Regge-Hanson formalism to curved spacetime, and its applications to perturbative binary dynamics first appeared in~\cite{Porto:2005ac}.    A related treatment of spinning particles from the point of view of non-linearly realized symmetries and Goldstone's theorem can be found in ref.~\cite{{Delacretaz:2014oxa}}.    While the inclusion of spin effects is crucial for constructing gravitational wave templates relevant to phenomenology, for space reasons we will omit any detailed discussion in this review article.   Up-to-date reviews of spin effects in the worldline EFT can be found in refs.~\cite{Porto:2016pyg,Levi:2018nxp}.

There is a gauge redundancy in the variables $(g_{\mu\nu},x^\mu(s),e^a{}_\mu(s))$ that define the EFT, namely:
\begin{enumerate}
\item Spacetime diffeomorphism invariance:   $\delta x^\mu = \xi^\mu(x)$, $\delta g_{\mu\nu} = 2\nabla_{(\mu}\xi_{\nu)},$ $\delta e^\mu{}_a(s) = [\xi,e_a]^\mu(x(s))$, \etc
\item Reparameterizations $s\rightarrow {\tilde s}={\tilde s}(s)$ of the worldline time coordinate.
\end{enumerate}
We assume that there are smooth limits $g_{\mu\nu}\rightarrow\eta_{\mu\nu}$, or  ${\cal R}\rightarrow 0$.    We also assume that Wilsonian decoupling of UV physics holds, guaranteeing that whatever the microscopic description, the low-frequency response of the isolated compact object is described by a local effective theory of the generic form
\begin{equation}
\label{eq:lights}
S=S_{EH}[g_{\mu\nu}] + S_{pp}[g_{\mu\nu},x^\mu(s),e^\mu{}_a(s)],
\end{equation}
where $S_{EH}$ is the bulk gravity theory Eq.~(\ref{eq:EH}) , and $S_{pp}[g,x(s),e(s)]$ is a term localized on the worldline.  It is an infinite sum of gauge invariants constructed from the spacetime curvature and its derivatives, with Wilson coefficients that, by dimensional analysis, scale as successively larger powers of the radius ${\cal R}$.

The Wilson coefficients in $S_{pp}$ are free parameters from the point of view of the low energy theory, to be determined by matching to the UV theory, as we describe in more detail below.   However, even if the microscopic description is unknown, the EFT still has predictive power.   Since spacetime or worldline derivatives are small,  $\partial_\mu,d/ds\ll {\cal R}^{-1}$, $S_{pp}$ may be truncated at a fixed order in the derivative expansion at the expense of introducing (presumably small) errors suppressed by powers of ${\cal R}/L\ll 1$.   Therefore to calculate an observable to finite precision in the EFT, one only needs to know a finite set of Wilson coefficients, which can be regarded as experimental inputs.

The expansion of $S_{pp}$ up to second order in spacetime derivatives takes the form 
\begin{equation}
\label{eq:pp}
S_{pp} = S^{(0)}_{pp} + S^{(1)}_{pp} + S^{(2)}_{pp} + \cdots,
\end{equation}
where the unique leading (zero derivative) term is proportional to the proper time $d\tau=\sqrt{g_{\mu\nu} dx^\mu dx^\nu}$ elapsed along the worldline,
\begin{equation} 
S^{(0)}_{pp}=-m\int d\tau,
\end{equation}
with $m>0$ a real parameter with dimensions of mass.   In the ${\cal R}\rightarrow 0$ limit, we ignore the backreaction of the compact object on the spacetime metric, in which case the parameter $m$ becomes irrelevant and the leading order equations of motion for $x^\mu$ is simply the geodesic equation
\begin{equation}
\delta S_{pp}^{(0)}=0\Rightarrow a^\mu = {dx^\nu\over d\tau} \nabla_\nu  {dx^\mu\over d\tau}=0,
\end{equation}    
as expect on the basis of the Einstein Equivalence Principle.

On the other hand when we include two-body interactions in sec.~\ref{sec:2B}, we will need to account for the gravitational field sourced by the compact object, in which case the parameter $m$ does not drop out of the dynamics.   To fix the precise dependence of the EFT mass parameter $m$ on the microscopic properties of the compact we must perform a \emph{matching calculation} to the full UV theory.    This is accomplished by computing the same physical observable in both the EFT and in the UV theory, adjusting the EFT parameters so that both results agree in the overlapping regime of validity of the two theories.   

For the case of the mass parameter $m$, a convenient quantity to match is simply the graviton one-point function sourced by the compact object, probed by observers at spatial infinity.   In full general relativity, we choose deDonder coordinates $x^\mu$ such that the metric takes the form
\begin{eqnarray}
  g_{\mu\nu}(x) = \eta_{\mu\nu}+h_{\mu\nu}(x), &  \partial_\nu h^{\mu\nu} = {1\over 2}\partial^\nu h,
  \end{eqnarray}
$(h=h^\rho{}_\rho=\eta^{\rho\sigma} h_{\rho\sigma})$ everywhere on spacetime.  The graviton field $h_{\mu\nu}(x)$ need not be small, but, by the Einstein equations, it falls off at long distances from the compact object~\cite{Weinberg:1972kfs}.  In particular, at $|{\vec x}|\rightarrow\infty$, the gravitational field of the compact object becomes
\begin{equation}
\label{eq:fullh}
\lim_{r\rightarrow\infty} h^{Full}_{\mu\nu}(x) = {2 G_N M_{ADM}\over |{\vec x}|} \left(\eta_{\mu\nu} + u_\mu u_\nu\right)
\end{equation}
in the center-of-mass frame (CM), in which the object's four-momentum is 
\begin{equation}
\label{eq:pcm}
P^\mu = (M_{ADM},0,0,0) \equiv M_{ADM} u^\mu,
\end{equation}
where $M_{ADM}$ is the ADM mass of the compact object.

In the EFT, we instead solve the deDonder gauge linearized Einstein equations, taking the source term to be
\begin{equation}
T^{\mu\nu}_{pp} = {2\over \sqrt{g}} {\delta\over\delta g^{\mu\nu}} S^{(0)}_{pp} = m u^\mu u^\nu \delta^3({\vec x}),
\end{equation}
which corresponds to a point defect at rest at the origin.  This yields the result,
\begin{equation}
\label{eq:1pt}
h_{\mu\nu}^{EFT}(x) =-{i\over 2 m_{\Pl}^2} \int d^4 x' D_R(x-x') P_{\mu\nu;\rho\sigma} T^{\rho\sigma}_{pp}(x') \approx  {2 G_N m\over |{\vec x}|} \left(\eta_{\mu\nu} + u_\mu u_\nu\right),
\end{equation}
where the graviton propagator in deDonder gauge is a product of the massless retarded Green's function,
\begin{equation}
-i D_R(x) =\int {d^4 k\over (2\pi)^4}  { e^{-i k \cdot x}\over (k^0+i0^+)^2 -{\vec k}^2}
\end{equation}
and the tensor structure is $P_{\mu\nu;\rho\sigma} = {1\over 2}\left[\eta_{\mu\rho}\eta_{\nu\sigma} + \eta_{\mu\sigma}\eta_{\nu\rho} - \eta_{\mu\nu}\eta_{\rho\sigma}\right]$ in four spacetime dimensions.   Eq.~(\ref{eq:1pt}) is the gravitational field in the EFT, valid up to corrections from higher derivative operators in $S_{pp}$ that fall off like more powers of $1/r$.

Because the full GR and EFT calculations are performed in the same coordinate system, in this example, matching is just the statement that, at long distances $|{\vec x}|\rightarrow\infty$, where both descriptions are equally valid, $h_{\mu\nu}^{Full}=h_{\mu\nu}^{EFT},$ which yields,
\begin{equation}
\label{eq:ADM}
m = M_{ADM}.
\end{equation}
We conclude that in EFT, the point particle mass should be identified with the ADM mass of the compact object in isolation, \ie~not subject to external fields.   This is of course what we would intuitively expect, but the point of this (perhaps somewhat pedantic) exercise is to illustrate that there is a systematic procedure for relating the physical properties of the compact object to the parameters of its worldline proxy in the EFT.

The matching condition Eq.~(\ref{eq:ADM}) can in principle receive corrections on the EFT side in powers of $G_N$, \eg~from the self-energy of the static field sourced by the worldline.    In the EFT, such effects in general carry short distance singularities, and depend on the choice of UV regulator, \eg~dimensional regularization sets such power UV divergent contributions to zero.   While the UV behavior of the EFT differs from that in the full theory, any discrepancy can be compensated by renormalizing the local counterterms (Wilson coefficients) in $S_{pp}$  order-by-order in perturbation theory

Because the leading order term in $S_{pp}$ is universal, to resolve the internal structure of the object we need to keep the higher derivative terms.   For compact objects, with $\kappa\gsim {\cal O}(1)$ we expect that by dimensional analysis, the  Wilson coefficients of terms with $n$-derivatives should scale as $m{\cal R}^n$ up to order unity numerical factors.    To simplify the discussion, I will assume in the rest of this section that the compact object does not carry any permanent multipole moments, encoded in the EFT as local Lorentz tensors on the worldline, in its equilibrium state.   For example, a Kerr black hole is characterized by an infinite tower of multipoles, all proportional to powers of its spin~\cite{Hansen:1974zz}, which imply a rich structure of worldline couplings that we ignore here, see~\cite{Levi:2018nxp} for a detailed review of such effects.

To construct the full set of invariants at a given order in the derivative expansion, one may use the equations of motion to eliminate ``redundant'' terms~\cite{Georgi:1991ch}.  For example, the zeroth order worldline equations of motion imply that $a^\mu=0$, so that terms involving the acceleration can be omitted from $S_{pp}$.   Similarly, the leading Einstein equations imply that the Ricci curvature $R_{\mu\nu}$ can be traded for a contact term localized on the particle worldline,
\begin{equation}
\label{eq:ric}
R_{\mu\nu}(x) = 8\pi G_N m \int d\tau {\delta^4(x-x(\tau))\over\sqrt{g}} \left(u^\mu u^\nu - {1\over 2} g^{\mu\nu}\right).
\end{equation}
As consequence of the equations of motion, we may therefore assume that $S^{(1)}_{pp}=0,$ while at the two-derivative level we can drop operators constructed out of the Ricci curvature, \eg
\begin{equation}
\int d\tau R(x(\tau)), \int d\tau u^\mu u^\nu R_{\mu\nu}(x(\tau)),
\end{equation}
since by Eq.~(\ref{eq:ric}) these can be absorbed into the definition of the EFT parameter $m$.    Thus by performing field redefinitions, which have no effect on the physical predictions, we can assume that $S^{(2)}_{pp}=0$ as well.

The first genuinely physical finite size corrections only appear at the fourth derivative order.   Using the algebraic properties of the Riemann tensor in four spacetime dimensions, the most general
 structure consistent with our assumptions can be expressed as
\begin{equation}
\label{eq:s4}
 S^{(4)}_{pp} = c_E m{\cal R}^4 \int d\tau E_{\mu\nu} E^{\mu\nu} + c_B m{\cal R}^4 \int d\tau B_{\mu\nu} B^{\mu\nu}  + c_{EB} m{\cal R}^4 \int d\tau E_{\mu\nu} B^{\mu\nu}
\end{equation}
for objects that do not carry spin.   In this equation $E_{\mu\nu}$ and $B_{\mu\nu}$ are the ``electric'' and ``magnetic'' components of the Weyl tensor relative to the particle worldline,
\begin{eqnarray}
E_{\mu\nu} &=& W_{\mu\rho\nu\sigma} u^{\rho} u^\sigma,\\
B_{\mu\nu} &=& {\tilde W}_{\mu\rho\nu\sigma} u^{\rho} u^\sigma,
\end{eqnarray}
where ${\tilde W}_{\mu\nu\rho\sigma} = {1\over 2} \epsilon_{\mu\nu\alpha\beta} W^{\alpha\beta}{}_{\rho\sigma}$ is the dual curvature.   (We will use the terms ``Weyl'' and ``Riemann" interchangeably in light of the fact that they are equivalent on-shell).   Using the algebraic properties of the Weyl tensor, one can show that $E_{\mu\nu}$, and $B_{\mu\nu}$ are symmetric, traceless $E^\mu{}_\mu=B^\mu{}_\mu=0$ and transverse to the velocity $E_{\mu\nu} u^\nu = B_{\mu\nu} u^\nu = 0,$ so that they indeed encode all ten independent components of the Weyl tensor.

We have defined the dimensionless Wilson coefficients in Eq.~(\ref{eq:s4}) in such a way that, for relativistic compact objects $c_E\sim c_B\sim {\cal O}(1)$.   On the other hand, it is possible to define a parity transformation, such that any given instant $\tau$ in a comoving frame  $u^{\hat\mu}=(1,0,0,0)$ centered at $x^{\hat\mu}(\tau)\equiv 0$, acts as
\begin{equation}
P: E_{{\hat i}{\hat j}}(0)\rightarrow +E_{{\hat i}{\hat j}}(0), B_{{\hat i}{\hat j}}(0)\rightarrow -B_{{\hat i}{\hat j}}(0).
\end{equation}
This first two terms in Eq.~(\ref{eq:s4}) are even under $P$ while the last term is parity odd.   For neutron stars, whose microscopic properties are well described by the physics of the SM, parity violating effects are expected to be highly suppressed, $|c^{NS}_{EB}|\ll 1.$.  On the other hand, more exotic compact objects predicted by extensions of the SM, \eg~containing a ``dark sector'' with sizeable parity violation, may be characterized by a Wilson coefficient  $c_{EB}\sim {\cal O}(1)$, see~\cite{Modrekiladze:2022ioh}.

If the coefficients in Eq.~(\ref{eq:s4}) are non-zero, the compact's object motion is no longer geodesic.   In fact $c_{E,B}\neq 0$ characterize the static tidal quadrupolar response of the compact object to external gravitational fields.  We can see this intuitively, by putting a small ``moon'' in orbit about the compact object, corresponding to a weak external Newtonian potential $\Phi_{ext}({\vec x},t)$.   This external potential induces tides on the object, distorting its shape away from perfect spherical symmetry.    This deformation can then be detected by far away inertial observers, by measuring the long distance gravitational field that it produces.

We write the total Newtonian gravitational potential as
\begin{equation}
\Phi = \Phi_{co} + \Phi_{ext} + \delta \Phi,
\end{equation}
where $\Phi_{co}=-G_N m/|{\vec x}|$ is the unperturbed long distance field produced by the compact object, and $\delta \Phi$ is the response, \ie~the potential generated by the tidal bulge generated by the moon orbiting the object.   Taking  the compact object to be at rest at ${\vec x}=0$, and that the moon's orbit is non-relativistic, with velocity $v\ll 1$ we have 
\begin{equation}
E_{ij}\approx W_{0i0j}=-\partial_i \partial_j \Phi+\cdots
\end{equation}
and $B_{ij}\sim {\cal O}(v)\cdot  E_{ij}$, so that, in the linear response approximation,
\begin{equation}
\label{eq:s4nr}
S^{(4)}_{pp}\approx m {\cal R}^4 c_E \int dt (\partial_i \partial_j\Phi(0,t))^2 \supset 2 m {\cal R}^4 c_E \int dt  \partial_i \partial_j\Phi_{ext}(0,t)) \partial^i \partial^j \delta\Phi(0,t)+\cdots
\end{equation}
We have dropped a (UV divergent) self-energy term induced by the insertion of  $\Phi_{co}({\vec x}=0)$ into $S^{(4)}_{pp}$, which can be be absorbed into the definition of $m$, as well as a term of order $(\delta\Phi)^2$ which is assumed to be small.

To observers far away, the coupling to the induced  field has the form of the Newtonian gravitational interaction,
\begin{equation}
{\cal L}_{Newt}=- \rho_{\ell=2}({\vec x},t) \delta\Phi, 
\end{equation}
sourced by a pointlike mass quadrupole distribution located at the origin
\begin{equation}
\rho_{\ell=2}({\vec x},t)={1\over 2} I^{ij}(t)\partial_i\partial_j \delta^3({\vec x}),
\end{equation}
%\Rightarrow \delta\Phi = -G_N\int d^3{\vec x}' {\rho_{\ell=2}({\vec x}',t)\over |{\vec x}-{\vec x}'|} = -{1\over 2} G_N Q^{ij}(t) \partial_i \partial_j {1\over |{\vec x}|}
which we can read off Eq.~(\ref{eq:s4nr}),
\begin{equation}
\label{eq:qe}
I^{ij}(t) = 4 c_E m {\cal R}^4  E_{ext}^{ij}(0,t)
\end{equation}
$E_{ext}^{ij}=-\partial^i \partial^j \Phi_{ext}$.   The induced quadrupole moment has the precise form that one would expect for the linear gravitational perturbation of a spherically symmetric distribution whose internal dynamics is gapped, so that the response to  slowly varying external fields is instantaneous and linearly proportional to the perturbing field.

 For a nearly static and weakly self-gravitating Newtonian mass distribution, it is conventional for historical reasons to denote the constant of proportionality between the external tidal field $\partial_i \partial_j \Phi_{ext}$ and the induced mass quadrupole as the ($\ell=2$) static \emph{Love number} $k_{\ell=2}$ of the system (following the definition in~\cite{Binnington:2009bb}),
\begin{equation}
\label{eq:inq}
I^{ij}(t) = {2\over 3} {{\cal R}^5\over G_N}k_{\ell=2}E_{ext}^{ij}(t,0).
\end{equation}
So by matching the linear response in the EFT to that predicted by Newtonian geo-elasticity theory (the full theory), we learn that the Wilson coefficient $c_E$ describes the tidal susceptibility in the worldline EFT,
\begin{equation}
c_E = {1\over 6} \left({{\cal R}\over G_N m}\right) k_{\ell=2}.
\end{equation}

This interpretation of the linear response generalizes straightforwardly to the case of $\ell>2$ induced mass multipole moments, as well as gravitomagnetic interactions, which in general induce ``current'' multipole moments (i.e, moments of the angular momentum density) for each $\ell\geq 2$.   In the worldline EFT, the static linear response of a fully relativistic spherically symmetric object is encoded in the higher-derivative interactions
\begin{eqnarray*}
m {\cal R}^{2 \ell} \int d\tau \left(\nabla^\perp{}^{\langle\alpha_1}\cdots  \nabla^\perp{}^{\alpha_{\ell-2}\rangle}E^{\mu\nu}\right)\left(
\nabla^\perp_{\langle\alpha_1}\cdots  \nabla^\perp_{\alpha_{\ell-2}\rangle}E_{\mu\nu}\right),\\
m {\cal R}^{2\ell} \int d\tau \left(\nabla^\perp{}^{\langle\alpha_1}\cdots  \nabla^\perp{}^{\alpha_{\ell-2}\rangle}B^{\mu\nu}\right)\left(
\nabla^\perp_{\langle\alpha_1}\cdots  \nabla^\perp_{\alpha_{\ell-2}\rangle}B_{\mu\nu}\right),\\
m {\cal R}^{2 \ell} \int d\tau \left(\nabla^\perp{}^{\langle\alpha_1}\cdots  \nabla^\perp{}^{\alpha_{\ell-2}\rangle}E^{\mu\nu}\right)\left(
\nabla^\perp_{\langle\alpha_1}\cdots  \nabla^\perp_{\alpha_{\ell-2}\rangle}B_{\mu\nu}\right),
\end{eqnarray*}
where the transverse covariant derivative is $\nabla^\perp_\mu = \nabla_\mu - u_\mu (u\cdot \nabla)$ and $\langle\cdots\rangle$ denotes the symmetric traceless projection of the enclosed indices.   

Going beyond the static approximation, the linear response is also characterized by terms involving time derivatives ${\dot E}_{\mu\nu} = u^\rho\nabla_\rho E_{\mu\nu}$, or ${\dot B}_{\mu\nu}$,  \eg~a term of the form
$$
m{\cal R}^6 \int d\tau {\dot E}_{\mu\nu} {\dot E}^{\mu\nu}
$$
acts as a high-pass filter for electric perturbations. It is even also possible to incorporate effects beyond the linear response approximation.   For instance a term $m{\cal R}^6\int d\tau E^\mu{}_\nu E^\nu{}_\rho E^\rho{}_\mu$ on the worldline reflects non-linear couplings between gravity and the UV modes that have been integrated out of the full theory.

For fully relativistic compact objects which have strong internal gravity, the notion of  induced multipole moments as in Eq.~(\ref{eq:inq}) in the full theory (GR) is not invariant under diffeomorphisms.   Instead, the Love numbers are \emph{defined} as the Wilson coefficients of an effective Lagrangian, \ie
\begin{equation}
\label{eq:ldef}
S^{Love}_{\ell=2}\equiv {1\over 6} k^E_{\ell=2} \left({{\cal R}^5\over G_N}\right)\int d\tau E_{\mu\nu} E^{\mu\nu}  +{1\over 6} k^B_{\ell=2} \left({{\cal R}^5\over G_N}\right)\int d\tau B_{\mu\nu} B^{\mu\nu} ,
\end{equation}
and similarly on for $\ell>2$.   This definition has the advantage of being fully gauge invariant under coordinate transformations.

Matching the Love numbers in the EFT Lagrangian to full GR is achieved by comparing gauge invariant observables.  For this purpose, a conceptually clean (in principle) observable is the quantum mechanical probability amplitude for elastic $1\rightarrow 1'$ elastic graviton scattering off the compact object~\cite{Goldberger:2007hy}.   One would calculate the amplitude in the full theory, by linearizing the Einstein equations around the background field ${\bar g}_{\mu\nu}$ sourced by the compact object, with asymptotic boundary conditions
\begin{equation}
\lim_{r\rightarrow\infty}h_{\mu\nu}(x)\rightarrow \epsilon^h_{\mu\nu}(k) e^{-ik\cdot x} + {{\cal A}_{\mu\nu}\over r} e^{-i\omega (t-r)},
\end{equation}
corresponding to an incoming plane wave with four-momentum $k^\mu$ and helicity $h$ from past null infinity ${\cal I}^-$ and an outgoing (scattered) spherical wave at future infinity ${\cal I}^+$ with frequency $\omega=k^0$.  Because the ingoing/outgoing waves have direct physical meaning as graviton asymptotic states, the scattering amplitude ${\cal A}_{\mu\nu}$ is gauge invariant, and can be compared to the same quantity in the EFT.

In the EFT, one calculates the amplitude  ${\cal A}_{\mu\nu}$ for an incoming plane wave of definite helicity $h$ using the Feynman rules from Eq.~(\ref{eq:s4}) expanded about flat space.   In addition to terms where the graviton scatters off the mass monopole, $m$, there is a term in the scattering amplitude from the contact terms in Eq.~(\ref{eq:s4}), \eg~the electric Love operator contributes a term of the form
\begin{equation}
{\cal A}_{\mu\nu} \supset k^E_{\ell=2} {\cal R}^5 \omega^4 \epsilon^{h}_{\mu\nu}(k),
\end{equation}
since each insertion of $E_{\mu\nu}$ acting on an asymptotic state brings down a factor of $\omega^2$.    The short distance part of the amplitude to scatter into a final state plane wave of helicity $h'$ is therefore proportional to $\omega^4\epsilon^{-h'}_{\mu\nu}(k')\epsilon^{h}{}^{\mu\nu}(k)$.

In general, the Love numbers are dependent on the form of the equation of state of the compact star.  They were studied first in refs.~\cite{Flanagan:2007ix} for the case of neutron stars, by analyzing the perturbations to a background Oppenheimer-Volkoff model of relativistic stellar structure.  See also~\cite{Hinderer:2007mb,Hinderer:2009ca}.   For Schwarzschild black holes, the situation is somewhat trickier, for two reasons:  (i).  Technically ${\cal R}^5/G_N\propto G_N^4 m^5$, so that to match the EFT one must in principle subtract contributions to the amplitude that involve multiple (up to five) insertions of the mass $m$. (ii).  Due to the presence of the event horizon, the low frequency linear response of a black hole is necessarily absorptive.   Because the local in-time point particle Lagrangian Eq.~(\ref{eq:pp}) cannot describe dissipative finite size effects, additional degrees of freedom need to be added to the worldline theory.   Therefore, we postpone the topic of black hole response until the last part of this article, see sec.~\ref{sec:horizon}.

Regardless of the specific numerical value of $k^{E,B}_{\ell=2}$, the observation that $\ell=2$ tidal deformability scales like ${\cal R}^5$ implies, by dimensional analysis, that tidal corrections in a compact binary in a non-relativistic orbital of radius $r\gg r_s$ do not enter until at least the relative order 
\begin{equation}
({\cal R}/r)^5\sim ({\cal R}/r_s)^5 \times (G_N m/r)^5\sim ({\cal R}/r_s)^5 v^{10},
\end{equation}
which is formally a 5PN effect, although possibly enhanced~\cite{Flanagan:2007ix} for compact objects such as neutron stars with ${\cal R}/G_N m\sim {\cal O}(10)$.    We have established, using EFT reasoning (symmetries, power counting)  the ``Effacement Principle''~\cite{blanchet} that non-dissipative finite size corrections cannot appear until 5PN order in the non-relativistic limit.   Thus in order to learn about the internal structure of compact objects during the adiabatic inspiral phase, one needs waveform templates that take into account the dynamics of point particles to 5PN beyond the quadrupole radiation formula.   We turn to a review of such corrections next.

\section{Perturbative binary dynamics}
\label{sec:2B}

\subsection{Setup}

We now consider a binary merger of compact objects, restricting ourselves to the regime in which the typical orbital distance $r$ is large, $r\gg {\cal R}\gsim r_s$.   In this phase, the system is well described by Einstein gravity coupled to the point particle action $S_{pp}$, truncated at some fixed order in the derivative expansion.  The goal is to predict the classical gravitational waveform $h_{\mu\nu}=g_{\mu\nu}-\eta_{\mu\nu}$ measured by detectors at ${\cal I}^+$, as well as the flux of energy, momentum and angular momentum as a perturbative expansion in the small quantities ${\cal R}/r$ and $r_s/r$.

Even though we are mainly interested in solving a classical problem, it turns out to be convenient to set it up as the $\hbar\rightarrow 0$ limit of a quantum field theory calculation.   The graviton $h_{\mu\nu}$ is treated as a propagating quantum field, but the binary constituents are classical worldline sources with zero quantum fluctuations.  In this picture, the waveform then corresponds to an expectation value 
$$
\langle in| h_{\mu\nu}(x)|in\rangle,
$$
evaluated in the initial vacuum state of the radiation field and of the binary constituents.     As was first emphasized in ref.~\cite{Galley:2009px}, the appropriate path integral formalism for computing this expectation value, one in which we hold the initial state fixed but sum over the final states of the radiation field, is the ``Schwinger-Keldysh''~\cite{Schwinger:1960qe,keldysh}  closed time path (CTP) or ``in-in" version of the functional integral.   (See~\cite{Calzetta:2008iqa} for a review of the in-in formalism for quantum mechanics and field theory).     This is analogous to the situation in cosmology, where late time correlations are measured in a given initial state~\cite{Weinberg:2005vy}.   

(As an aside, it is also possible to formulate the radiation problem as an $S$-matrix calculation~\cite{Kosower:2018adc}, in which both the graviton and the massive particles are dynamical, and one uses time-ordered propagators in the Feynman rules.   As in the worldline EFT, such an approach is restricted to observables that are calculable in perturbation theory, \ie~to times well before the coalescence of the binary into (presumably) a final state black hole.   The IR safe observables are not the individual $S$-matrix elements but rather semi-inclusive averages that sum over all the unobserved final asymptotic states of the system.   Evaluating such sums via unitarity cuts is equivalent to using in-in Feynman rules, where the cuts correspond to Wightman propagator exchange between sources on opposite sides of the closed time path, see Eq.~(\ref{eq:wight}).)

To calculate observables, we introduce a generating function, the in-in effective action $\Gamma[x_A,{\bar g};{\tilde x}_A {\tilde{\bar g}}]$, defined by the path integral expression
 \begin{equation}
\label{eq:inin}
e^{i\Gamma[x_A,{\bar g};{\tilde x}_A {\tilde{\bar g}}]} =\int {\cal D} h_{\mu\nu}(x) {\cal D}{\tilde h}_{\mu\nu}(x) e^{i S[{\bar g},h,x_A] - i S[{\tilde {\bar g}},{\tilde h},{\tilde x}_A]}.
\end{equation}
Here, the classical action is a functional $S[{\bar g},h,x_A]=S_{EH}[{\bar g}+h] + S_{pp}[{\bar g}+h,x_A]+S_{GF}[{\bar g},h]$ of the integration variable $h_{\mu\nu}$, as well as the worldlines $x^\mu_{A=1,2}(\tau)$, which are held fixed to arbitrary values in the computation of the integral.   The action also depends on a $c$-number background gravitational field ${\bar g}_{\mu\nu}$ which is also held fixed in the course of evaluating Eq.~(\ref{eq:inin}).  

 It is convenient to employ the background field method~\cite{DeWitt:1967ub,Abbott:1980hw}, in which the gauge fixing term $S_{GH}[{\bar g},h]$ is  chosen to preserve diffeomorphisms acting on ${\bar g}_{\mu\nu}$.   For instance, the choice $S_{GF}=m_\Pl^{d-2}\int d^d x \sqrt{\bar g}{\bar g}^{\mu\nu} \Gamma_\mu \Gamma_\nu,$ with $\Gamma_\mu ={\bar\nabla_\rho} h^\rho{}_\mu-{1\over 2} {\bar\nabla}_\mu h^\rho{}_\rho$ is background diffeomorphism invariant, and generates graviton propagators whose Lorentz tensor structure in $d$ spacetime dimensions is the standard one
\begin{equation}
P_{\mu\nu;\rho\sigma}= {1\over 2} \left[\eta_{\mu\rho}\eta_{\nu\sigma} + \eta_{\mu\sigma}\eta_{\nu\rho} - {2\over d-2} \eta_{\mu\nu}\eta_{\rho\sigma}\right].
\end{equation}
In the gauge specified by $S_{GF}$, it is also in principle necessary to introduce Fadeev-Popov ghost fields to ensure gauge invariance of the effective action.   However, in the $\hbar\rightarrow 0$ limit which is the main focus of this review, ghost loop contributions to the path integral are subleading, and may be ignored.

 Implicit in the integration measure of Eq.~(\ref{eq:inin}) is a choice of boundary conditions, corresponding to the vacuum in the far past for both $h_{\mu\nu},{\tilde h}_{\mu\nu}$.   In the far future, the boundary condition is that $h_{\mu\nu}={\tilde h}_{\mu\nu}$, so roughly speaking, we may think of the arrow of time as starting at $t=-\infty,$ going forward in time to $t=+\infty={\tilde t}$ and then back to ${\tilde t}=-\infty$.   From this point of view, the path integrals is over fields that propagate on the closed time contour from $-\infty$ back to $\infty$ in the forward sense, with ${\tilde h}_{\mu\nu}$ field insertions occurring at times later than those of $h_{\mu\nu}$, at a time $t<{\tilde t}$.

 The path integral Eq.~(\ref{eq:inin}) computes expectation values of operators which are time-ordered ($\tau$-ordered) with respect to the closed time contour, \eg~at zero external field, an insertion of fields $h,{\tilde h}$,
 $$
 {\cal T} \left[{\tilde h}(x_1)\cdots {\tilde h}(x_k) h(x_{k+1})\cdots h(x_n) \right]
 $$ 
 corresponds in canonical quantization to the operator product 
 $$
 T^*\left[{\hat h}(x_1)\cdots {\hat h}(x_k)\right] \cdot T\left[{\hat h}(x_{k+1})\cdots {\hat h}(x_n)\right],
 $$
 so that operator products are time-ordered (Feynman) or anti-time-ordered (Dyson) depending on which branch of the closed time path they lie on.

 In addition of the doubling of the integration variables relative to the standard time-ordered path integral for $|in\rangle\rightarrow |out\rangle$ transition matrix elements, we also need to double the external classical sources in Eq.~(\ref{eq:inin}).    By construction, the integral is normalized as $\Gamma[x_A,{\bar g},x_A {\bar g}]=0$, so to get any use out of the effective action we need to first differentiate it with respect to $x_A{}^\mu$ or ${\bar g}_{\mu\nu}$ before setting ${\tilde {\bar g}}_{\mu\nu}={\bar g}_{\mu\nu}$ or ${\tilde x}_A=x_A$.   
 
 In fact, all the relevant observables for radiation from the binary are obtained by functional differentiation of the in-in effective action, as we now explain.  First, we solve the equations of motion for the worldlines $x^\mu_A$, defined as the extrema of the in-in action
 \begin{equation}
\label{eq:nleom}
\left.{\delta\over\delta x^\mu_A}\Gamma[x_A,{\bar g},{\tilde x}_A {\tilde{\bar g}]}\right|_{x_A={\tilde x}_A;{\bar g},={\tilde{\bar g}}=\eta}\equiv 0,
\end{equation}
subject to a suitable set of initial data (particle positions and momenta).    Because the in-in action is formally the result of integrating out gravitons that interact with the worldline and with themselves, the equations of motion are (at least at sufficiently high order in powers of $G_N$) in general non-local in time (see sec.~\ref{sec:rr}).    They include both conservative and radiative corrections to the gravitational two-body dynamics.   The solution of these equations of motion is then re-inserted into $\Gamma\left[x_A,{\bar g},{\tilde x}_A,{\tilde{\bar g}}\right]$, yielding an ``on-shell'' effective action for the background metric ${\bar g}_{\mu\nu}$.   

Next, we vary this on-shell action with respect ${\bar g}_{\mu\nu}$ to obtain the total energy-momentum pseudotensor $\tau_{\mu\nu}(x)$ of the binary system:
\begin{equation}
\label{eq:pemt}
\tau_{\mu\nu} = {2\over \sqrt{{\bar g}}} {\delta\over \delta {\bar g}^{\mu\nu}} \left.\Gamma[x_A, {\bar g}; {\tilde x}_A \tilde{\bar g}]\right|_{x_A={\tilde x}_A;{\bar g}={\tilde{\bar g}}=\eta}.
\end{equation}
Because of the choice of background field gauge, this pseudo-tensor is conserved on-shell, $\partial_\nu \tau^{\mu\nu}=0$, but dependent on the gauge fixing term $S_{GF}[{\bar g},h]$.

Even though it depends on the choice of gauge, $\tau^{\mu\nu}$ has physical content, as the quantum mechanical amplitude for the binary system to emit an on-shell graviton of definite momentum $k^\mu$  out to ${\cal I}^+$, 
\begin{equation}
\label{eq:amp}
{\cal A}(k) = \epsilon_{\mu\nu}(k) {\cal A}^{\mu\nu}(k) =  -{1\over 2 m_{Pl}} \int d^4 x e^{ik\cdot x} \epsilon_{\mu\nu}(k) \tau^{\mu\nu}(x).
\end{equation}
In turn, this on-shell amplitude has a simple relation to the waveform, once a gauge for the background field ${\bar h}_{\mu\nu}\equiv{\bar g}_{\mu\nu}-\eta_{\mu\nu}$ has been chosen.   For instance, in deDonder gauge, the waveform at $|{\vec x}|\rightarrow\infty$ and fixed retarded time $u=t-|{\vec x}|$ is (setting $d=4$)~\cite{Weinberg:1972kfs} 
\begin{equation}
\label{eq:wave}
\left.\langle in| h_{\mu\nu}(x)|in\rangle\right|_{{\cal I}^+} = {4 G_N\over |{\vec x}|} \int_{-\infty}^{\infty} {d\omega\over 2\pi} e^{-i\omega u} \left[{\cal A}^{\mu\nu}(k)-{1\over 2} \eta^{\mu\nu} {\cal A}^\rho{}_\rho(k)\right],
\end{equation}
where the on-shell momentum is $k^\mu=\omega(1,{\vec x}/r)$.    Finally, the pseudotensor can also be used to calculate the flux of energy, momentum and angular momentum radiated to ${\cal I}^+$, as well as the conserved ADM charges $P^\mu, J^{\mu\nu}$ of the system at spatial infinity $i^0$, \eg
\begin{equation}
\Delta P^\mu =  \int {d^d k\over (2\pi)^d} (2\pi) \delta(k^2) \theta(k^0) k^\mu |{\cal A}(k)_{\rho\sigma}|^2
\end{equation}
is the energy-momentum deposited in a detector placed at ${\cal I}^+$, summed over polarizations.

In addition to being a useful generating function for observables,  $\Gamma\left[x_A,{\bar g};{\tilde x}_A,{\tilde{\bar g}}\right]$ also admits a convenient perturbative expansion in powers of $G_N$,  in terms of Feynman diagrams.    The Feynman rules for the graviton propagators and self-interaction vertices of the theory are derived in exactly the same way as in Yang-Mills theory (see \eg~\cite{Donoghue:1995cz} for a review of the Feynman rules for gravity).   There are also vertices generated by the coupling of gravitons to the classical worldlines in $S_{pp}$, for example the term $-m\int d\tau$ in the point particle action generates a vertex with a single off-shell graviton of momentum $k$,
\begin{equation}
\label{eq:ffg}
\includegraphics[width=0.2\hsize,valign=c]{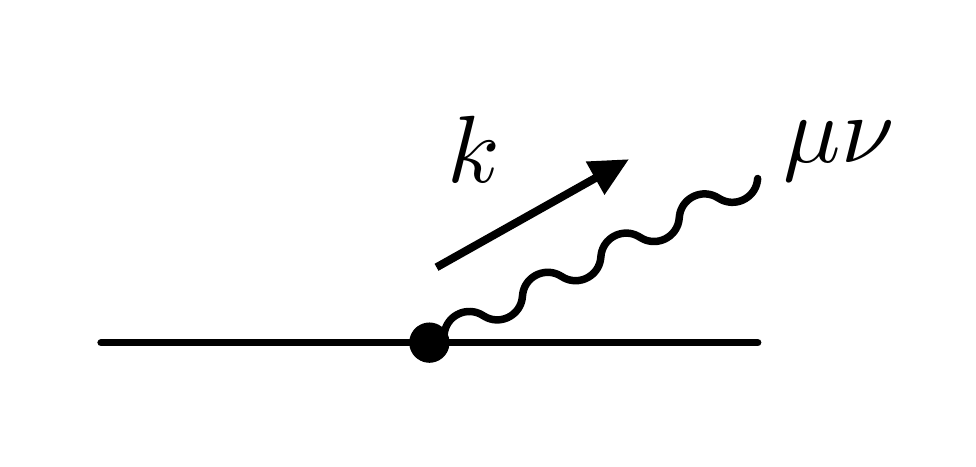}=-{i m\over 2 m_{\Pl}^{d-2}} \int d\tau e^{i k\cdot x(\tau)} u^\mu(\tau) u^\nu(\tau).
\end{equation}

To calculate the functional integral, we also have keep track of field insertions on either side of the closed time contour, which requires a doubling of the fields in the theory, as mentioned above.      Insertions of $h_{\mu\nu}$ and ${\tilde h}_{\mu\nu}$ are time ordered with respect to an integration contour that starts at $t=-\infty$ and passes through $t=+\infty$ before bending back and ending at ${\tilde t}=-\infty$, as outlined above.   Therefore the free field two-point functions are parsed as follows~\cite{Calzetta:1986ey,Calzetta:2008iqa}
\begin{eqnarray}
\langle h_{\mu\nu}(x) h_{\rho\sigma}(y)\rangle &=& \langle 0|T \left[{\hat h}_{\mu\nu}(x) {\hat h}_{\rho\sigma}(y)\right]|0\rangle =P_{\mu\nu;\rho\sigma} D_F(x-y),\\
\langle {\tilde h}_{\mu\nu}(x) h_{\rho\sigma}(y)\rangle &=&  \langle 0| {\hat h}_{\mu\nu}(x) {\hat h}_{\rho\sigma}(y)|0\rangle =P_{\mu\nu;\rho\sigma} W(x-y),\\
 \langle {\tilde h}_{\mu\nu}(x) h_{\rho\sigma}(y)\rangle &=&  \langle 0|T^* \left[{\hat h}_{\mu\nu}(x) {\hat h}_{\rho\sigma}(y)\right]|0\rangle = P_{\mu\nu;\rho\sigma} D_D(x-y),
\end{eqnarray}
where 
\begin{equation}
D_F(x-y) = \int {d^d k\over (2\pi)^d} e^{-i k\cdot (x-y)} {i\over k^2+ i 0^+},
\end{equation}
is the propagator defined by the usual  (time ordered) Feynman-$i\epsilon$ contour prescription, $D_D(x-y)=[D_F(x-y)]^*$ is the Dyson anti-time-ordered propagator, and 
\begin{equation}
\label{eq:wight}
W(x-y) = \langle 0| \phi(x) \phi(y)|0\rangle=  \int {d^d k\over (2\pi)^d} (2\pi) \delta(k^2) \theta(k^0) e^{-i k\cdot (x-y)}
\end{equation}
is the Wightman function of a free massless field, which is evidently the Green's function that propagates positive energy on-shell signals forwards in time.

%%%%%%%%%%%%%%%%%%%%%%%%%%%%%%%%%%%%%%%%%%%%%%%%%%%%%%%%%%%%%%%%%%%%%%%%%
\begin{figure}[t!]
\begin{center}
\includegraphics[width=0.50\hsize]{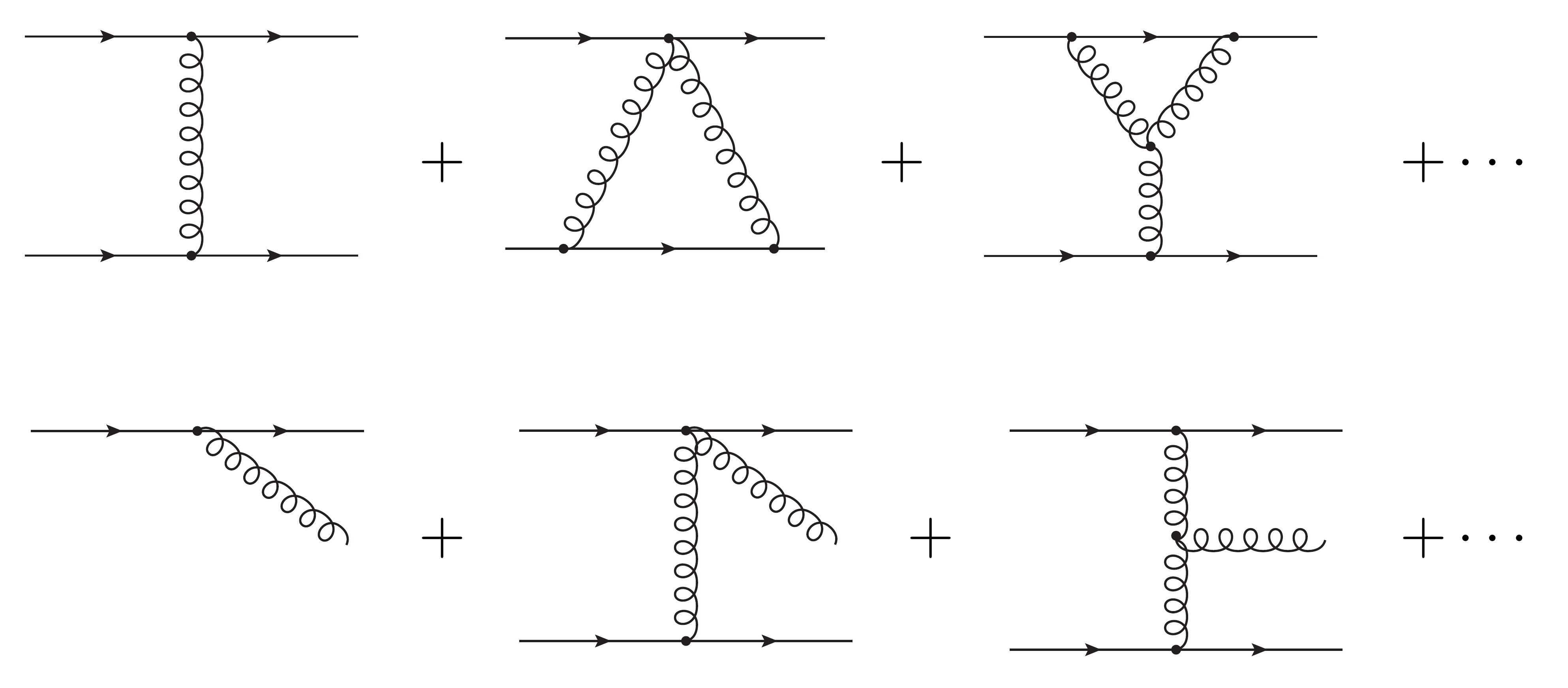}
\end{center}
\caption{Feynman diagram expansion of the in-in action.   External lines correspond to insertions of the background field ${\bar h}_{\mu\nu}$,  ${\tilde {\bar h}}_{\mu\nu}$.   Each diagram shown stands for a sum of contributions from insertions on either side of the Schwinger-Keldysh closed time contour.}
\label{fig:LOdiags}
\end{figure}
%%%%%%%%%%%%%%%%%%%%%%%%%%%%%%%%%%%%%%%%%%%%%%%%%%%%%%%%%%%%%%%%%%%%%%%%%%%

The Feynman diagrams that contribute to the effective action are ones that have internal $h_{\mu\nu},{\tilde h}_{\mu\nu}$ graviton lines coupled to classical worldline sources, as in Eq.~(\ref{eq:ffg}), as well as diagrams with external (background gravitons) ${\bar h}_{\mu\nu},{\tilde {\bar h}}_{\mu\nu}$, as depicted in fig.~\ref{fig:LOdiags}.    In the $\hbar\rightarrow 0$ limit, we can drop graphs that have internal graviton loops or more than one insertion of ${\bar h},{\tilde {\bar h}}$.

The framework outlined so far is well suited to calculate ``post-Minkowskian" (PM) corrections to two-body dynamics, in which the compact objects are fully relativistic, as a formal expansion in powers of $G_N$.   Because weakly coupled relativistic particles do not form bound states, PM calculations are typically relevant for scattering processes at CM energies $E_{\CM}$ larger than the ADM masses of the compact objects, and large impact parameter $b$ such that the expansion parameter $G_N E_{\CM}/b\ll 1$ is suitably small.   In this limit  the Feynman rules of the EFT provide a systematic simultaneous expansion in powers of $G_N E/b\ll 1$ (PM effects),  $\hbar /L\ll 1$ (quantum corrections), where $L\sim E b$ is the orbital angular momentum scale of the binary, and $({\cal R}/b)\ll 1$ if one includes the finite size vertices in Eq.~(\ref{eq:pp}).   The only physical scale appearing in any Feynman integral over internal graviton momenta is the separation $b$, so the effective theory has manifest powers counting in $G_N E/b \ll 1$, with $\hbar /L\ll 1$ serving to count the number of internal graviton loops in a given graph.  Such PM scattering calculations have been the focus of intense study in recent years.   For a recent review of recent developments and a more complete guide to the literature, see~\cite{Buonanno:2022pgc}.

We will instead focus on the applications of Eq.~(\ref{eq:inin}) to PN calculations, where the expansion in powers of $G_N$ and in powers of velocity become correlated.     The formalism is not yet optimized for the computation of gravitational radiation from objects in bound non-relativistic orbits.    Rather than integrating out the graviton in one fell swoop, in the non-relativistic case it is useful instead to perform the  in-in path integral in successive stages, by factorizing the integration measure into modes with support near the orbital distance scale $\sim r$ and those localized around IR momentum scales $k^\mu\sim v/r$ corresponding to the frequency of the outgoing radiation.   This Wilsonian, multi-step approach to bound state dynamics is described in detail starting in the next section.

\subsection{NRGR}
\label{sec:NRGR}

When  $v\ll 1$, there exists a hierarchy of scales between orbital dynamics at distances $\sim r$ and radiation emission at wavelengths $\lambda\sim r/v\gg r$.    While the Feynman diagrams described in the previous section scale as definite powers of $G_N M/r$ and of $\hbar/L$, they also contribute, after expanding in $v\ll 1$, at all orders in powers of velocity.   Because in the PN expansion we are trying to determine the gravitational wave observables up to a fixed finite order in powers of $v$, it is desirable to have a perturbative scheme in which the Feynman rules are homogeneous in velocity, so that each diagram scales as a fixed power of$v$ that can be determined from a simple set of power counting rules.

The lack of manifest low velocity scaling in the PM expansion can be traced to the presence of multiple scales in the  momentum space Feynman integrals of the theory.   When defined using dimensional regularization in $d=4-\epsilon$ dimensions, momentum space integrals receive non-vanishing contributions from two types of kinematic regions~\cite{Goldberger:2004jt}
\begin{itemize}
\item \emph{Potential}:   $(p^0,{\vec p})\sim (v/r,1/r)$,
\item \emph{Radiation}:   $(k^0,{\vec k})\sim (v/r,v/r)$.
\end{itemize}
The potential region corresponds to off-shell gravitons which are exchanged between the point sources.   In position space, they generate the nearly instantaneous in time, long range conservative gravitational forces responsible for binding the particles into closed orbits.    On the other hand, radiation gravitons can go on-shell, propagating out to the detector, or remain off-shell, generating both ``dissipative" (time reversal odd) and ``conservative" ($T$-even) radiation reaction forces.

In dimensional regularization, a given Feynman integral can be ``threshold expanded"~\cite{Beneke:1997zp} around the various kinematic configurations of potential and radiation regions set by the external momenta (``method of regions'').   The expanded Feynman integral is equal to a sum of simpler integrals, each characterized by a single physical scale.   These simplified integrals can now be calculated for arbitrary (bound or unbound) non-relativistic trajectories ${\vec x}_{1,2}(t)$, as is necessary for the inspiral problem, and scale homogeneously as definite powers of the expansion parameter $v$.

Rather than expanding out the PM momentum integrals in powers of $v$, we perform the expansion at the level of the Lagrangian, so as to obtain vertices with definite velocity scaling.    By assumption, in the PN limit there is a nearly inertial frame (\eg~the CM frame) where both sources are non-relativistic.   Working in such a frame, we expand the worldline action explicitly in powers of velocity, \eg 
\begin{eqnarray}
\nonumber
-m\int d\tau &\supset& -{m\over 2 m_{\Pl}} \int d\tau h_{\mu\nu} u^\mu u^\nu+\cdots\\
&\approx& -{m\over 2 m_{\Pl}}\int dt\left[h_{00} + 2 h_{0i} {\vec v}^i +h_{ij} {\vec v}^i {\vec v}^j -{1\over 2} h_{00} {\vec v}^2\right] +{\cal O}(v^3).
\end{eqnarray}

More crucially, we also perform at the level of the Lagrangian an explicit mode decomposition of the graviton into distinct potential and radiation fields, ($\int_{\vec p}\equiv \int d^{d-1}{\vec p}/(2\pi)^{d-1}$)
\begin{equation}
\label{eq:modes}
 h_{\mu\nu}(x)= {\bar h}_{\mu\nu}(x) + \int_{\vec p} e^{i{\vec p}\cdot {\vec x}}  H_{\mu\nu;\vec p}(x^0).
\end{equation}
The radiation field ${\bar h}_{\mu\nu}$ has momentum scaling $k^\mu\sim (v/r,v/r)$ and thus spacetime derivatives $\partial_\mu$ acting on such fields are power counted as 
\begin{equation}
\partial_\rho {\bar h}_{\mu\nu}\sim {v\over r} {\bar h}_{\mu\nu}
\end{equation}
at the level of the Lagrangian.   The potential graviton field $H_{\mu\nu;\vec p}(x^0)$, is defined in a mixed time/spatial momentum representation, such that the large momentum scale ${\vec p}\sim 1/r$ associated with virtual exchange between the particle sources is explicit.    This way, spacetime derivatives acting on the re-phased potential mode $H_{\mu\nu;\vec p}(x^0)$ also scale as $\partial_\mu\sim v/r$ and therefore every derivative appearing in the Lagrangian of the theory is treated on the same footing.

Inserting the mode expansion Eq.~(\ref{eq:modes}) into the action, we can read off the potential propagator from the ${\cal O}(H^2)$ part of the Einstein-Hilbert Lagrangian
\begin{equation}
{\cal L}_{H^2} = -{1\over 2}\int_{\vec p}\left[{\vec p}^2 H_{\vec p\mu\nu}  H^{\mu\nu}_{-\vec p} -{{\vec p}^2\over 2} H_{\vec p} H_{-\vec p}-\partial_0 H_{\vec p\mu\nu}\partial_0  H^{\mu\nu}_{-\vec p} + {1\over 2} \partial_0 H_{\vec p} \partial_0 H_{-\vec p}\right].
\end{equation}
The last two terms are $O(v^2)$ relative to the first two and are treated perturbatively, as insertions, in the path integral.   They encode retardation effects due to the finite speed of light.   On the other hand, the first two terms determine the propagator
\begin{equation}
\langle H_{{\vec p}\mu\nu}(x^0) H_{{\vec q}\rho\sigma}(0)\rangle = -(2\pi)^3\delta^3({\vec p}+{\vec q}) {i\over {\vec p}^2} P_{\mu\nu;\rho\sigma}\delta(x^0),
\end{equation}
which is instantaneous in time.   Note that because the potential modes do not go on-shell, ${\vec p}^2\neq 0$, an $i\epsilon$ contour prescription is not needed to define the free two-point function.   Given that $x^\mu\sim v/r$ and ${\vec p}\sim 1/r$, this equation tells us that we should assign the scaling rule
\begin{equation}
H_{\vec p}\sim r^2\sqrt{v},
\end{equation}
 at the level of the EFT Lagrangian.

The scaling rule for radiation is more straightforward.    The radiation graviton propagator scales as $1/k^2\sim (r/v)^2$ in momentum space so that in position space $\langle h(x) h(0)\rangle\sim\int d^4 k/k^2\sim (v/r)^2$, and therefore we assign the rule
 \begin{equation}
{\bar h}_{\mu\nu}(x)\sim v/r
\end{equation}
to insertions of the radiation field.

The final step in constructing an EFT with definite velocity scaling is to multipole expand the radiation field at the level of the action, either in its couplings to the particle sources or to the potentials~\cite{Grinstein:1997gv}.   To see why this is necessary, consider \eg~the amplitude for a particle to absorb a single radiation graviton, given in Eq.~(\ref{eq:ffg}).   In the non-relativistic limit, the exponential phase $e^{i k\cdot x}$ in the amplitude contains a factor, 
\begin{equation}
e^{-i{\vec k}\cdot {\vec x}}\sim e^{{\cal O}(v)},
\end{equation}
that does not scale homogeneously in the expansion parameter, given that the particle orbits  ${\vec x}\sim r$ emit radiation of typical momentum  ${\vec k}\sim v/r$.  Similarly,  consider a generic Feynman diagram that includes absorption of radiation by an internal potential graviton, \eg~
\begin{equation}
\includegraphics[width=0.25\hsize,valign=c]{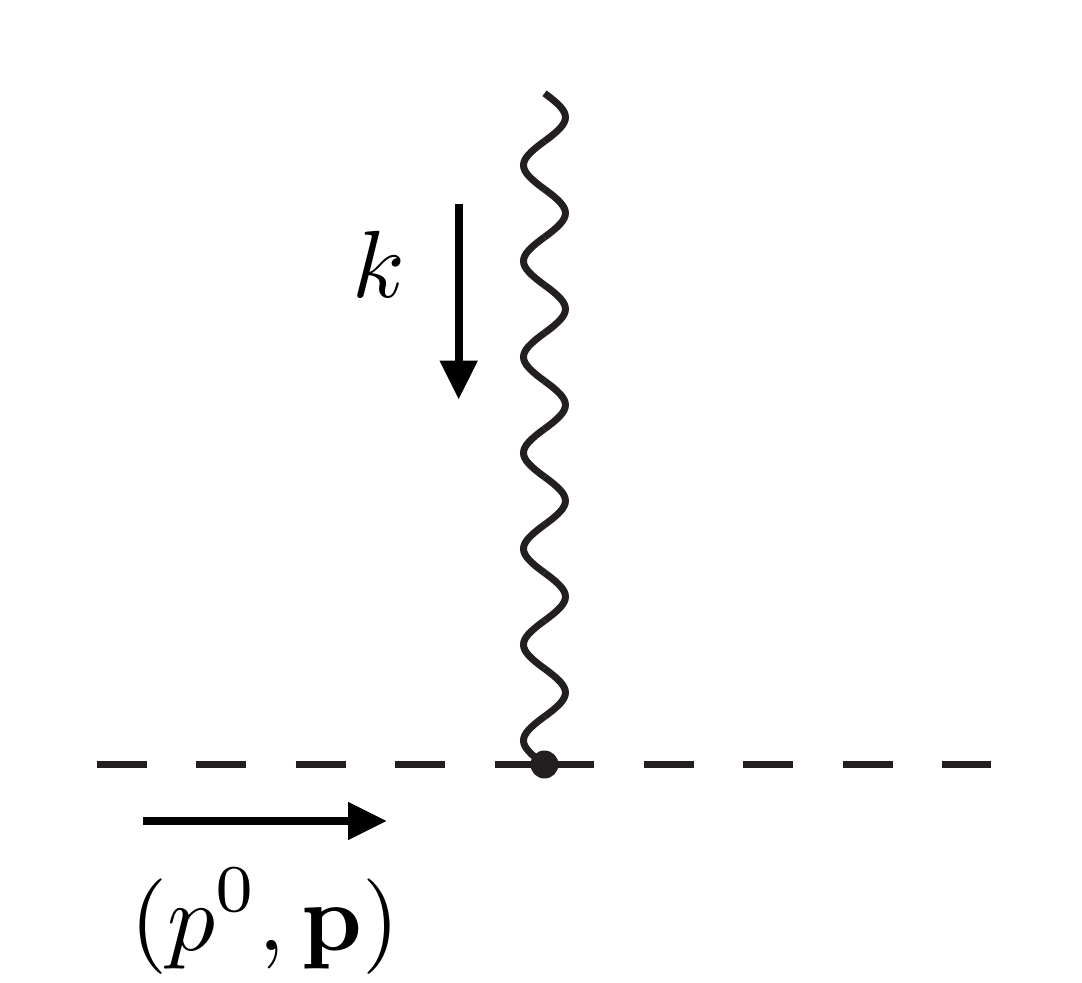}
\end{equation}
In this diagram, the outgoing potential graviton comes with a propagator $\propto {1/ ({\vec p}+{\vec k})^2}$, so given that ${\vec p}\sim 1/r$, ${\vec k}\sim v/r$, it gives contributions at all orders in powers of $v$.

The remedy in either case is to substitute the multipole expanded radiation field
\begin{equation}
\label{eq:taylor}
{\bar h}_{\mu\nu}(x)\mapsto \sum_{\ell=0}^\infty {1\over \ell!} {\vec x}^{i_1}\cdots {\vec x}^{i_\ell} \partial_{i_1}\cdots \partial_{i_\ell} {\bar h}_{\mu\nu}(x^0,0),
\end{equation}
into the terms in the Lagrangian involving radiation couplings to either potentials or the worldlines (the radiation mode self-interactions do not get multipole expanded).  Here, we have chosen coordinates such that ${\vec x}=0$ lies somewhere near the center of mass of the binary system, and ${\vec x}\sim r$ is a point somewhere inside the binary system.   (The precise choice of center for the the multipole decomposition is arbitrary, and chosen for computational convenience.   It has no effect on physical observables.)     The $\ell$-th order term in the multipole expansion is suppressed by $v^\ell$ relative to the monopole ($\ell=0$) term, and in practical calculations at fixed PN accuracy the expansion may be truncated at a finite value of $\ell$.

Given the decomposition of the graviton into potentials and radiation, together with the multipole expansion of radiation, we now have a set of rules that allow us to count powers of $v$ either at the level of the action or inside correlation functions.   These rules are summarized in the following table:
$$
\begin{array}{c|c|c|c|c|c|c|c}
\partial_\mu & x^\mu & {\vec x}_a & {\vec p} & k^\mu & H_{\vec p} & {\bar h} & m/m_{\Pl}\\
\hline
v/r& r/v & r & 1/r & v/r& r^2 \sqrt{v} & v/r & \sqrt{L v}.
\end{array}
$$
Using these rules, one finds by dimensional analysis that every Feynman diagram scales as a definite power of the angular momentum scale $L= m v r$ times a definite power of $v$.   For example a tree level diagram (no internal graviton loops) gives a contribution to the effective action Eq.~(\ref{eq:inin}) of order $\sim L^{1-n/2}$, where $n$ is the number of external radiation graviton lines.    Each additional internal graviton loop costs a factor of $1/L$.   The classical limit corresponds to orbital angular momenta  $L\gg \hbar$, in which case one can simply ignore graphs with internal loops or more than one external leg.

For example, the leading order coupling of the worldline to potential (dashed) or radiation (wavy) gravitons scale as
\begin{eqnarray}
\includegraphics[width=0.18\hsize,valign=c]{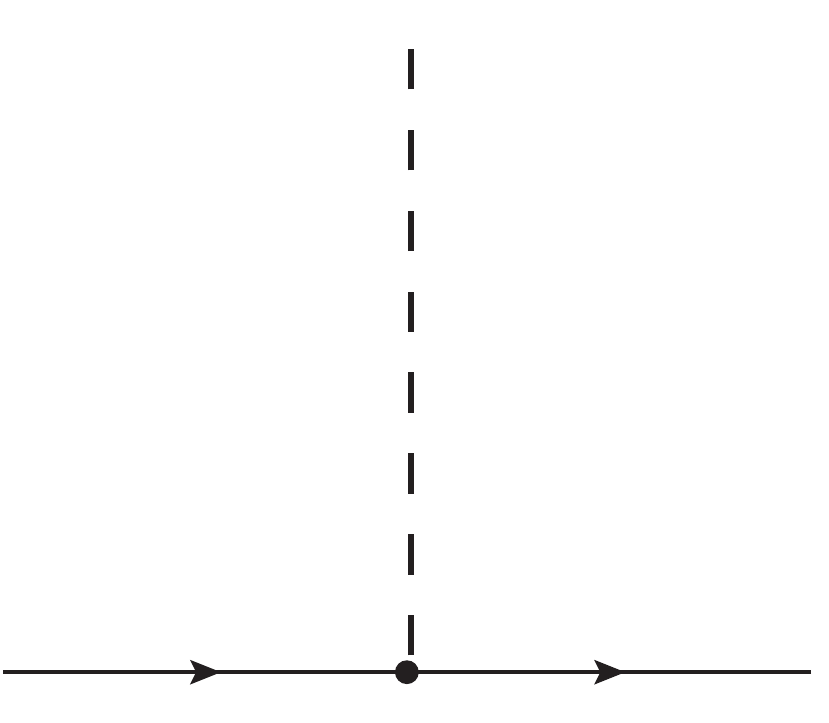}= \sqrt{L} v^0  & \includegraphics[width=0.18\hsize,valign=c]{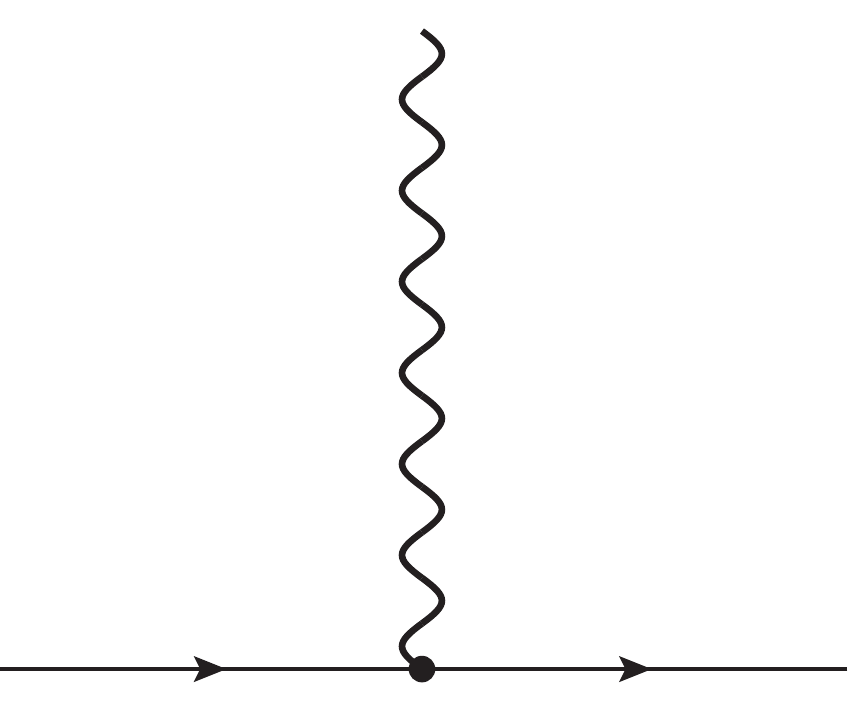}=\sqrt{L} v^{1/2},
\end{eqnarray}
and therefore the leading order Newton potential exchange between two particles, involving two insertions of the vertex on the left, is of order $\sim L v^0$ in the PN expansion.   Similarly, the $H^3$, $H^2 {\bar h}$  interaction vertices correspond to
\begin{eqnarray}
\includegraphics[width=0.18\hsize,valign=c]{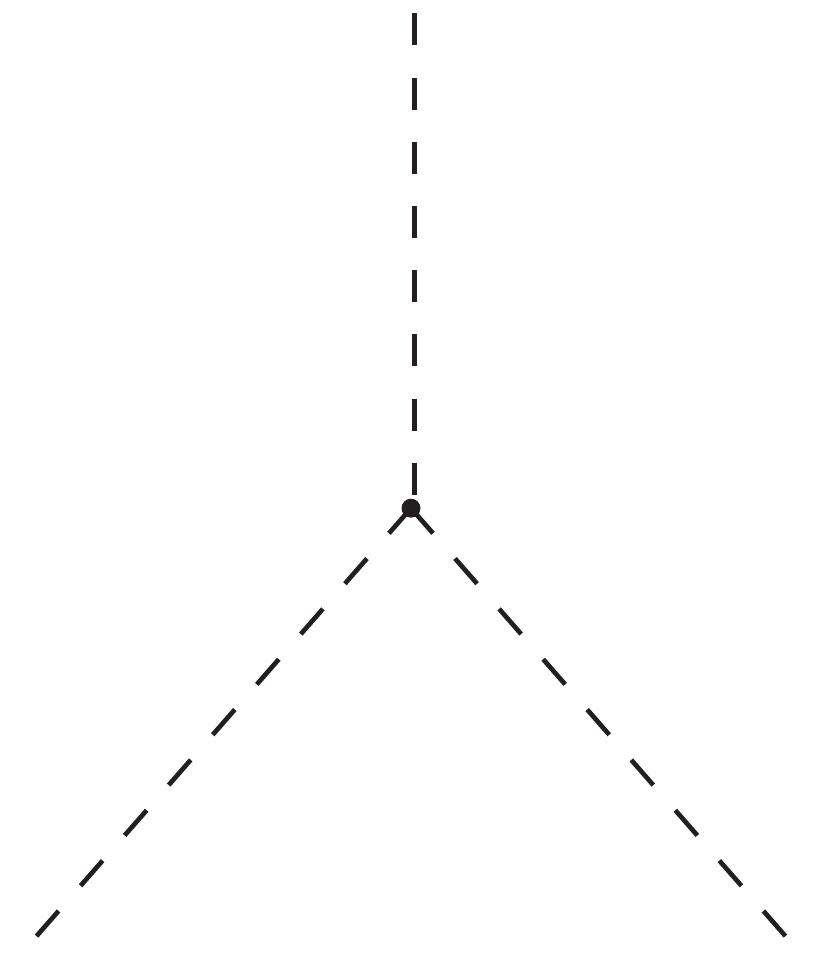}= {v^{2}\over\sqrt{L}}& \includegraphics[width=0.18\hsize,valign=c]{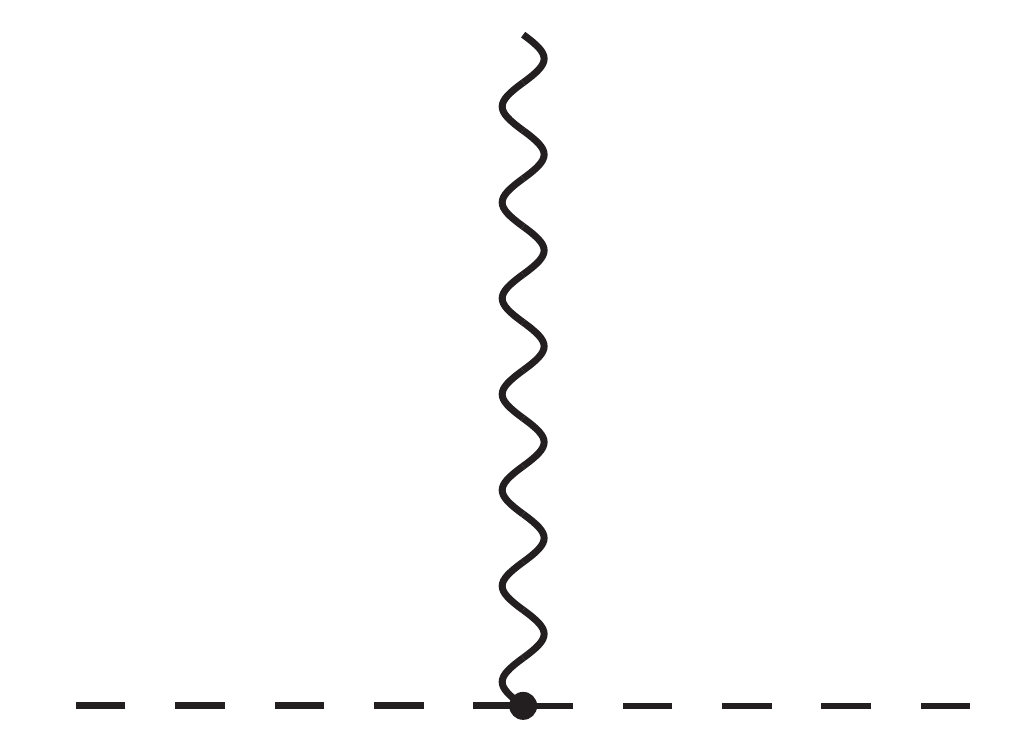}=\normalsize{{v^{5/2}\over \sqrt{L}},
}\end{eqnarray}
and so on.

We will refer to the theory with both potentials and radiation as NRGR~\cite{Goldberger:2004jt} to emphasize the analogy to NRQCD~\cite{Caswell:1985ui}, the EFT description of non-relativistic heavy quark bound states $Q{\bar Q}$  ($M_Q\gg \Lambda_{QCD}$), which employs a similar mode decomposition~\cite{Luke:1999kz} and multipole expansion~\cite{Grinstein:1997gv} of the gluon fields in full QCD.   In practical calculations, NRGR is used to interpolate between the theory of relativistic particles in Eq.~(\ref{eq:lights}) in the UV and another EFT in the IR that results from integrating out the potential gravitons.   This EFT streamlines the calculation of radiative corrections to bound state dynamics, as described in the next section.

\subsection{Radiative corrections and \zosoir}
\label{sec:zosoir}

Because the potential gravitons cannot go on-shell, it is possible to integrate them out to obtain a local EFT of self-interacting radiation gravitons coupled to the bound state.    We refer to this EFT, valid at distances longer than the orbital scale $r$, by the acronym ``\zosoir'' since it encodes the interactions of a \emph{Zoomed Out {S}ingle Object} whose internal structure, \ie~the binary constituents, cannot be directly resolved by long wavelength radiation modes (the subscript ``IR'' is to distinguish this EFT from a similar theory of black hole horizon fluctuations in the UV that will be discussed in sec.~\ref{sec:horizon}).

Like the gapped point particle of sec.~\ref{sec:1B}, the composite object is defined in terms of a center-of-mass variable $x^\mu(\tau)$, and an orthonormal frame $e^a{}_\mu$ that accounts for the spatial orientation relative to asymptotic inertial observers.   In addition, it contains an infinite tower of multipole moments $I_{a_1\cdot a_\ell}$, $J_{a_1\cdot a_\ell}$ of parity $(-1)^\ell$, $(-1)^{\ell+1}$, respectively, coupled to the Weyl curvature of the radiation field.   The form of the worldline action follows~\cite{Goldberger:2005cd,Porto:2005ac,Goldberger:2009qd}
 from diffeomorphism invariance:
\begin{eqnarray}
\label{eq:zoso}
\nonumber
S_{\includegraphics[width=0.05\hsize,valign=c]{zosoir}} &=&S_{EH} +\int d\tau L(X,x(\tau),{\bar g}) +{1\over 2}\int d\tau S_{ab}(\tau) \Omega^{ab} +{1\over 2} \int d\tau I_{ab}(\tau) E^{ab}(x(\tau))\\
& & {} -{2\over 3} \int d\tau J_{ab}(\tau) B^{ab}(x(\tau))+{1\over 6}\int d\tau I_{abc}(\tau) \nabla^c E^{ab}(x(\tau))+\cdots.
\end{eqnarray}   
We proceed to explain the meaning of the various terms in this equation.

First, the term 
\begin{equation}
S_X=\int d\tau L(X,x(\tau),{\bar g})
\end{equation}
provides the dynamics of the internal degrees of freedom ``$X$'' of the composite system in the absence of radiation, $E_{ab}=B_{ab}=0$.  For gapped constituents in a non-relativistic bound state, ``$X$'' stands in for the coordinates ${\vec x}_{1,2}$ and the spins ${\vec S}_{1,2}$, but we can consider a more general situation in which each particle carries additional low-lying modes (see sec.~\ref{sec:horizon}).   The Lagrangian $L(X,x^\mu(\tau),{\bar g}_{\mu\nu})$ determines the ADM mass of the composite object, obtained by expanding to linear order in ${\bar h}_{\mu\nu}$.   Even though we have chosen to parameterize the action Eq.~(\ref{eq:zoso}) in terms of the proper time along $x^\mu(\tau)$, the Hamiltonian for the system is non-zero
\begin{equation}
H_X=\left[{dX\over d\tau} {\partial \over \partial {\dot X}} L(X,x(\tau),{\bar g}) -  L(X,x(\tau),{\bar g})\right],%_{{\bar g}_{\mu\nu}=\eta_{\mu\nu}},
\end{equation}
since excitations localized on the composite worldline acts like an internal clock that spontaneously break the time reparameterization gauge symmetry of the underlying theory.

Ignoring spin for simplicity (and still keeping $E_{ab}=B_{ab}=0$), the equations of motion for  $x^\mu$ imply that $p^\mu = H_X(\tau) dx^\mu/d\tau$ follows a geodesic of  ${\bar g}_{\mu\nu}$, while the Euler-Lagrange equations for $X$ imply that $H_X$ is a constant on-shell.  By comparing the asymptotic (linearized) gravitational field of the composite particle, sourced by 
\begin{equation}
\label{eq:tx}
T^{\mu\nu}_X(x) = {2\over \sqrt{{\bar g}}} {\delta\over \delta g^{\mu\nu}(x)} S_X = \int d\tau H_X {\delta^4(x-x(\tau)) \over{\sqrt{\bar g}}} {dx^\mu\over d\tau} {dx^\nu\over d\tau},
\end{equation}
with the full theory, Eq.~(\ref{eq:fullh}), we can identify $H_X$, evaluated on-shell, with the ADM mass of the system in the CM frame, Eq.~(\ref{eq:pcm}).

Thus, in flat space, the motion of the worldline is simply $x^\mu(\tau) = u^\mu \tau + x^\mu{}_\CM,$ with $x^\mu_{\CM}$ and $u^\mu$ are constants.        Including also the spin,~\cite{Hanson:1974qy} the term $-\int d\tau S_{ab} \Omega^{ab}$ (the angular velocity $\Omega_{ab}$ is defined in Eq.~(\ref{eq:om})) then gives the ADM angular momentum of the system as
\begin{equation}
J^{\mu\nu} = x^\mu_{\CM} P^\nu - x^\mu_{\CM} P^\nu + S^{\mu\nu},
\end{equation}
$S^{\mu\nu} = e^\mu{}_a e^\nu{}_b S^{ab}$, which is also conserved by the equations of motion.

One can think of the moments $S_{ab},I_{ab},J_{ab},\ldots$ as (time-dependent) Wilson coefficients that encode the internal structure of the radiating system.   They are, in general, functions of the internal variables $X$, and are obtained by matching to the UV theory, as we will describe in more detail below.     We have chosen to define these moments relative to the local frame $e^a_\mu$, and to classify them in terms of representations of a $SO(3)\subset SO(3,1)$ subgroup of the local Lorentz transformations that leaves $P^a=e^a{}_\mu P^\mu$ invariant\footnote{The $\ell$-th order moments $I_{a_1\cdots a_\ell},J_{a_1\cdots a_\ell}$ fit together into a representation $(\ell,0)\oplus (0,\ell)$ os $SO(3,1)$.}.    Similarly the electric and magnetic curvatures, $E_{ab} = e^\mu{}_a e^\nu E_{\mu\nu}$, $\nabla_a E_{bc} = e^\mu{}_a e^\nu{}_b e^\rho{}_c \nabla_\mu E_{\nu\rho}$ \etc~are projections onto the rotating frame $e^a{}_\mu$ .

The full dynamics of the non-relativistic bound state consists of Eq.~(\ref{eq:zoso}) coupled to the Einstein-Hilbert term for the radiation metric $g_{\mu\nu}=\eta_{\mu\nu} + {\bar h}_{\mu\nu}$.   It is worth noting that Eq.~(\ref{eq:zoso}) is actually universal, in the sense that it gives a description of soft gravitational radiation from a completely generic self-gravitating object of finite spatial extent $\sim b$, and ADM energy $\sim E$.   For such a system, we necessarily\footnote{Assuming that any physically reasonable system that is is squeezed down to a size smaller than its Schwarzschild radius will inevitably collapse gravitationally into a black hole~\cite{Penrose:1969pc}.} have $G_N E\lsim b$, and multipole moments of generic size $\sim E b^\ell$.   Therefore the emission or absorption of multipole radiation with $\omega b \ll 1$ can be described systematically by this EFT.     

\begin{figure}[t!]
\begin{center}
\includegraphics[width=0.50\hsize]{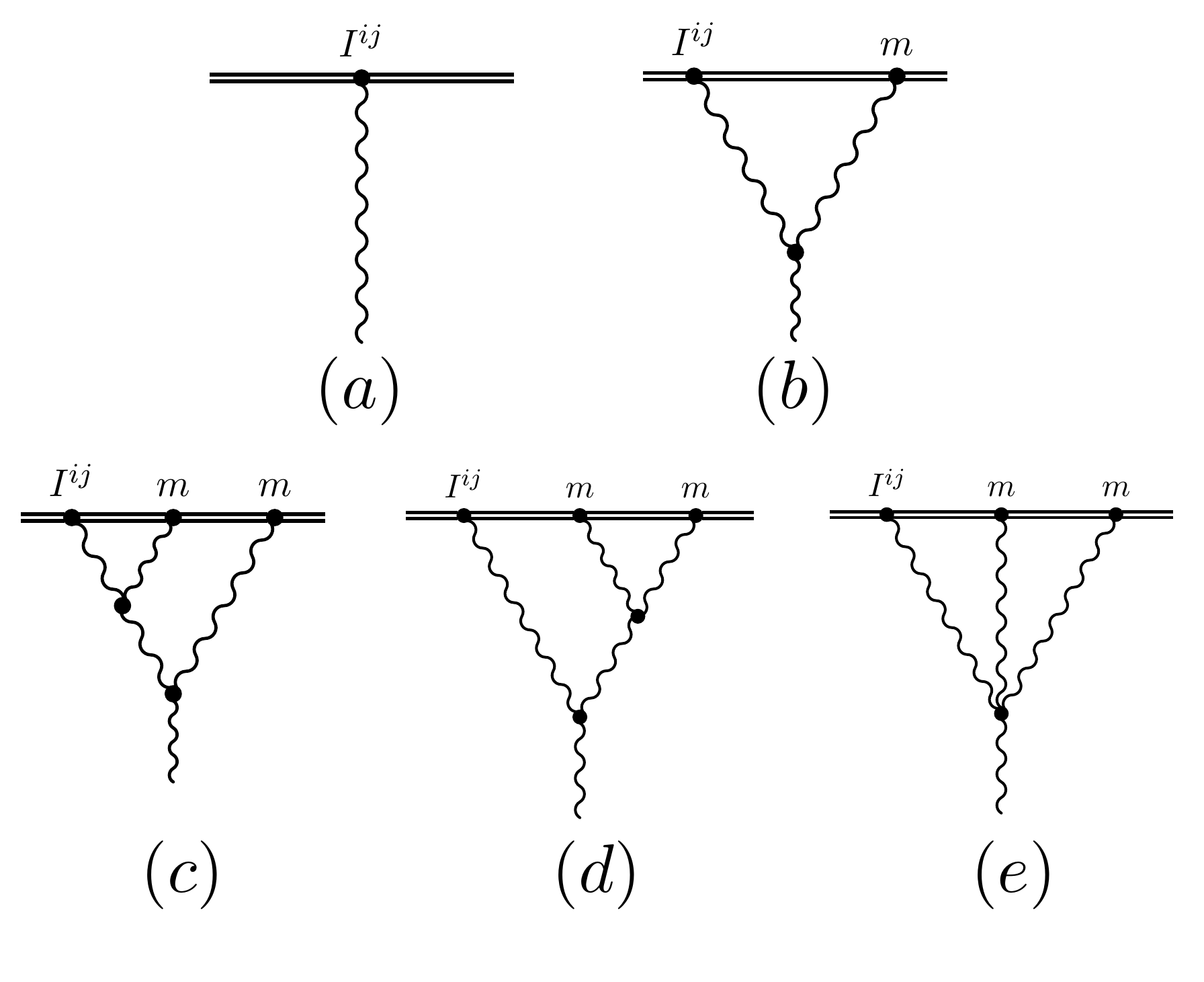}
\end{center}
\caption{Leading order quadrupole emission (a) in \zosoir~and perturbative ``tail'' corrections at ${\cal O}(G_N E\omega)$ (b) and ${\cal O}(G_N E\omega)^2$ (c)-(e).   Diagram (b) has a $1/\epsilon_{IR}$ singularity in dimensional regularization, while (c)-(e) contain both IR and UV poles in $d=4-\epsilon$ spacetime dimensions.}
\label{fig:tails}
\end{figure}

At the classical level, the EFT computes perturbative corrections as a double expansion in powers of $G_N E \omega\ll 1$, which controls effects due to graviton propagation in the curved spacetime sourced by the object, and the multipole expansion parameter $\omega b\ll 1$.   In principle, the EFT can also calculate quantum corrections due to graviton loops, which are suppressed by powers of $\hbar/E b\ll1$ and completely negligible for astrophysical applications.  Because $G_N E\omega\ll 1$, they Feynman rules are those of flat space gravity, coupled to sources localized on a worldline.   By power counting, one finds that in the classical limit $\hbar/E b\ll 1$ only the diagrams with at most a  single external graviton survive,  of the form shown in fig.~\ref{fig:tails}.   The resulting Feynman integrals are tractable by well-established techniques~\cite{Weinzierl:2022eaz} and, at least at sufficiently small orders in $G_N E/b\ll 1$ are calculable analytically for arbitrary time-dependent source moments $I_{a_1\cdots a_\ell}$, $J_{a_1\cdots a_\ell}$.

%%%%%%%%%%%%%%%%%%%%%%%%%%%%%%%%%%%%%%%%%%%%%%%%%%%%%%%%%%%%%%%%%%%%%%%%%
%%%%%%%%%

It is convenient to compute radiative corrections to binary dynamics , \eg~effects from radiation graviton exchange, directly in the radiation EFT of Eq.~(\ref{eq:zoso}) rather than in NRGR.   The advantage of doing so is that the the factorization of contributions from the UV (the multipole moments), which depend on the specific source and IR (radiation), which are universal, is manifest.   For astrophysical applications, the relevant quantities are the zero point function, which encodes the radiative corrections to the equations of motion (radiation reaction forces), and the one-point function, which determines the waveform measured at the detector as a function of the time-dependent moments evaluated on the solutions to the PN equations of motion for the orbits.  

In such calculations, one encounters both UV and IR logarithmically divergent Feynman diagrams\footnote{The relevant Feynman integrals are isomorphic to those of a fictitious 3D Euclidean field theory of particles with propagators $1/({\vec \ell}^2-\omega^2)$ and (complex valued) ``masses'' related to the frequencies $\omega$ of the external radiation gravitons.  Restricting the EFT to the sector with at most one external  radiation graviton implies that at most one external momentum can show up in the propagators.}, even at the classical level.  In order to preserve manifest diff invariance, these are defined via dimensional regularization, where the log divergences correspond to poles in $\epsilon=4-d$.   We describe the resolution of such IR and UV singularities in the next two sections.

\subsubsection{Infrared divergences}

The IR divergences arise from so-called ``gravitational wave tails,''~\cite{blanchet} which refer to the distortion of the outgoing graviton wavefunctions by the $1/r$ gravitational potential sourced by the mass monopole, as depicted in Fig~\ref{fig:tails}(b)-(e).     They are analogous to the IR divergences found in non-relativistic Coulomb scattering, and in the gravitational context appear first at order $G_N M\omega\ll 1$ beyond leading order radiation emission.

 As an example, consider Fig~\ref{fig:tails}(b), which contains (after tensor reduction into scalar master integrals) a Feynman integral of the form
\begin{eqnarray}
\label{eq:lotail}
\nonumber
\includegraphics[width=0.25\hsize,valign=c]{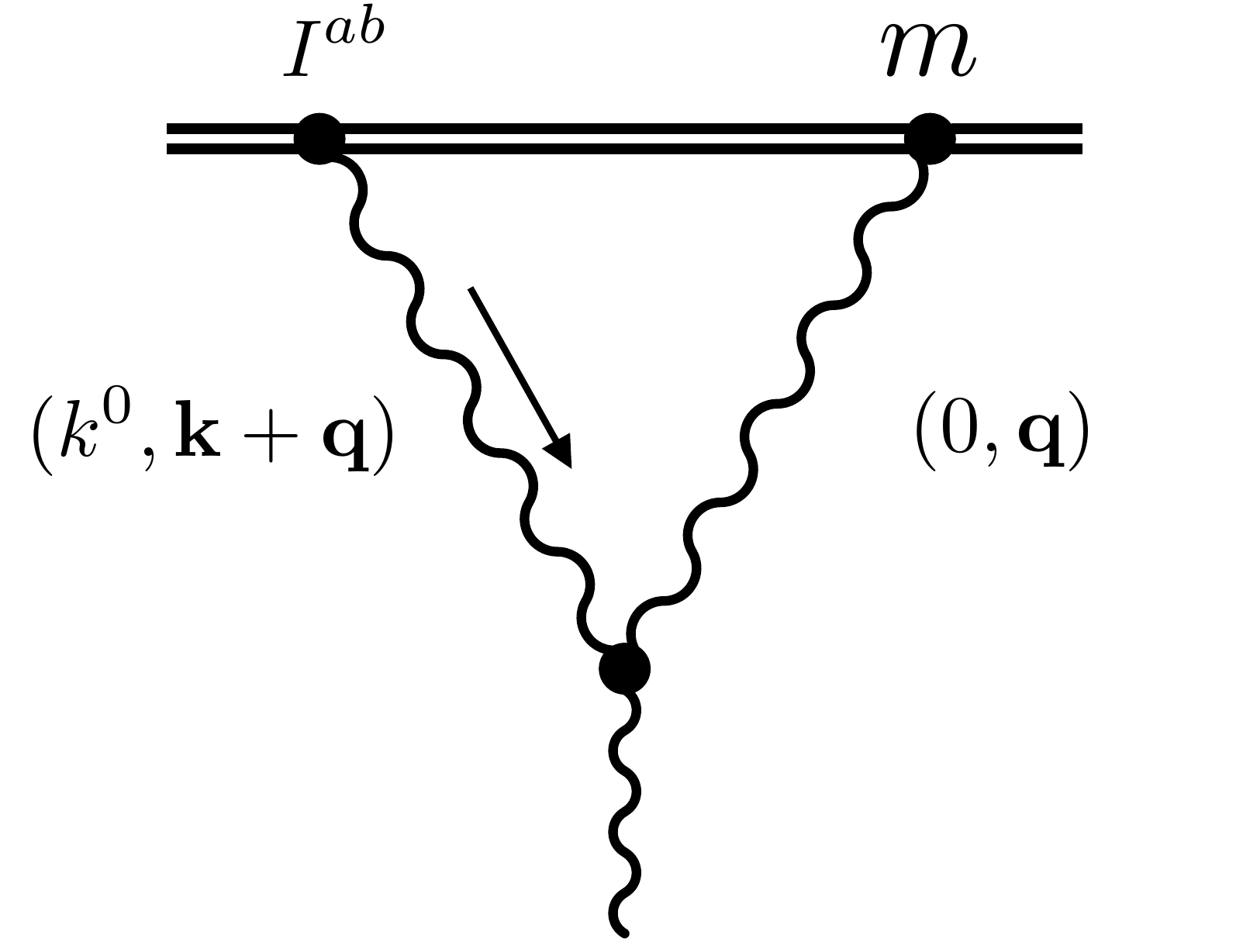} &\sim& {\cal A}^{\ell=2}_{LO}(k^0) \times \int_{\vec q} {1\over {\vec q}^2} {1\over k_0^2 - ({\vec k}+{\vec q})^2}\\
\nonumber
& & {}\stackrel{\longrightarrow}{{{\vec q}\rightarrow 0}}  {\cal A}^{\ell=2}_{LO}(k^0)\times \int_{\vec q} {1\over {\vec q}^2} {1\over 2 {\vec q}\cdot {\vec k}} = -{iG_N m |k^0|\over\epsilon_{IR}} \times {\cal A}_{LO}+\cdots,\\
\end{eqnarray}
where the LO quadrupole emission amplitude fig.~\ref{fig:tails} is given by
\begin{equation}
\label{eq:al2}
\includegraphics[width=0.2\hsize,valign=c]{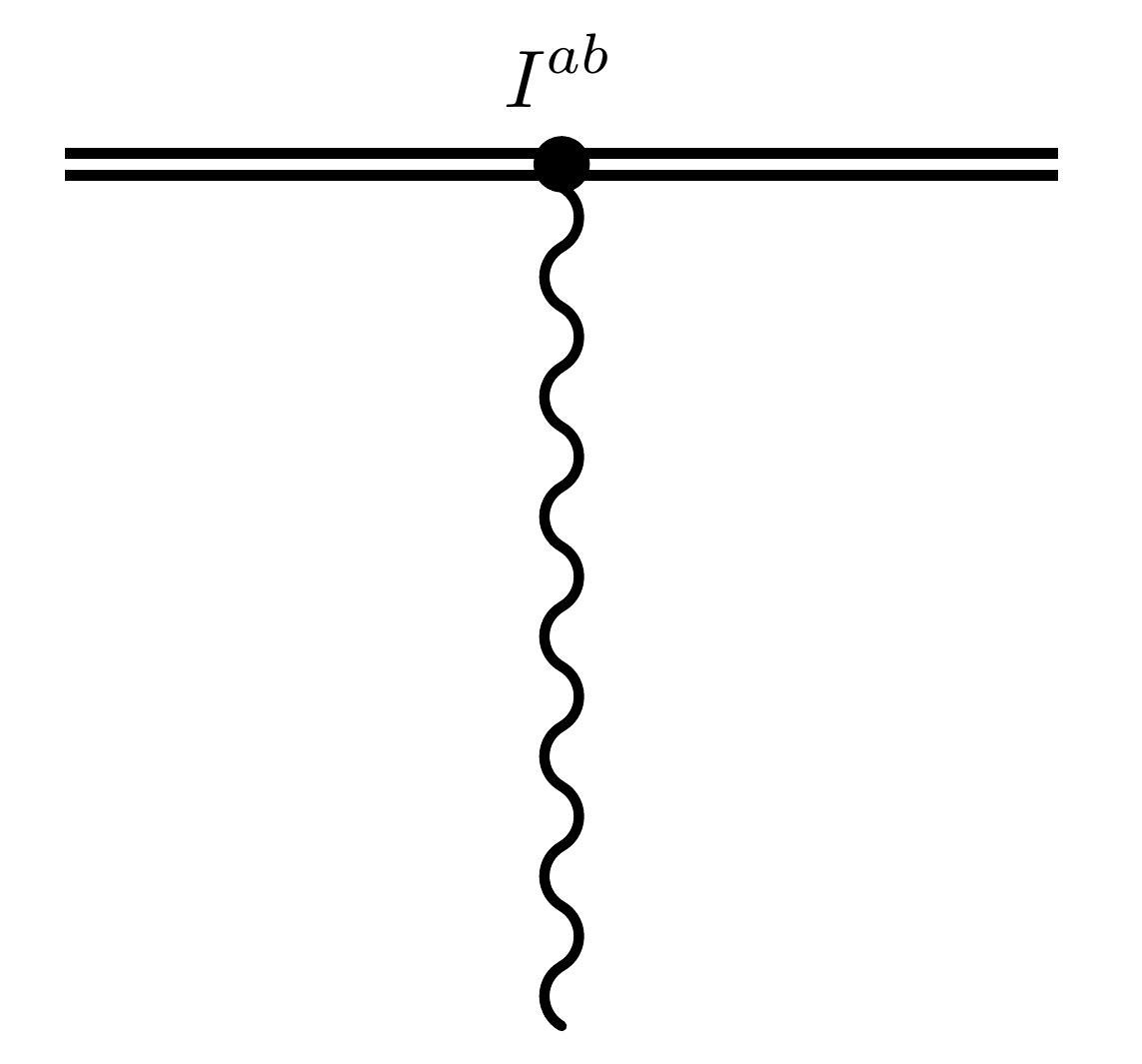}=i{\cal A}^{\ell=2}_{LO}(\omega) = {i\omega^2\over 4 m_{Pl}} \int_{-\infty}^{\infty} dt e^{i\omega t} \epsilon^*_{ab}(k) I^{ab}(t)
\end{equation}
in the CM frame of the composite object, $P^\mu=(m,0,0,0)$ and $x^\mu_\CM  = 0$.  

Eq.~(\ref{eq:lotail}) has a logarithmic $1/\epsilon_{IR}$ IR pole when the external graviton goes on-shell, $k_0^2\rightarrow {\vec k}^2$, from the region of soft loop momentum ${\vec q}\rightarrow 0$.    Similarly, going next-to-leading order, there is a $1/\epsilon^2_{IR}$ IR divergence in the diagram Fig~\ref{fig:tails}(c), 
\begin{equation}
\includegraphics[width=0.2\hsize,valign=c]{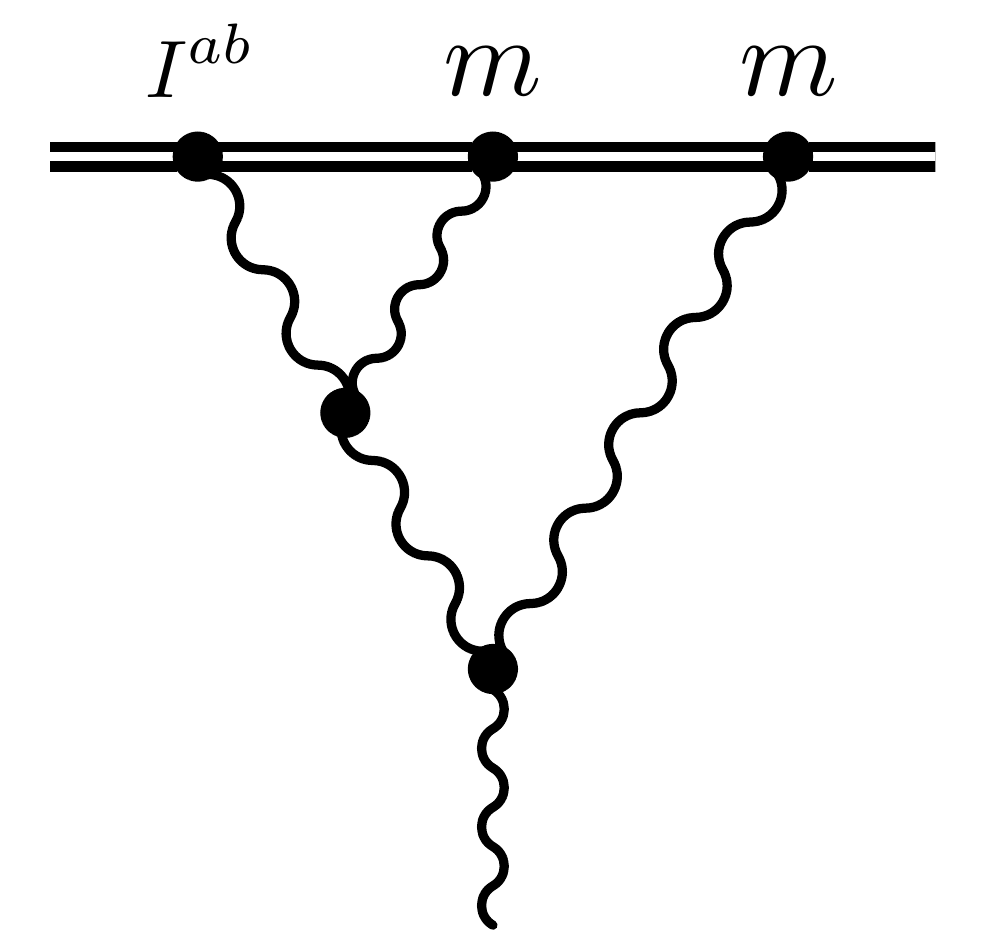}\sim {1\over 2!} \left[ -{iG_N m |k^0|\over\epsilon_{IR}}\right]^2 \times {\cal A}^{\ell=2}_{LO}, 
\end{equation}
when the external graviton goes on-shell, from the region where all the internal momenta go to zero.  More generally, at order $(G_N M\omega)^n$, there is  $1/\epsilon^n_{IR}$ pole from the ``ladder'' diagram containing $n$ internal gravitons, each sourced by a mass monopole, connect with the radiation graviton emitted by the quadrupole source.   In addition, graviton emission in the higher $\ell$-th multipole channels receives corrections from $n$-the order ladder graphs analogous to those shown in fig.~\ref{fig:tails}.

The resolution~\cite{Goldberger:2009qd,Porto:2012as} of these IR divergences is similar to what happens in QED~\cite{Weinberg:1965nx}:    in frequency space,  the series of $G_N M\omega/\epsilon_{IR}$ poles exponentiates into an overall phase factor in the graviton emission amplitude~\cite{Goldberger:2009qd}.  This phase then cancels in the gravitational energy flux (emitted power), 
\begin{equation}
\label{eq:powah}
{d^3 P^\mu\over d\Omega d\omega} = {\omega^2\over 8\pi^2} k^\mu \left|{\cal A}(\omega)\right|^2, 
\end{equation}
where $k^\mu=(\omega,\omega{\vec n})$ points from the source in the direction of the gravitational wave detector at ${\cal I}^+$.   Similarly in the flux of angular momentum to infinity, which depends only on $\left|{\cal A}\right|^2$, is free of IR divergences.

One can also show that the IR divergences do not affect the waveform seen at infinity:  upon transforming the amplitude to the time domain, the IR divergent phase has the effect of shifting the argument of the gravitational wave signal $h(t)$, Eq.~(\ref{eq:wave}), recorded at the detector.  This shift is arbitrary, and is absorbed into the definition of the (experimentally determined) ``initial time'' when the signal first enters the detector's frequency band~\cite{Porto:2012as}.

Note that despite the disappearance of the IR regulator $1/\epsilon_{IR}$ from infrared safe physical observables, the gravitational wave tails leave a measurable imprint on the waveform at ${\cal I}^+$.  For example, refs.~\cite{Asada:1997zu,Khriplovich:1997ms} have shown that the entire series of powers of $4\pi G_N M\omega$ in the graviton emission amplitude squared is given by the Sommerfeld factor~\cite{Sommerfeld:1931qaf} 
\begin{equation}
S(\omega)=4\pi G_N M\omega/(1-e^{-4\pi G_N M \omega}),
\end{equation}
familiar from Coulomb scattering in non-relativistic quantum mechanics.

\subsubsection{UV divergences and renormalization group evolution}

In addition to the usual graviton loop UV divergences of effective quantum gravity~\cite{tv}, suppressed by powers of $\hbar/L\ll 1$, the theory is also afflicted by short distance singularities which persist even in the classical limit.   These classical UV divergences are generic in field theories that are coupled to defects, \ie~Dirac delta function sources of non-zero codimension.    Such divergences are resolved by the finite transverse size of the defect in the full theory, but, in the EFT, can be absorbed into local counterterms on the defect worldvolume, generating in some cases non-trivial renormalization group (RG) flows for the Wilson coefficients, \emph{even at the classical level}~\cite{Goldberger:2001tn,Goldberger:2009qd}.   

Power UV divergences are generated already at leading in perturbation theory (for instance from the energy stored in the composite objects $1/r$ Newtonian gravitational field), but dimensional regularization simply defines these to be zero.   More interesting logarithmic divergences arise in  \zosoir  starting at order $(G_N M \omega)^2$ in the expansion.  They appear for instance in the order  $(G_N M \omega)^2$ corrections to graviton emission in the mass quadrupole channel, as in fig.~\ref{fig:tails}(c)-(e).    For example, 
\begin{eqnarray}
\label{eq:uvpole}
\nonumber
\includegraphics[width=0.25\hsize,valign=c]{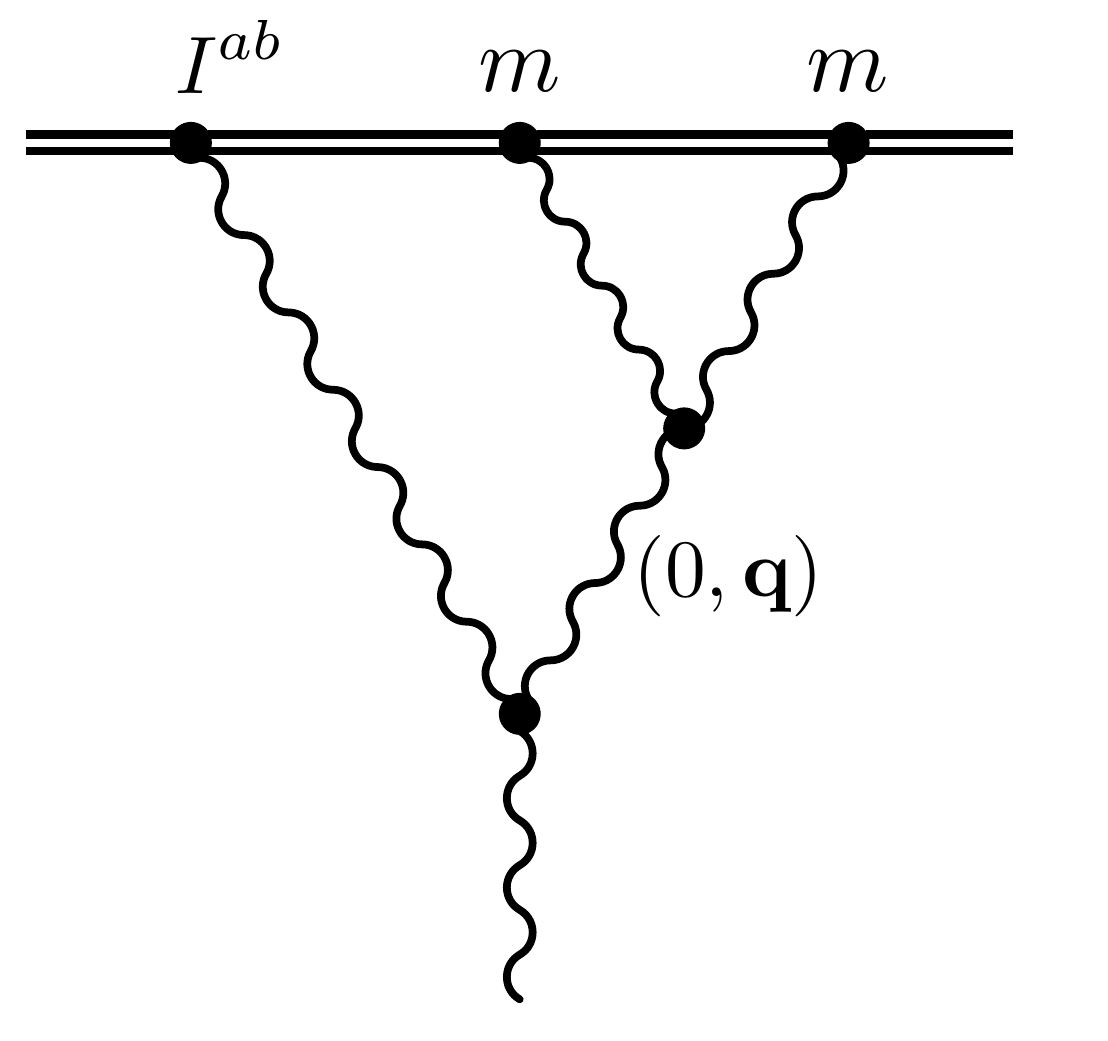} &\sim& {\cal A}^{\ell=2}_{LO}(k^0) \times \int_{\vec q} {1\over |{\vec q}|} {1\over k_0^2 - ({\vec k}+{\vec q})^2}  \stackrel{\longrightarrow}{{{\vec q}\rightarrow \infty}}  {\cal A}^{\ell=2}_{LO}(k^0)\times \int_{\vec q} {1\over |{\vec q}|^3} \\
& & \sim {(G_N m k^0)^2\over\epsilon_{UV}}\times {\cal A}^{\ell=2}_{LO}(k^0),
\end{eqnarray}
has a $1/\epsilon_{UV}$ pole from the $|{\vec q}|\gg k^0$ region of loop momentum.

Physically, the log divergence in Eq.~(\ref{eq:uvpole}) is generated by the propagation of the emitted radiation mode in the short distance part of the source's static gravitational potential, namely the order $(G_N M/ r)^2$ relativistic correction to the metric sourced by the mass monopole in Eq.~(\ref{eq:zoso}).   This singular $\sim 1/r^2$ potential is an artifact of the EFT.  It begins to dominate at distance scales where the multipole expansion is no longer a good description of the internal structure of the binary in the full theory.   For example, in the diagram  fig.~\ref{fig:tails}(d), the Fourier transform of the $1/r^2$ potential is embedded in the subdiagram
\begin{equation}
\includegraphics[width=0.25\hsize,valign=c]{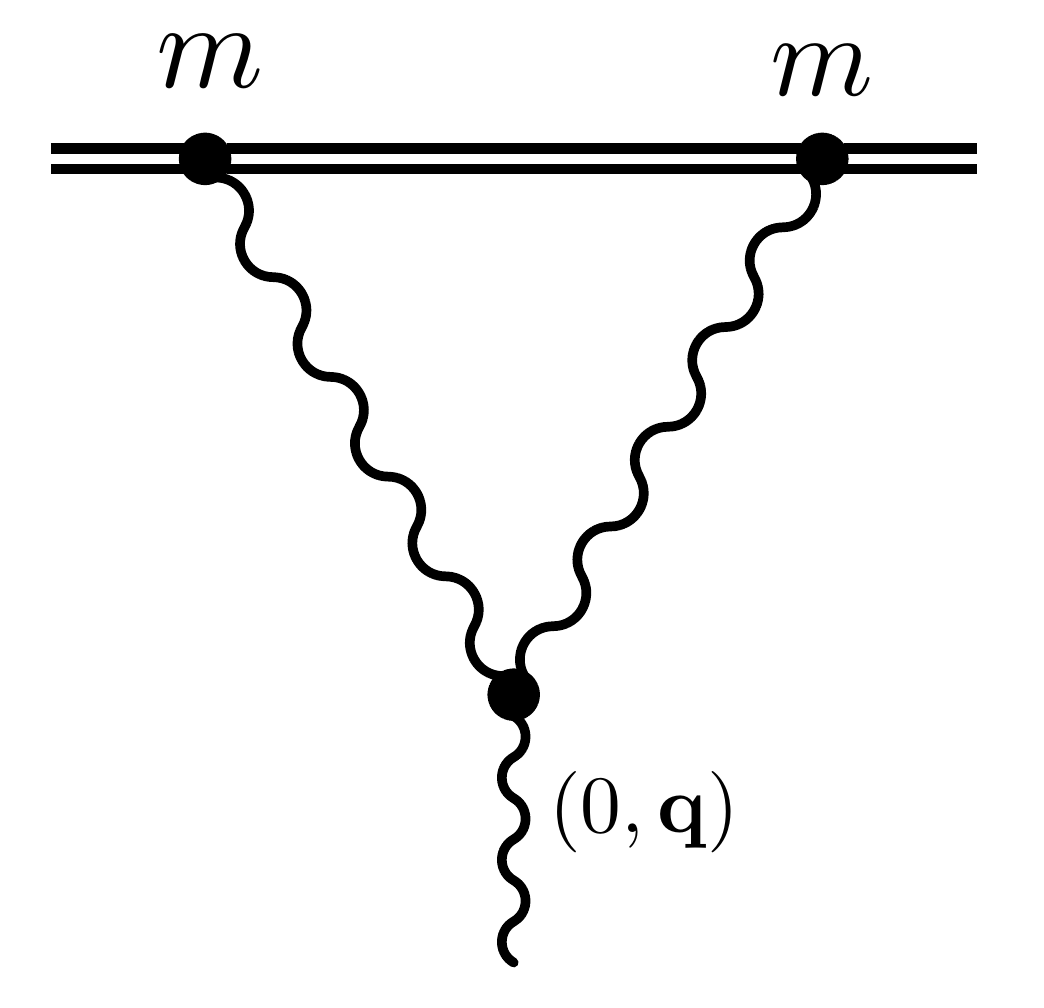}\sim {1\over |{\vec q}|}, 
\end{equation}
which accounts for the factor ${1/ |{\vec q}|}$ in the Feynman integral over the momentum flowing out of the quadrupole vertex in fig.~\ref{fig:tails}(d).

The sum of the diagrams in figs.~\ref{fig:tails}(c)-(e) contains a UV divergent\footnote{The sum of figs.~\ref{fig:tails}(c)-(e) also contains $1/\epsilon_{IR}$ and $1/\epsilon_{IR}^2$ singularities, as discussed in the previous section.   It is possible~\cite{Goldberger:2009qd} to use the method of regions~\cite{Beneke:1997zp} to disentangle the UV and IR contributions to the coefficient of the $1/\epsilon$ pole.} term
\begin{equation}
{1046\over 315} {\left(G_N M\omega\right)^2\over \epsilon_{UV}}\times {\cal A}^{\ell=2}_{LO}(\omega)
\end{equation}
which is analytic in the frequency of the emitted graviton, and can be absorbed into a local counterterm on the zoomed out object's worldline.  Specifically,  the UV pole renormalizes the electric quadrupole Wilson coefficient $I_{ab}(\tau)$ in \zosoir.  

The graviton emission matrix element is finite when expressed in terms of the renormalized quadrupole, at the expense of introducing an arbitrary renormalization scale $\mu$.   Explicit logarithmic dependence on $\mu$ arising from the evaluation of the dimensionally regularized Feynman integrals cancels against the subtraction scale dependence of the renormalized moment $I_{ab}(\mu,\tau)$, ensuring that the emission amplitude (a physical observable) is insensitive to the choice of $\mu$.    This requires that, in frequency space, the electric quadrupole satisfies the RG equation (RGE) ~\cite{Goldberger:2009qd}
\begin{equation}
\label{eq:rg}
\mu {d\over d\mu} I_{ab}(\omega,\mu) = -{214\over 105} (G_N M\omega)^2 I_{ab}(\omega,\mu).
\end{equation}

The RGE is universal, so it can be used to predict the pattern of logarithms of the frequency $\omega$ in the matrix element for the emission of soft graviton radiation from any localized source of finite size.    By running Eq.~(\ref{eq:rg}) from $\mu_{UV}\sim 1/b$ to $\mu_{IR}\sim \omega,$ it is possible to ``resum'' the series of powers $(G_N m \omega)^{2n} \ln^n (\omega b)$ in quadrupole emission.   Similar RG flows to Eq.~(\ref{eq:rg}) occur for multipole moments beyond $\ell=2$, \cite{Goldberger:2012kf,Galley:2015kus,Almeida:2021jyt}, so all terms of the form $(G_N m\omega)^n \ln^m\omega b$ induced by soft graviton radiation are in principle known.

As an application of Eq.~(\ref{eq:rg}), we can use it to predict the pattern of PN logarithms in the $\ell=2$ electric radiated power from a non-relativistic binary, by inserting the renormalized quadrupole into Eq.~(\ref{eq:al2}) and running the RG from a the UV at a scale $\mu_{UV}\sim 1/r$ where the binary matches onto \zosoir down to the IR scale $\mu_{IR}\sim \omega$.   For example, the entire series of integer powers $(v^6\ln v)^n$ in the energy flux of $\ell=2$ radiation from a binary in a circular orbit with velocity $v\ll 1$, normalized to the leading order quadrupole radiation formula,
\begin{eqnarray}
\label{eq:v6ln}
\nonumber
{{\dot E}^{\ell=2_E}_{\mbox{log}}\over {\dot E}^{\ell=2_E}_{\mbox{LO}}}=\left[\mu\over \mu_0\right]^{-{428\over 105} (G_N M\omega)^2}&=&1-{428\over 105} v^6 \ln v + {91592\over 11025} v^{12} \ln^2 v\\
& & {} -{39201376\over 347287} v^{18}\ln^3 v +\cdots,
\end{eqnarray}
is fully determined by RG evolution.

EFT reasoning predicts that the same pattern of logarithms should also appear in other kinematic regimes.   For example, consider binary black holes in the ``EMRI'' limit of hierarchical masses, so that the smaller constituent can be treated as a perturbation of the Kerr geometry sourced by the heavier one.    A semi-analytic treatment~\cite{Fujita:2011zk} of the energy flux from non-circular $v\ll 1$orbits around a black hole up to  14PN (!) order found logarithmic terms that match those predicted by Eq.~(\ref{eq:v6ln}).    The fact that the non-analytic low frequency behavior of waves propagating in curved spacetime can be obtained from $1/\epsilon_{UV}$ poles of Feynman diagrams in flat spacetime provides a sharp example of the universality of the worldline EFT predictions.

While the RG can efficiently generate the coefficients of logarithms, by itself cannot fix the precise UV scale $\mu_{UV}\sim 1/r$ where we define the Wilson coefficients.   That must be determined by performing a matching calculation to the more UV complete theory that resolves the internal structure of the radiating object.   For non-relativistic applications, with $G_N M\omega\sim v^3$ , a 3PN matching calculation to NRGR is needed to fix the relation between the renormalized moments at $\mu_{UV}\sim 1/r$ and the microscopic (orbital) degrees of freedom of the binary system.    The basic procedure for matching  \zosoir to NRGR is the subject of the next section.

\subsection{Matching NRGR to \zosoir}
\label{sec:matching}

Focusing now on non-relativistic binaries, we assign power counting $I_{a_1\cdots a_\ell}\sim M r^\ell$, $J_{a_1\cdots a_\ell}\sim M v r^\ell$, in which case the two expansion parameters of \zosoir control effects which are down by different powers in $v$:   multipole corrections in powers of $\omega r\sim v$ and gravitational wave tails such as those in fig.~\ref{fig:tails} in powers of $G_N M\omega\sim v^3$.   In the PN regime, it is possible to fix the Wilson coefficients in Eq.~(\ref{eq:zoso}) by matching to NRGR at distance scales $\gsim r$ where the two theories are both valid.   

Since the potential modes in NRGR do not go on-shell, the effective Lagrangian for the radiation field coupled to the worldline degrees of freedom ${\vec x}_{1,2}$ (and the spins ${\vec S}_{1,2}$ which we have been largely ignoring) is local at length scales larger than the orbital scale $r$.   Formally, we can obtain this effective Lagrangian by integrating out $H_{{\vec p};\mu\nu}$,
\begin{equation}
\label{eq:NRGRPI}
e^{i S_{\includegraphics[width=0.05\hsize,valign=c]{zosoir}}[{\vec x}_{1,2},{\bar g}]} = \int {\cal D} H_{{\vec p};\mu\nu} e^{i S_{NRGR}},
\end{equation}
without the need to impose in-in boundary conditions.  The long distance theory can be organized as an expansion in powers of the radiation field,
\begin{equation}
\label{eq:zoso1}
S_{\includegraphics[width=0.05\hsize,valign=c]{zosoir}}  = S_{EH}[{\bar g}]  + \Gamma^{(0)}[{\vec x}_{1,2}] + \Gamma^{(1)}[{\vec x}_{1,2},{\bar h}]+\cdots, 
\end{equation}
where $\Gamma^{(0)}[{\vec x}_{12}]$ is independent of ${\bar h}_{\mu\nu}$, while $\Gamma^{(1)}$ depends linearly,
\begin{equation}
\label{eq:pseudo}
\Gamma^{(1)}[{\vec x}_{1,2},{\bar h}]=-{1\over 2 m_\Pl} \int d^4 x {\bar \tau}^{\mu\nu}(x) {\bar h}_{\mu\nu}.
\end{equation}
for some function ${\bar \tau}^{\mu\nu}(x)$ of the orbital degrees of freedom.

In perturbation theory, ${\bar \tau}^{\mu\nu}(x)$ can be identified with the sum over Feynman diagrams that only carry internal potential graviton lines and a single external radiation graviton.    We perform the path integral over $H_{{\vec p};\mu\nu}$ in background field gauge~\cite{DeWitt:1967ub,Abbott:1980hw}, in which the gauge fixing term is invariant under diffeomorphisms $\delta{\bar h}_{\mu\nu} = \nabla_\mu\xi_\nu+ \nabla_\nu\xi_\mu$ acting on the background filed.   This implies in particular the conservation law $\partial_\nu {\bar \tau}^{\mu\nu}=0$.   For applications to classical binary inspirals it is only necessary to match to NRGR in the sectors with zero or one external radiation gravitons, as subsequent emissions bring down powers of $\hbar/L\rightarrow 0$ .  By diffeomorphism invariance, knowing the vacuum and single-graviton matrix elements in Eq.~(\ref{eq:zoso1}) is sufficient to determine the relevant non-linear couplings of the radiation mode as well.

\subsubsection{Two-body potentials}
\label{sec:pot}

%%%%%%%%%%%%%%%%%%%%%%%%%%%%%%%%%%%%%%%%%%%%%%%%%%%%%%%%%%%%%%%%%%%%%%%%%
\begin{figure}[t]
\begin{center}
\includegraphics[width=0.90\hsize]{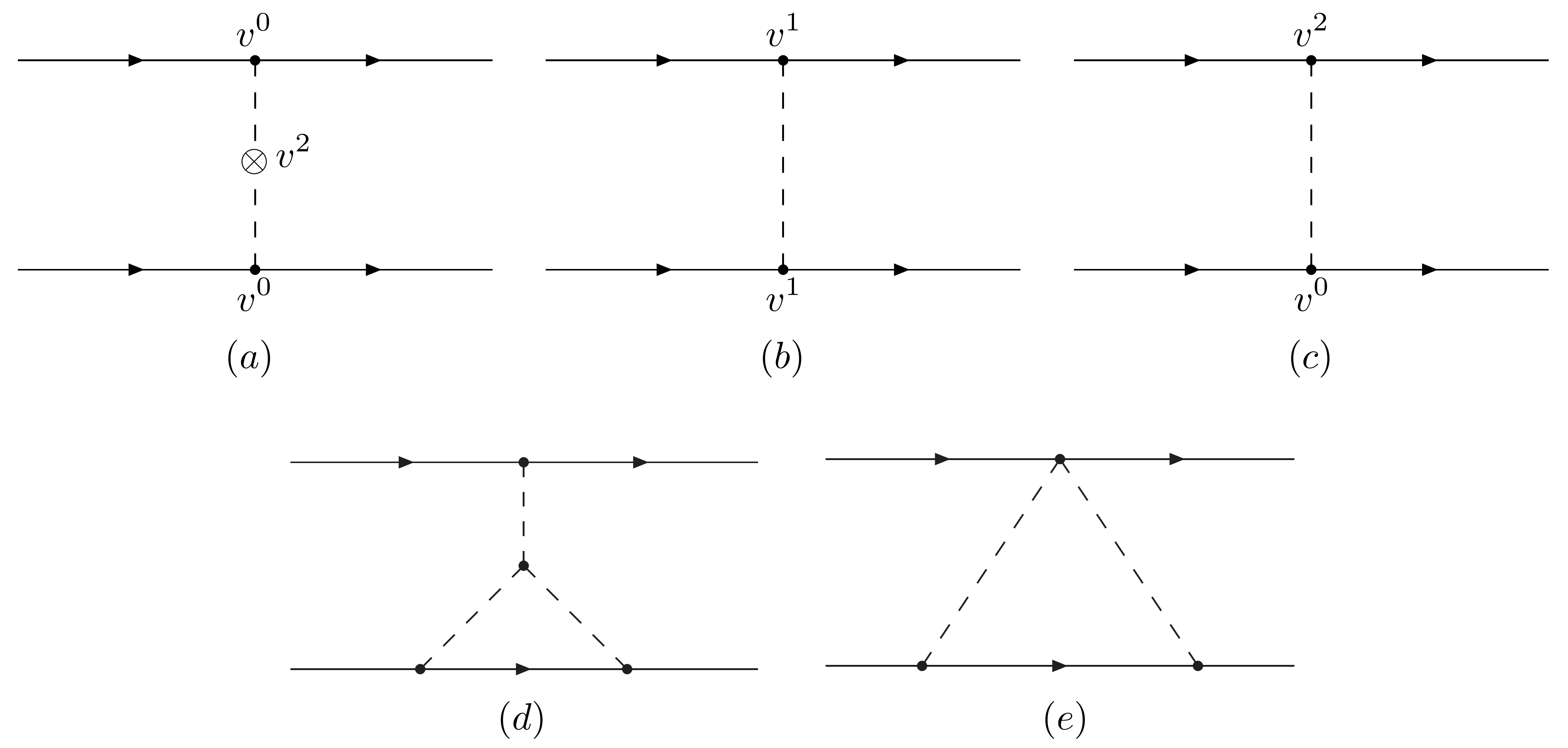}
\end{center}
\caption{1PN corrections to conservative two-body dynamics.   In addition to the diagrams shown, there are also mirror $(1\leftrightarrow 2)$ diagrams to (c),(d),(e) which also contribute at ${\cal O}(v^2)$.}
\label{fig:EIH}
\end{figure}
%%%%%%%%%%%%%%%%%%%%%%%%%%%%%%%%%%%%%%%%%%%%%%%%%%%%%%%%%%%%%%%%%%%%%%%%%%%

In NRGR, the term $\Gamma^{(0)}[{\vec x}_{12}]$ corresponds to Feynman diagrams involving arbitrary insertions of potential and worldline vertices, but no internal or external radiation lines.   As such, it is a functional only of the worldline variables, and therefore defines the \emph{conservative} dynamics of the two-body system, in the sense that the Euler-Lagrange equations for $\Gamma^{(0)}[{\vec x}_{12}]=\int dt L[{\vec x}_{1,2}]$, with 
\begin{equation}
\label{eq:pot}
L[{\vec x}_{1,2}] = -\sum_A m_A\sqrt{1-{\vec v}_A^2} +\Delta L[{\vec x}_{1,2}],
\end{equation}
define the equations of motion for the trajectories ${\vec x}_{1,2}$ including all possible short distance gravitational interactions, but neglecting radiation (radiation reaction effects will be discussed below in sec.~\ref{sec:rr}).    For example, using  the integral$\int_{\vec p} e^{i {\vec p}\cdot{\vec x}}/{\vec p}^2 = (4\pi|{\vec x}|)^{-1}$ as well as the definition  $m_\Pl^{-2}=32\pi G_N$, the leading order term,
\begin{eqnarray*}
\nonumber
\includegraphics[width=0.25\hsize,valign=c]{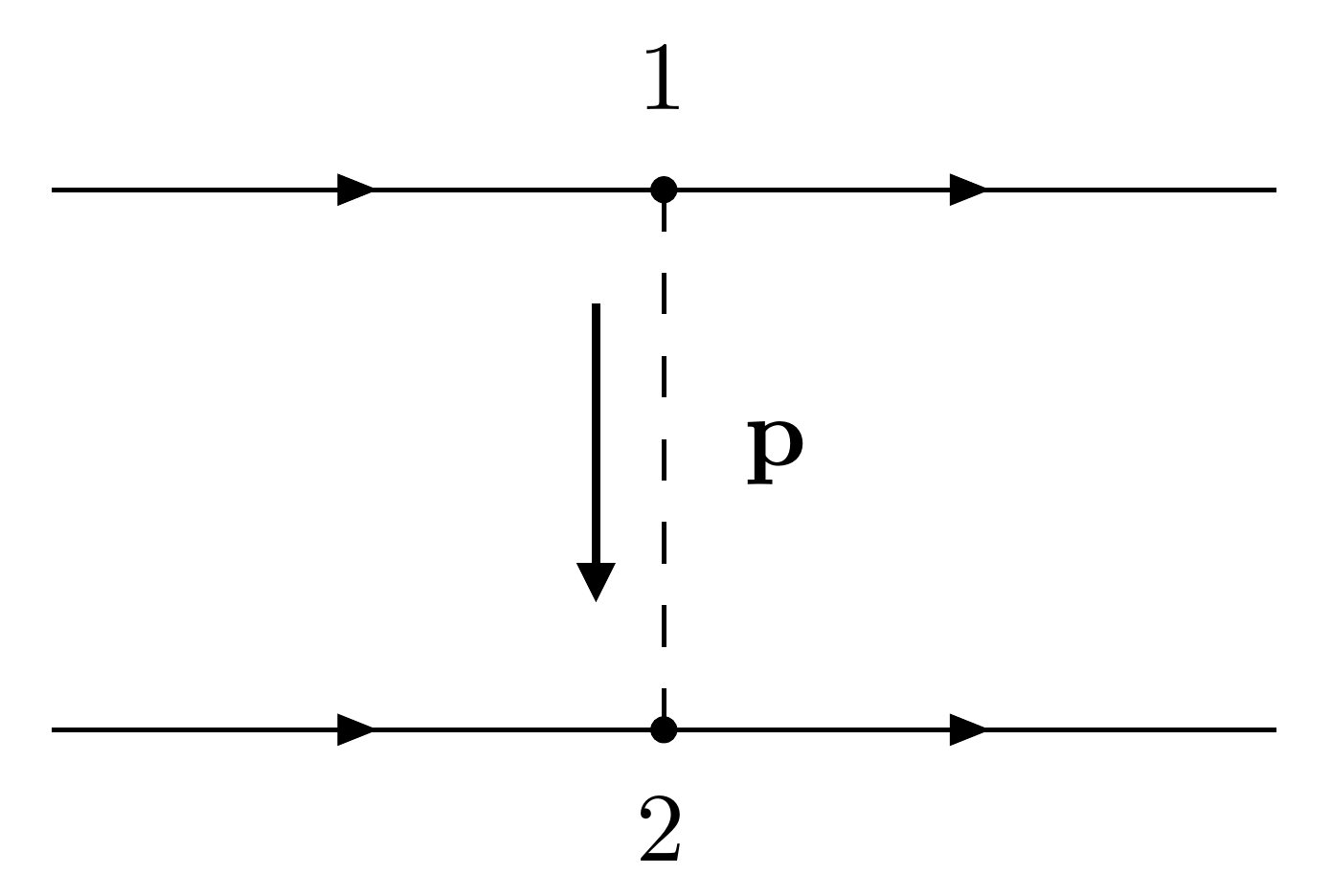} &=& \int dt_1 dt_2\int_{{\vec p},{\vec q}} \left(-{i m_1\over 2 m_{\Pl}} e^{i{\vec p}\cdot {\vec x}_1(t_1)}\right)   \left(-{i m_2\over 2 m_{\Pl}} e^{i{\vec q}\cdot {\vec x}_2(t_2)}\right)\\
\nonumber
& & {}\times  \langle H_{{\vec p};00}(t_1)  H_{{\vec q};00}(t_2)\rangle  = {i m_1 m_2\over 4m_\Pl^2} P_{00,00}\int dt \int_{{\vec p}}  {e^{i{\vec p}\cdot {\vec x}_{12}} \over {\vec p}^2}\\
&=&i  \int dt {G_N m_1 m_2\over |{\vec x}_1-{\vec x}_2|}=i\Delta L_{0PN},
%=\int dt_1 dt_2\int_{{\vec p},{\vec q}} \left(-{i m_1\over 2 m_{\Pl}} e^{i{\vec p}\cdot {\vec x}_1(t_1)}\right) \langle H_{{\vec p};00}(t_1)  H_{{\vec q};00}(t_2)\rangle   \left(-{i m_2\over 2 m_{\Pl}}\right } e^{i{\vec q}\cdot {\vec x}_2(t_2)}\right) = % {i m_1 m_2\over 8 m_\Pl^2} \int dt \int_{\vec p}  e^{i{\vec p}\cdot {\vec x}_{12}}\in
\end{eqnarray*}
 reproduces the Newtonian gravitational interaction, as one would hope.
 
Similarly, the sum of the diagrams fig.~\ref{fig:EIH} yields the so-called Einstein-Infeld-Hoffman (EIH) correction to non-relativistic gravitational motion,
\begin{eqnarray}
\nonumber
\Delta L_{1PN}={ L}_{EIH} &=&{1\over 8}\sum_A m_A {\vec v}_A^4 + {G_N m_1 m_2\over 2 |{\vec x}_1-{\vec x}_2|}\left[3({\vec v}_1^2+{\vec v}_2^2) - 7 {\vec v}_1\cdot  {\vec v}_2 - {({\vec v}_1\cdot  {\vec x}_{12})({\vec v}_2\cdot  {\vec x}_{12}) \over {\vec x}_{12}^2}\right]  \\
& & {}- {G_N^2 m_1 m_2 (m_1+m_2)\over 2 |{\vec x}_1-{\vec x}_2|^2}
\end{eqnarray}
where we have also included the first relativistic correction to the kinetic energy of the particles, from expanding the first term in Eq.~(\ref{eq:pot}).   The  EIH Lagrangian has implications for planetary motion in the solar system, so that, \eg~the perihelion advance of Mercury's orbit can be regarded as a probe of the cubic self-interaction vertex of the graviton, see fig.~\ref{fig:EIH}(d).

For two-body systems, at higher PN orders, one encounters momentum space Feynman integrals that are formally the same that would appear in the radiative corrections to propagators (two-point functions) in a massless Euclidean QFT living in $3-\epsilon$ spatial dimensions.   Such multi-loop integrals are tractable by standard techniques of perturbative QFT (see~\cite{Smirnov:2006ry,Weinzierl:2022eaz} for comprehensive reviews).    In a generic gauge,  the potentials at the $n$PN order require the evaluation of $n$-loop Feynman integrals, but by exploiting a convenient field redefinition of the graviton that is well suited to the non-relativistic limit, introduced in refs.~\cite{Kol:2007rx,Kol:2007bc,Kol:2010ze}, it is possible to postpone the number of loops by one order in perturbation theory.  Within the EFT approach, the non-relativistic spin-independent potentials at 2PN order where first tackled in ref.~\cite{Gilmore:2008gq}, which introduced some of the tools necessary to carry out higher order PN loop diagrams.  The systematic study of higher order spinless PN potentials was initiated in~\cite{Foffa:2011ub} and extended~\cite{Foffa:2012rn,Foffa:2013gja,Foffa:2016rgu} to 4PN in~\cite{Foffa:2016rgu}, which is the current state-of the-art in NRGR.

\subsubsection{Radiation and multipole moments}

%%%%%%%%%%%%%%%%%%%%%%%%%%%%%%%%%%%%%%%%%%%%%%%%%%%%%%%%%%%%%%%%%%%%%%%%%
\begin{figure}[t]
\begin{center}
\includegraphics[width=0.90\hsize]{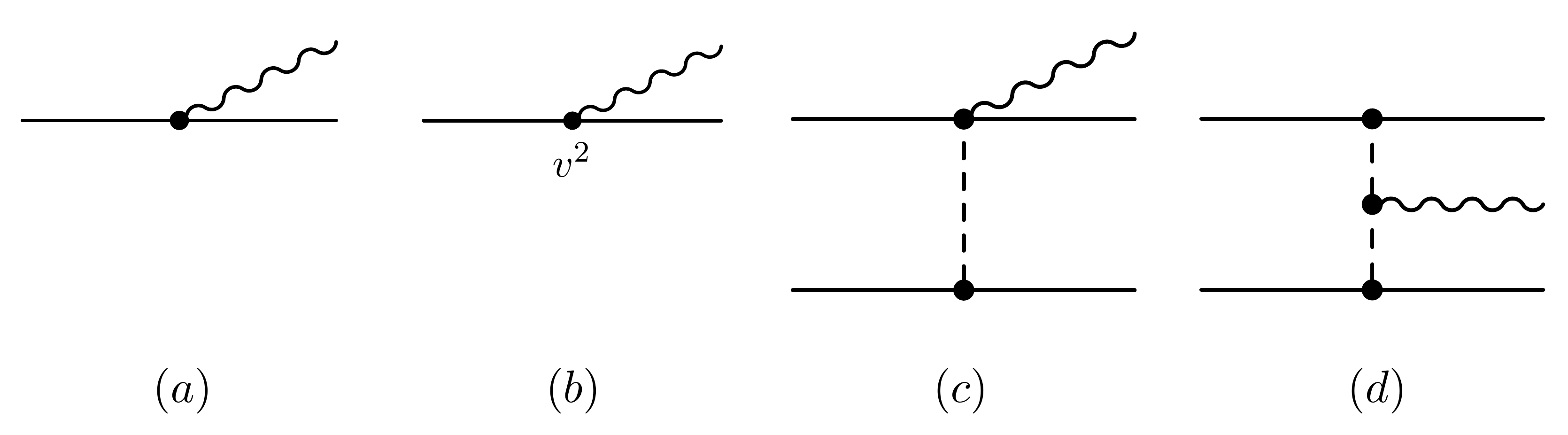}
\end{center}
\caption{Diagrammatic expansion of ${\bar\tau}^{\mu\nu}$ in NRGR.  (a) is order 0PN (b)-(d) are 1PN.}
\label{fig:rad1pt}
\end{figure}
%%%%%%%%%%%%%%%%%%%%%%%%%%%%%%%%%%%%%%%%%%%%%%%%%%%%%%%%%%%%%%%%%%%%%%%%%%%

In light of Eq.~(\ref{eq:NRGRPI}), the function ${\bar \tau}^{\mu\nu}$ can be calculated by summing all Feynman diagrams in NRGR containing only internal potential lines and a single off-shell external radiation line, see fig.~\ref{fig:rad1pt}.   Even though in background field gauge,  ${\bar \tau}^{\mu\nu}$ is conserved, $\partial_\nu{\bar \tau}^{\mu\nu}=0$, it should not be confused with the energy-momentum pseudo-tensor $\tau^{\mu\nu}$ of the entire system, which receives both radiative and potential contributions.   Nevertheless, the total four-momentum of the composite system can be calculated directly from ${\bar \tau}^{\mu\nu}$, 
\begin{equation}
P^\mu = \int d^{d-1}{\vec x}  \,{\bar \tau}^{0\mu}(x)= \int d^{d-1}{\vec x}  \,\tau^{0\mu}(x),
\end{equation}
since radiative corrections in $\zosoir$ do not renormalize\footnote{In dimensional regularization, where scaleless momentum space integrals are defined to be zero.} the static part of $\tau^{\mu\nu}(k)=\int d^4 x e^{ik\cdot x} \tau^{\mu\nu}(x)$, \ie~the part proportional to a Dirac delta function of the external graviton frequency $k^0$.   Similarly, the system's center of mass worldline,
\begin{equation}
\label{eq:xcm}
{\vec X}_{\CM}(x^0) = {\int d^{d-1}{\vec x}\,\,  {\vec x} \tau^{00}(x^0,{\vec x})\over P^0}
\end{equation}
does not get renormalized by radiation, and may be evaluated by replacing ${\tau}^{\mu 0}\mapsto {\bar\tau}^{\mu0}$ inside the integral.   Because ${\bar\tau}^{\mu\nu}$ is conserved, the CM coordinate drifts with uniform velocity $d{\vec X}_{\CM}/dx^0={\vec P}/P^0$ in any asymptotic Lorentz frame.

If ${\bar \tau}^{\mu\nu}$ is given, one can extract the moments in \zosoir by inserting Eq.~(\ref{eq:taylor}) into Eq.~(\ref{eq:zoso1}) and decomposing the coefficients into representations of the $SO(3)$ that preserves the CM four-momentum of the composite system.   It is convenient to do this in the CM frame $u^\mu = P^\mu/P^0=(1,{\vec 0})$ and ${\vec X}_{\CM}=0$.   In this frame, the leading term in the Taylor expansion,
%$$
\begin{equation}
\label{eq:lomult}
-{1\over 2 m_{\Pl} }\int dx^0 \left[\int d^{d-1}{\vec x} {\bar \tau}^{\mu\nu}(x^0,{\vec x})\right] {\bar h}_{\mu\nu}(x^0,0),
\end{equation}
%$$
receives contributions from terms with $\mu\nu=00,ij$ only.

By Eq.~(\ref{eq:tx}), the $\mu\nu=00$ part is supposed to match the expansion of the term $\int d\tau L(X,x(\tau),{\bar g}) \subset S_{\includegraphics[width=0.05\hsize,valign=c]{zosoir}}$ in the EFT to linear order in ${\bar h}_{\mu\nu},$ \ie
\begin{equation}
H_X=\int d^{d-1}{\vec x} {\bar\tau}^{00}.
\end{equation}
On the other hand, the $\mu\nu=ij$ part of Eq.~(\ref{eq:lomult}) does not contribute at leading order in the multipole expansion, since the conservation law $\partial_{\nu} {\bar \tau}^{\mu\nu}=0,$ implies ``moment relations'' identical to those obeyed by the full pseudotensor  (see any textbook on gravitational radiation, \eg~\cite{Weinberg:1972kfs,Maggiore:2007ulw})
\begin{equation}
\label{eq:mr}
\int d^{d-1}{\vec x} {\bar \tau}^{ij}(x^0,{\vec x}) = {1\over 2}{d^2\over {dx^0}^2} \int d^{d-1}{\vec x} {\bar \tau}^{00}(x^0,{\vec x}) x^i x^j.
\end{equation}
Therefore, after integrating by parts, the ${\bar \tau}^{ij} {\bar h}_{ij}$ term in Eq.~(\ref{eq:lomult}) is proportional on-shell to $\partial_0^2 {\bar h}_{ij}=\nabla^2 {\bar h}_{ij}$, which is order $\ell=2$ in the multipole expansion.

In a general frame, there are two independent $SO(3)$ moments at order $\ell=1$, both of which are contained in the decomposition of the term
\begin{equation}
\label{eq:so}
-{1\over 2 m_{\Pl} }\int dx^0 \left[\int d^3{\vec x} {\bar \tau}^{\mu\nu}(x^0,{\vec x}) x^j\right] \partial_ j{\bar h}_{\mu\nu}(x^0,0)
\end{equation}
in Eq.~(\ref{eq:pseudo}).    The electric dipole moment is simply the CM coordinate, contained in the  $\mu\nu=00$ part,
\begin{equation}
-{1\over 2 m_{\Pl}} P^0 \int dx^0{X}^i_{\CM}\partial_i {\bar h}_{00}.
\end{equation}
which we set to zero by our choice of coordinates.  To extract the magnetic dipole, project the  $\mu\nu=0i$ part of Eq.~(\ref{eq:so}) into even or odd parts under permutations $i\leftrightarrow j$.   The symmetric part,  
\begin{equation}
\int d^3 {\vec x} x^{(i} {\bar \tau}^{j)0}= {1\over 2} \int d^{3} {\vec x} \partial_k \left(x^i x^j\right){\bar\tau}^{0k}={1\over 2} {d\over dx^0}\int d^{3} {\vec x} x^i x^j {\bar\tau}^{00},
\end{equation}
is, after integrating parts and using the conservation of ${\bar\tau}^{\mu\nu}$, higher order in the multipole expansion.    Likewise the various projections of $\int d^3{\vec x} {\bar\tau}^{ij} x^k$ onto irreps of the permutation group do not generate moments with $\ell<2$.   This leaves behind a coupling of the form
\begin{equation}
 -{1\over 4 m_{\Pl}} \int dx^0 L^{ij} \left(\partial_ i{\bar h}_{0 j}-\partial_ j{\bar h}_{0 i}\right),
\end{equation}
to the angular momentum of the system, \ie the antisymmetric moment 
\begin{equation}
L^{ij}=\int d^3{\vec x} \left(x^i \tau^{0j}-x^j \tau^{0i}\right).
\end{equation}
(As for $P^\mu$ and ${\vec X}_{\CM}$, we can replace ${\bar \tau}^{0\mu}\mapsto{\tau}^{0\mu}$ in this expression in dimensional regularization).

Both $\ell=1$ moments  fit into the same coupling  $-\int d\tau S^{ab}\Omega_{ab}$ in the worldline EFT.    To match to the worldline, we have chosen the comoving frame to be trivial $e^a{}_\mu = \delta^a{}_\mu,$ so that, in the rest frame $u^\mu=(1,{\vec 0})$
\begin{equation}
{1\over 2}\int d\tau S_{ab} \Omega^{ab} = {1\over 2} \int dx^0 S_\mu{}^\nu {\bar \Gamma}^\mu{}_{0\nu}\approx -{1\over 4 m_{\Pl}} \int dx^0 S^{\mu\nu} \left(\partial_\mu{\bar h}_{0\nu}-\partial_\nu{\bar h}_{0\mu}\right),
\end{equation}
so that 
\begin{equation}
{1\over 2} \int d\tau S_{ab} \Omega^{ab} \approx {1\over 4 m_{\Pl}} \int dx^0\left[2(\partial_0 S^{0i}){\bar h}_{0i}+2 S^{0i}\partial_i {\bar h}_{00} - S^{ij} \left(\partial_i{\bar h}_{0j}-\partial_j{\bar h}_{0i}\right)\right]. 
\end{equation}
This matches the multipole expansion of ${\bar \tau}_{\mu\nu}$, provided that we identify
\begin{eqnarray}
S^{0i} = -P^0 X^i_{\CM},  & S^{ij} = L^{ij},
\end{eqnarray}
in which case  $\partial_0 S^{0i} = -P^i$ vanishes in the rest frame.   Therefore, in any coordinate system where ${\vec X}_{\CM}=0$ (but ${\vec P}$ is not necessarily zero), the spin  satisfies the  ``covariant spin supplementary condition'' $S_{ab} P^b{}=0$.

Because the pseudotensor is conserved, none of the multipole terms considered so far can act as a source of gravitational radiation.   Rather, they simply encode the (ADM) mass and angular momentum of the system as determined by measurements of the object's gravitational field at spatial infinity.   The radiative moments instead show up at second order in the multipole expansion and higher.   

The strategy for finding the relation between the moments in Eq.~(\ref{eq:zoso}) and the weighted integrals $\int d^3{\vec x} x^{i_1}\cdots x^{i_\ell} {\bar\tau}^{\mu\nu}$ at order $\ell\geq 2$ follows along the same lines as above.    In the rest frame, we decompose $\mu\nu$ into time and space components, and project onto objects lying in irreducible representations of the permutation group.  By the use of moment relations such as Eq.~(\ref{eq:mr}), and by removing traces, the resulting tensors are then projected onto irreducible tensors which correspond, in a general frame, to the irreducible representations of the $SO(3)$ little group of the composite object.   For example, the averages $\int d^3{\vec x} x^{i_1}\cdots x^{i_\ell} {\bar\tau}^{\mu\nu}$ contain representations of dimensions $\leq 2\ell+1$ if $\mu\nu=00$, for $\mu\nu=0i$ we find representations with dimension $\leq 2\ell+3$, and for $\mu\nu=ij,$ all dimensions $\leq 2\ell+5$ can in principle appear.     By diffeomorphism invariance, these $SO(3)$ irreps couple linearly to the derivatives of the Riemann tensor, projected orthogonal to $P^\mu$.  (Couplings to the Ricci tensor also appear in the decomposition, but such ``non-radiative'' moments vanish on-shell and can be removed by field redefinitions of ${\bar g}_{\mu\nu}$ without any effect on physical observables.)

This procedure has been worked out in full generality~\cite{Ross:2012fc}.   One finds that an infinite number of coefficients from Eq.~(\ref{eq:taylor}) contribute at each multipole order $\ell$.   For generic systems, each one of these contributions to $I^{a_1\cdots a_\ell}$, $J^{a_1\cdots a_\ell}$ is comparable magnitude, but if the sources are non-relativistic this series of terms are suppressed by further powers of $v$ and can be truncated at finite order.    For example, for non-relativistic systems one finds that in the CM frame, the multipole expanded action takes the form
\begin{equation}
\Gamma^{(1)}\supset \int dx^0\left[  {1\over 2} I_{ij} E^{ij} -{2\over 3} J_{ij} B^{ij} + {1\over 6} I_{ijk} \nabla^i E^{ij}+\cdots\right],
\end{equation}
where the electric quadrupole moment is (in $d=3$)~\cite{Goldberger:2009qd}
\begin{equation}
\label{eq:2e}
I^{ij} = \int d^3{\vec x} \left[\tau^{00} + \tau^{kk} -{4\over 3} \partial_0\tau^{0k} {x}^k +{11\over 42}\partial_0^2 \tau^{00}{\vec x}^2+\cdots \right][{x}^i { x}^j]_{STF},
\end{equation}
whose  non-trivial numerical coefficients reflect the underlying $SO(3)$ representation theory\footnote{''STF'' denotes the operation of tensor symmetrization followed by the subtraction of traces with respect to the spatial metric in the comoving frame.}.    Similarly the magnetic quadrupole and electric octopole ($\ell=3$) are 
\begin{eqnarray}
\label{eq:2m}
J^{ij} &=& -{1\over 2}  \int d^3{\vec x} \left[\epsilon^{ikl} \tau^{0k} { x}^j { x}^l+\epsilon^{jkl} \tau^{0k} { x}^i { x}^l+\cdots\right],\\
\label{eq:3e}
I^{ijk} & =& \int d^3{\vec x} \left[\tau^{00} + \tau^{ll} +\cdots\right]\left[{ x}^i { x}^j\right]_{{STF}},
\end{eqnarray}
respectively.

Eqs.~(\ref{eq:2e})-(\ref{eq:3e}) are sufficient to compute gravitational radiation to 1PN beyond the leading order quadrupole formula, given the components of ${\bar\tau}^{\mu\nu}$ to at least that accuracy.   To organize the calculation, we work in a mixed momentum-time representation ${\bar\tau}^{\mu\nu}(x^0,{\vec k}) = \int d^3 {\vec x} e^{-i{\vec k}\cdot {\vec x}} {\bar\tau}^{\mu\nu}(x^0,{\vec x})$, so that we can read off the moments by taking the soft graviton limit
\begin{equation}
{\bar\tau}^{\mu\nu}(x^0,{\vec k}\rightarrow 0) = \sum_{\ell=0}^\infty {(-i)^\ell\over \ell!}\left[ \int d^3 {\vec x} {\bar\tau}^{\mu\nu}(x^0,{\vec x}) { x}^{i_1}\cdots { x}^{i_\ell}\right]  { k}_{i_1}\cdots { k}_{i_\ell}.
\end{equation}

As an example, we determine the non-zero multipole moments of a compact binary system up to 1PN order in the velocity expansion.     First, there are non-vanishing contributions to the electric moments at ${\cal O}(v^0)$, from the $\mu\nu=00$ component
\begin{equation}
\includegraphics[width=0.25\hsize,valign=c]{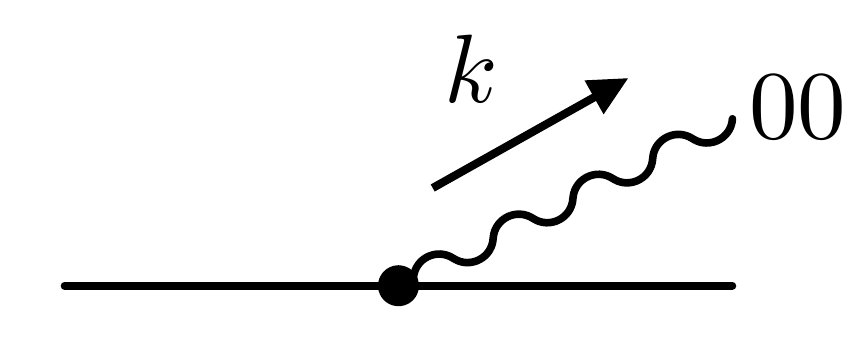}= \sum_A m_A \left[1-i{\vec k}\cdot {\vec x}_A +{1\over 2!} \left(-i{\vec k}\cdot {\vec x}_A\right)^2++{1\over 3!} \left(-i{\vec k}\cdot {\vec x}_A\right)^3+\cdots\right],
\end{equation}
and therefore, in the CM frame, with $M=\sum_A m_A$,
\begin{eqnarray}
P^0 &=&\sum_A m_A +{\cal O}(v^2),\\
{\vec X}_{\CM} &={1\over M} &\sum_A m_A {\vec x}_A +{\cal O}(v^2),\\
I^{ij} &=& \sum_A m_A \left[{ x}^i_A { x}^j_A\right]_{STF}+{\cal O}(v^2),\\
\label{eq:iijk}
I^{ijk}&=& \sum m_A A \left[{ x}^i_A { x}^j_A x_A^k\right]_{STF}+{\cal O}(v^2),
\end{eqnarray}
\etc~As expected, the leading order multipole moments agree with the predictions of Newtonian theory.    For example, the total mass of the system is just the sum $M=\sum_A m_A$ of the ADM masses of the binary constituents.

Similarly, the magnetic-type moments arise first at ${\cal O}(v^1)$, sourced by the $\mu\nu=0i$ components of ${\bar \tau}^{\mu\nu}$,
\begin{equation}
\includegraphics[width=0.25\hsize,valign=c]{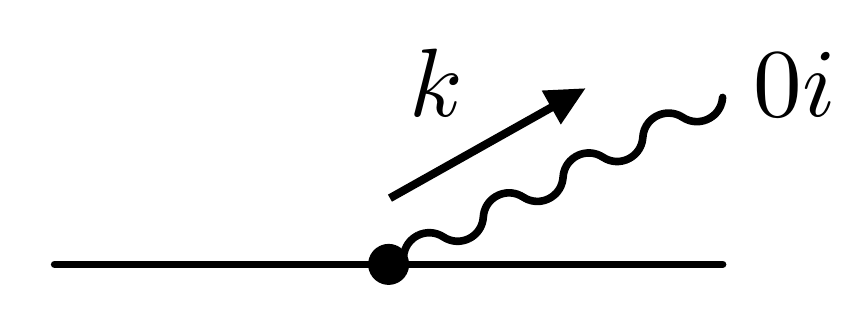}= \sum_A m_A {\vec v}_A^i\left[1-i{\vec k}\cdot {\vec x}_A +{1\over 2!} \left(-i{\vec k}\cdot {\vec x}_A\right)^2++{1\over 3!} \left(-i{\vec k}\cdot {\vec x}_A\right)^3+\cdots\right],
\end{equation}
and therefore,
\begin{eqnarray}
P^i &=&\sum_A m_A { v}_A^i+{\cal O}(v^3),\\
L^{ij} &= &\sum_A m_A \left({ x}^i_A { v}_A^j-{ x}^j_A { v}_A^i\right)+{\cal O}(v^3),\\
\label{eq:jij}
J^{ij} &=& {1\over 2} \sum_A m_A \left[\left({\vec x}_A\times {\vec v}_A\right)^i v_A^j+ (i\leftrightarrow j)\right],
\end{eqnarray}
and so on at $\ell>2$.

At ${\cal O}(v^2)$, ${\bar \tau}^{\mu\nu}$ is calculated from the sum of the Feynman diagrams in fig.~\ref{fig:rad1pt}(b)-(d).    These depict both special relativistic corrections, as in fig.~\ref{fig:rad1pt}(b) as well as the effects of gravitational interactions between the compact objects, fig.~\ref{fig:rad1pt}(c),(d).   To evaluate the diagrams, (c), (d) a complete table of integrals is provided by the standard one-loop integral
\begin{eqnarray}
\int_{\vec p} {1\over \left({\vec p}^2\right)^\alpha \left[\left({\vec p}+{\vec k}\right)^2\right]^\beta}&=&{1\over (4\pi)^{{d-1}\over 2}} {\Gamma(\alpha+\beta-{d-1\over 2})\over \Gamma(\alpha)\Gamma(\beta)}\\
\nonumber
& & \times{\Gamma({d-1\over 2}-\alpha)\Gamma({d-1\over 2}-\beta)\over \Gamma(d-1-\alpha-\beta)} ({\vec k}^2)^{{d-1\over 2}-\alpha-\beta},
\end{eqnarray}
and the Fourier transform
\begin{equation}
\int_{\vec p} {e^{i{\vec p}\cdot {\vec x}}\over \left({\vec p}^2\right)^\alpha}={1\over (4\pi)^{d-1\over 2}} {\Gamma({d-1\over 2}-\alpha)\over \Gamma(\alpha)}\left({{\vec x}^2\over 4}\right)^{\alpha-{d-1\over 2}}.
\end{equation}

One finds that including effects up to 1PN order, the electric-type $\ell<2$ moments are
\begin{eqnarray}
P^0 &=& \sum_A {\bar m}_A + {\cal O}(v^4)\\
{\vec X}_{\CM} &={1\over M} &\sum_A {\bar m}_A {\vec x}_A +{\cal O}(v^4),
\end{eqnarray}
which are averages weighted over a ``renormalized'' particle mass defined as
\begin{equation}
{\bar m}_A = m_A\left[1+{1\over 2} {\vec v}_A^2 -{1\over 2}\sum_B {G_N m_B\over |{\vec x}_A-{\vec x}_B|}\right]
\end{equation}
(We do not quote here the order $v^2$ corrections to magnetic moments with $\ell<2$ since these are not necessary  to predict the 1PN accurate observables in the CM frame.   Note that, in any case,  given $P^0,{\vec X}_{\CM}$ to 1PN order, the spatial momentum can be obtained from the moment relation ${\vec P}^i/P^0= d {\vec X}^i_{\CM}/dx^0$ without the need to explicitly calculate the 1PN corrections to ${\bar \tau}^{0i}$.).   The relevant $\ell\geq 2$ moments at 1PN are then $I^{ijk}$ in Eq.~(\ref{eq:iijk}), $J^{ij}$ in Eq.~(\ref{eq:jij}) and the electric quadrupole moment to ${\cal O}(v^2)$, obtained by inserting ${\bar\tau}^{\mu\nu}$ into Eq.~(\ref{eq:2e}),
\begin{equation}
I^{ij} =\sum_A m_A\left(1+{3\over 2} {\vec v}_A^2 -\sum_B {G_N m_B\over |{\vec x}_A-{\vec x}_B|} +{11\over 42} {d^2\over dt^2}{\vec x}_A^2-{4\over 3} {d\over dt}{\vec x}_A{\cdot} {\vec v}_A\right)\left[{ x}_A^i { x}_A^j\right]_{STF},
\end{equation}
where the time derivatives act on anything to their right.

Having obtained explicit formulae for the multipole moments in terms of the microscopic degrees of freedom ${\vec x}_A$, it is now possible to predict the radiative corrections to binary dynamics.    For example, the time averaged power into gravitons in the $\ell=2$ partial wave can be calculated by inserting the expression for $I_{ij}$ into Eq.~(\ref{eq:powah}) and summing over final state graviton polarizations/integrating over phase space (useful formulas for performing polarization sums and angular integrals can be found \eg~in~\cite{Weinberg:1972kfs}), leading to the classic formula
\begin{equation}
\label{eq:quadrad}
P^{\ell=2_E}_{GW} = {G_N\over 5}\left\langle {d^3 I_{ij}\over dt^3} {d^3 I_{ij}\over dt^3}\right\rangle,
\end{equation}
while the leading PN waveform at ${\cal I}^+$ follows from Eq.~(\ref{eq:wave}) and Eq.~(\ref{eq:al2}) 
\begin{equation}
h_{ij}(u) = {4 G_N\over |{\vec x}|} {d^2\over du^2} I_{ij}(u),
\end{equation}
where $I_{ij}$ is evaluated on-shell, \ie on the solution of the PN equations of motion for the orbits ${\vec x}_A$.   These equations of motion include ``conservative'' terms from integrating out potential modes, as discussed in sec.~\ref{sec:pot}, as well as radiative corrections due to the backreaction of the emitted gravitational waves on the particle trajectories.    The latter can be calculated directly in \zosoir as we explain in the next section.

\section{Time non-locality:   Radiation reaction, black hole event horizons}
\label{sec:nonlocal}

In this section we provide a brief overview of two applications of the worldline EFT Eq.~(\ref{eq:zoso}) to dissipative processes in binary dynamics.   In sec.~\ref{sec:rr} we discuss radiation damping of the particle orbits at leading PN order, following~\cite{Galley:2009px}.    Even though the effects of radiation reaction are local in time at the level of the equations of motion (at least at leading order), they cannot be obtained by variation of a local Lagrangian.   It is essential to employ Schwinger-Keldysh contours in the path integral in order to ensure causal time evolution.

In sec.~\ref{sec:horizon}, we incorporate the effects of black hole horizon dynamics.   The absorption of energy and momentum by the horizon horizon implies the existence of massless degrees of freedom on the worldline EFT, which are described by UV version of Eq.~(\ref{eq:zoso}).   When integrated out, these modes generat a non-local in time effective action for the particle trajectories, but local (and time reversal odd) friction forces at the level of the equations of motion.

\subsection{Radiation reaction in \zosoir}
\label{sec:rr}

Ultimately, the goal of the PN expansion is to calculate both the gravitational waveform as seen by observers at infinity.   At sufficiently high orders in velocity, this involves determining how the particle orbits ${\vec x}_A(t)$ are affected by the emission of radiation.    Such radiation reaction effects are encoded in the in-in effective action $\Gamma[x_A,{\bar g};{\tilde x}_A,{\tilde{\bar g}}]$ in Eq.~(\ref{eq:inin}) and the resulting equations of motion Eq.~(\ref{eq:nleom}).   Because the in-in action is generated by integrating out radiation it is in general non-local in time.

Since this non-locality in time occurs over scales of order the radiation wavelength, it is practical to compute  $\Gamma[x_A,{\bar g},{\tilde x}_A;{\tilde{\bar g}}]$ by first constructing the worldline theory \zosoir and then calculating vacuum Feynman graphs in that theory, as first advocated in ref.~\cite{Galley:2009px}.   To illustrate this procedure, we consider the effects of radiation reaction on a composite system described by the action in Eq.~(\ref{eq:zoso}).    In that theory, integrating out radiation generates a correction $\Delta\Gamma[x_A{\bar g};{\tilde x}_A,{\tilde{\bar g}}]$ to the in-in effective action which is in general non-local in time.     For example, the self-energy diagram,
\begin{equation}
i\Delta\Gamma[x_A,{\bar g};{\tilde x}_A,{\tilde{\bar g}}] =\includegraphics[width=0.35\hsize,valign=c]{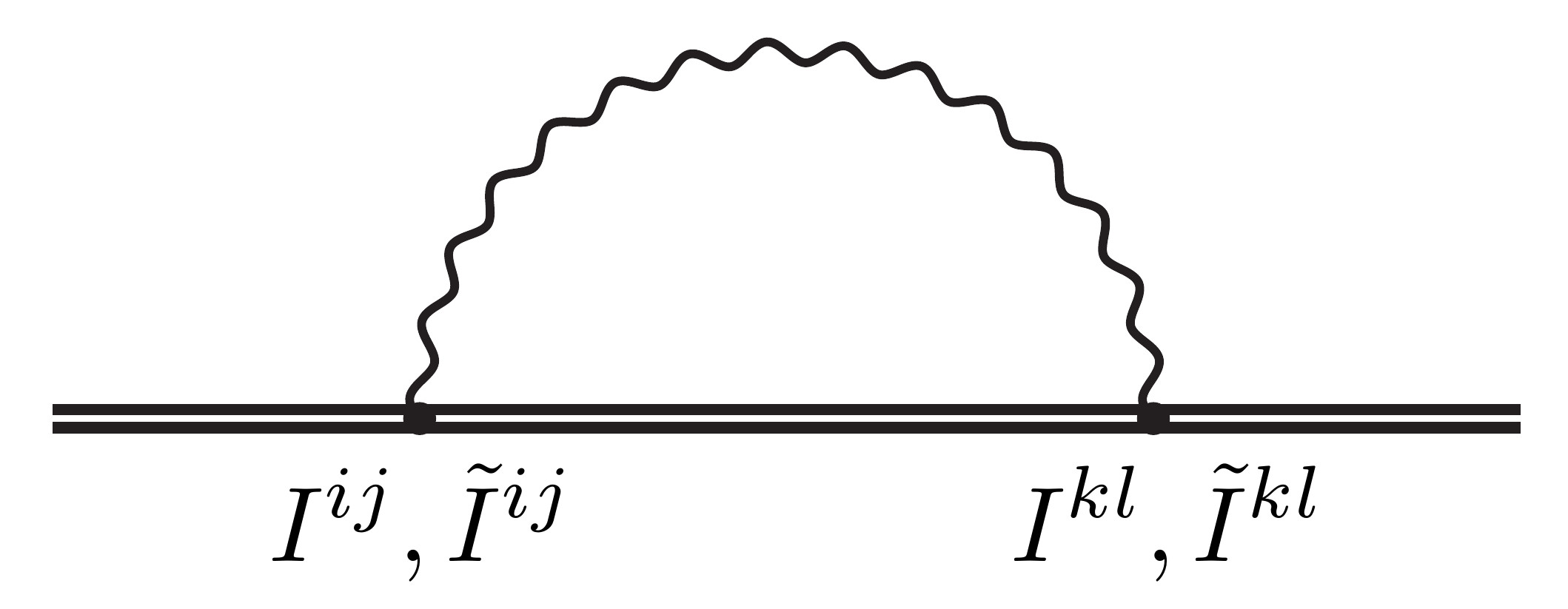}\sim L v^5,
\end{equation}
which represents a sum over four independent in-in Wick contractions, gives the leading order radiation reaction effect due to quadrupole emission,
\begin{eqnarray}
\nonumber
i\Delta\Gamma[x_A,{\bar g};{\tilde x}_A,{\tilde{\bar g}}] &=&{1\over 2!} \int dt_1 dt_2 \left({i\over 2} I^{ij}(t_1)\right) \langle E_{ij}(t_1,0) E_{kl}(t_2,0)\rangle\left ({i\over 2} I^{kl}(t_2)\right)\\
\nonumber
& &{} \hspace{0.15cm} + \int dt_1 dt_2 \left({i\over 2} I^{ij}(t_1)\right) \langle E_{ij}(t_1,0) {\tilde E}_{kl}(t_2,0)\rangle\left (-{i\over 2} {\tilde I}^{kl}(t_2)\right)\\
\nonumber
& & {} \hspace{0.15cm} + {1\over 2!} \int dt_1 dt_2 \left(-{i\over 2} {\tilde I}^{ij}(t_1)\right) \langle {\tilde E}_{ij}(t_1,0) {\tilde E}_{kl}(t_2,0)\rangle\left (-{i\over 2} {\tilde I}^{kl}(t_2)\right),\\
\end{eqnarray}
(recall from sec.~\ref{sec:2B}, that $\langle E_{ij}(t_1,0) E_{kl}(t_2,0)\rangle$ is the time-ordered expectation value while $\langle E_{ij}(t_1,0) {\tilde E}_{kl}(t_2,0)\rangle=\langle 0|E_{kl}(t_2,0) E_{ij}(t_1,0)|0\rangle$ is a Wightman two-point function) and $\langle {\tilde E}_{ij}(t_1,0) {\tilde E}_{kl}(t_2,0)\rangle$ is anti- time-ordered (Dyson)).   The equation of motion for each particle ${\vec x}_A$ is the extremum of $\Gamma[x_A,{\bar g};{\tilde x}_A,{\tilde{\bar g}}]$, after setting ${\tilde x}_A=x_A$ and setting to zero the background field, ${\bar g}_{\mu\nu}={\tilde{\bar g}}_{\mu\nu}=\eta_{\mu\nu}$.   In particular, the correction $\Delta\Gamma[x_A,{\bar g};{\tilde x}_A,{\tilde{\bar g}}] $ calculated above contributes to the radiation damping force on particle $A$,
\begin{eqnarray}
{\vec F}_A(t) &=& \left.{\delta\over \delta {\vec x}_A(t)} \Delta\Gamma[x_A,\eta;{\tilde x}_A,\eta] \right|_{{\tilde{\vec x}}_A={\vec x}_A}\\
\nonumber
&=&{i\over 4} \int dt_1 dt_2\left( {\delta I^{ij}(t_1)\over\delta {\vec x}_A(t)}\right)  \left[\langle E_{ij}(t_1,0) E_{kl}(t_2,0)\rangle- \langle E_{ij}(t_1,0) {\tilde E}_{kl}(t_2,0)\rangle\right] I^{kl}(t_2).
\end{eqnarray}

It is built into the in-in formalism that the action is a real quantity and that the equations of motion are causal.   In particular, the expression $\langle E_{ij}(t_1,0) E_{kl}(t_2,0)\rangle- \langle E_{ij}(t_1,0) {\tilde E}_{kl}(t_2,0)\rangle$, is proportional to the retarded two-point Green's function of the electric curvature operator, as  a consequence of the general operator identity
\begin{eqnarray}
\label{eq:opid}
T\left[{\cal O}_1(t_1) {\cal O}(t_2)\right] - {\cal O}_1(t_1) {\cal O}_2(t_2) = \theta(t_1-t_2) [{\cal O}_1(t_1),{\cal O}_1(t_1)].
 \end{eqnarray}
 Inserting the linearized Riemann tensor into the correlator,
 \begin{equation}
E_{ij}=R_{0i0j}\approx -{1\over 2} \partial_0^2 {\bar h}_{ij} + {1\over 2}\partial_0\partial_i {\bar h}_{0j} +{1\over 2}\partial_0\partial_j {\bar h}_{0i}-{1\over 2}\partial_i \partial_j {\bar h}_{00},
 \end{equation}
 and using rotational symmetry to reduce the angular integrals, we have
\begin{eqnarray}
\nonumber
\langle 0|\left[E_{ij}(t_1,0),E_{kl}(t_2,0)\right]|0\rangle&=&\int {d^3{\vec q}\over (2\pi)^3 2|{\vec q}|}  {|{\vec q}|^4\over 20 m_\Pl^2} \left[\delta_{ik}\delta_{jl}+\delta_{il}\delta_{jk}-{2\over 3}\delta_{ij}\delta_{kl}\right]\\
& & \times\left(e^{-i|{\vec q}|(t_1-t_2)}-e^{i|{\vec q}|(t_1-t_2)}\right)\\
\nonumber
&=& {4 G_N\over 5} \left[\delta_{ik}\delta_{jl}+\delta_{il}\delta_{jk}-{2\over 3}\delta_{ij}\delta_{kl}\right]\left(i{d\over dt_1}\right)^5\delta(t_1-t_2).
\end{eqnarray}

Because of the step function $\theta(t_1-t_2)$ in Eq.~(\ref{eq:opid}),  encoding causality, integrating out the radiation results in a non-local effective action.   However, at this PN order, the equations of motion are still local,
\begin{equation}
{\vec F}^i_A(t) = \left.{\delta\over \delta {\vec x}^i_A(t)} \Delta\Gamma[x_A,{\tilde x}_A] \right|_{{\tilde{\vec x}_A={\vec x}_A}}=-{2G_N\over 5} m_A {\vec x}_A^j{d^5\over dt^5} I^{ij}(t),
\end{equation}
as well as odd under time reversal, as is characteristic of dissipation.  By employing Schwinger-Keldysh boundary conditions in the EFT, we have reproduced~\cite{Galley:2009px},  the 2.5PN  Burke-Thorne radiation reaction force, derived originally by purely classical methods~\cite{BT}.    The instantaneous loss of mechanical energy and angular momentum into radiation are then
\begin{eqnarray}
{d\over dt} E &=& \sum_A {\vec v}_A \cdot {\vec F}_A = -{G_N\over 5} {d{I}^{ij}(t)\over dt} {d^5 I^{ij}(t)\over dt^5},\\
{d\over dt} J^i &=&\sum \left({\vec x}_A\times {\vec F}_A\right)^i = -{2 G_N\over 5}\epsilon_{ijk} {I}^{jl}(t) {d^5 I^{kl}(t)\over dt^5}.
\end{eqnarray}
Note in particular the time average over many orbital cycles agrees with the standard quadrupole radiation formula in Eq.~(\ref{eq:quadrad}).

The procedure outlined in this example is completely systematic, and has been extended to higher orders in the PN expansion.   For example, at 4PN order, there is a tail correction to the in-in action, from diagrams such as
\begin{eqnarray}
\includegraphics[width=0.3\hsize,valign=c]{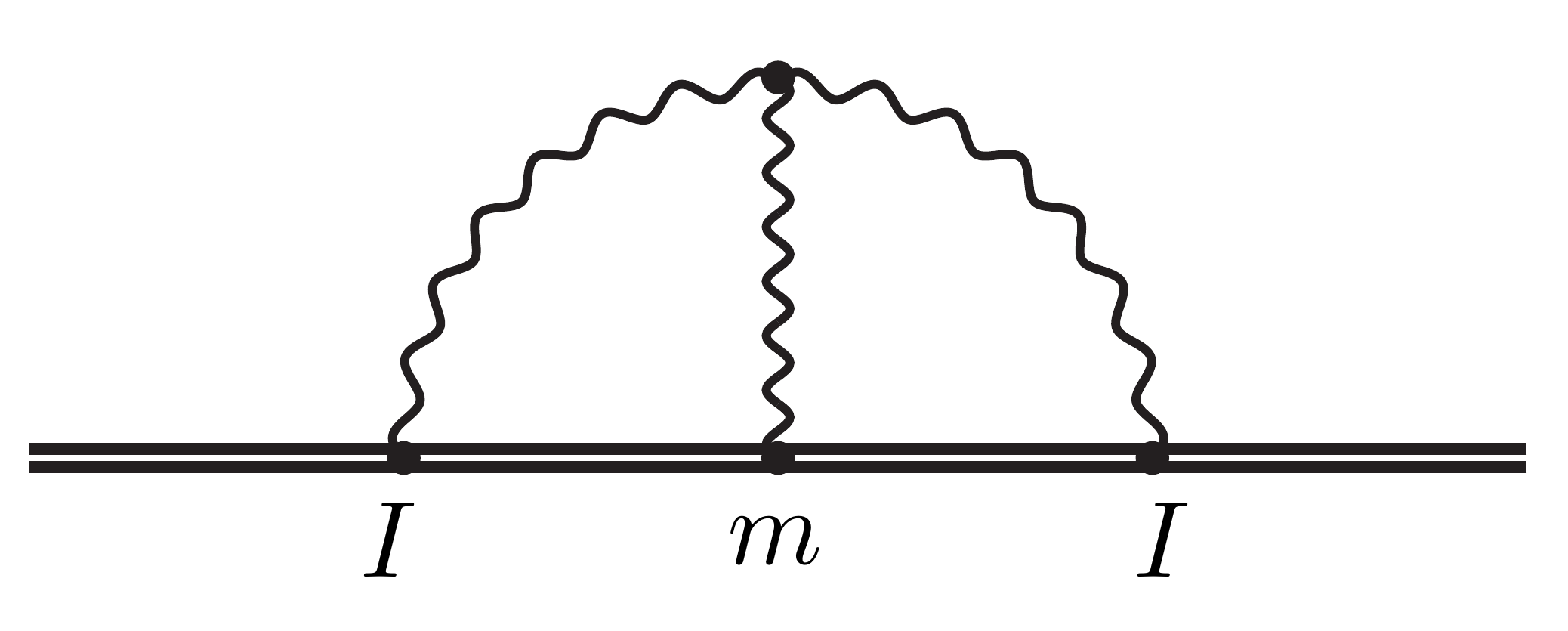}\sim L v^8
\end{eqnarray}
while at 5PN, there are non-linear ``hereditary'' or ``memory effects'' in which the equations of motion at a given time depend on the quadrupole moment of the system at earlier times, such as the term
\begin{eqnarray}
  \includegraphics[width=0.3\hsize,valign=c]{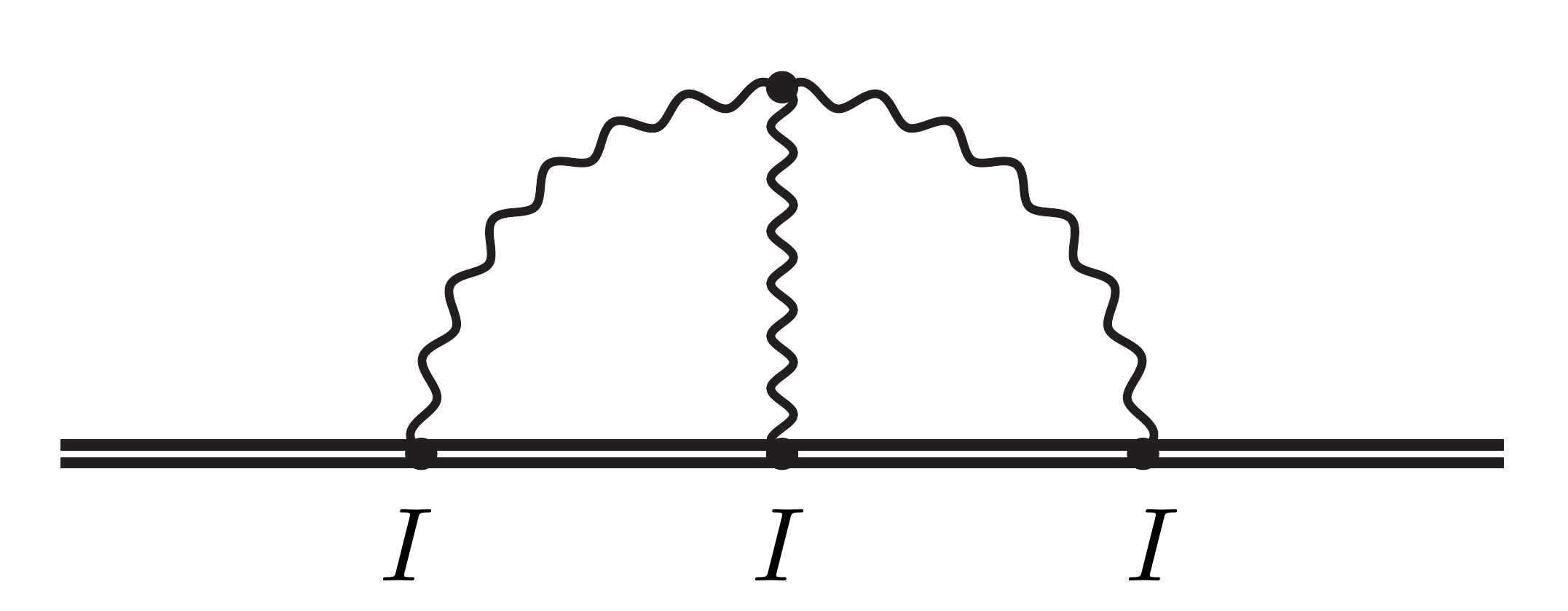}\sim L v^{10}.
\end{eqnarray}
In these calculations, a combination of field theoretic techniques (EFTs, dimensional regularization, the Schwinger-Keldysh path integral, \etc) play an essential role in obtaining self-consistent, ambiguity-free results for dynamical observables.   A review of these intricate higher order effects is beyond the scope of this article, and the reader is referred to the original literature, compiled in ref.~\cite{Goldberger:2022ebt}, for the technical details.

\subsection{Event horizon dynamics in \zosouv}
\label{sec:horizon}

The formalism as described so far is adequate for compact objects whose internal structure is gapped, so that gravitational interactions at scales longer than the orbital radius cannot irreversibly modify the intrinsic properties of the object.   For black holes, though, this frequency gap is of order $1/r_s$, so that finite size effects will come in at some finite order in $r_s/r\sim v^2$ in the PN expansion.   In order to have a fully systematic treatment of PN black hole binary dynamics, such dissipative effects cannot be neglected.

For instance, from black hole perturbation theory~\cite{Sasaki:2003xr} it is known that the change in mass due to tidal heating induced by a small binary companion in a bound orbit appears at order 4PN~\cite{Poisson:1994yf} in the Schwarzschild case, and becomes enhanced to 2.5PN~\cite{Tagoshi:1997jy} for (near extremal) Kerr black holes.    In the latter case, the tidal interactions can actually \emph{decrease}~\cite{DEath:1975jps,Poisson:2004cw} the mass of the black hole as a consequence of stimulated emission (``rotational superradiance")~\cite{zeldovich,staro}, a field theoretic realization of the Penrose process~\cite{Penrose:1971uk} of energy extraction from the black hole's ergosphere.  

On general grounds~\cite{Callen:1951vq}, dissipation, \eg~ flux of energy and angular momentum across the surface of the compact star, signals the presence of a continuum spectrum of localized degrees of freedom that couple to gravity in the bulk spacetime.  For a neutron star, the additional degrees of freedom correspond to the low-lying hydrodynamic modes of nuclear matter, while for classical black holes, the horizon fluctuations are presumably related to the quasinormal mode solutions of the Teukolsky equation.   Regardless of the microscopic origin of the internal degrees of freedom, their presence has an effect on the binary inspiral dynamics at some order in the PN expansion.    

It is therefore useful to have a way of incorporating the effects of dissipation directly in the worldline description of the compact objects, without explicitly having to track the evolution of the internal modes themselves.    An EFT framework for this was first introduced in ref.~\cite{Goldberger:2005cd}, which describes the long wavelength dissipative response of compact objects to external gravitational perturbations, by ``integrating in''  a \emph{quantum mechanical} $0+1$-dimensional defect field theory of degrees of freedom localized on the worldline.

Independent of their UV origin, in the long distance limit these modes have local diff and reparameterization invariant couplings to the spacetime curvature.   Organizing the algebra of defect operators in terms of the linearly realized rotations about the objects spatial location, the symmetries of the EFT guarantee that the Lagrangian must be of identical form to Eq.~(\ref{eq:zoso}), where now the multipole moments $I_{a_1\cdots a_\ell},J_{a_1\cdots a_\ell},$ should be regarded as a set of composite operators constructed out of the microscopic degrees of freedom, acting on some internal Hilbert space of physical states.    As long as we probe the system with slowly varying fields, the compact object itself also appears as a \emph{Zoomed Out Single Object}, even if its internal structure is arbitrarily complicated.

We do not need to know what the internal modes are in order to make  predictions in the infrared.   In this case, long distance observables can be calculated in terms of the correlation functions of the multipole operators, which in turn are determined by a matching calculation to the UV theory.   The power counting of the EFT indicates that at long distances, the leading contribution is from the two-point correlators of the electric and magnetic quadrupole operators
$$
\langle  I_{ab}(\tau) I_{cd}(0)\rangle, \langle  J_{ab}(\tau) J_{cd}(0)\rangle
$$
evaluated in the equilibrium (pure or mixed) state of the object.   Predictions in the EFT, in powers of $\omega {\cal R}\ll 1$ are systematically improvable by including more multipoles, higher-point correlators, or perturbative graviton interactions which scale as powers of  $G_N M\omega\lsim \omega{\cal R}\ll 1$.  

To match to this \zosouv~we compute on-shell graviton scattering off an isolated object in the EFT, using Eq.~(\ref{eq:zoso}), and compare it to the low frequency limit of the corresponding observable in the full theory.  As an example, consider graviton absorption by a Schwarzschild black hole, which in the EFT has the matrix element
\begin{eqnarray}
\nonumber
i{\cal A}(M\rightarrow X) &\approx& \includegraphics[width=0.3\hsize,valign=c]{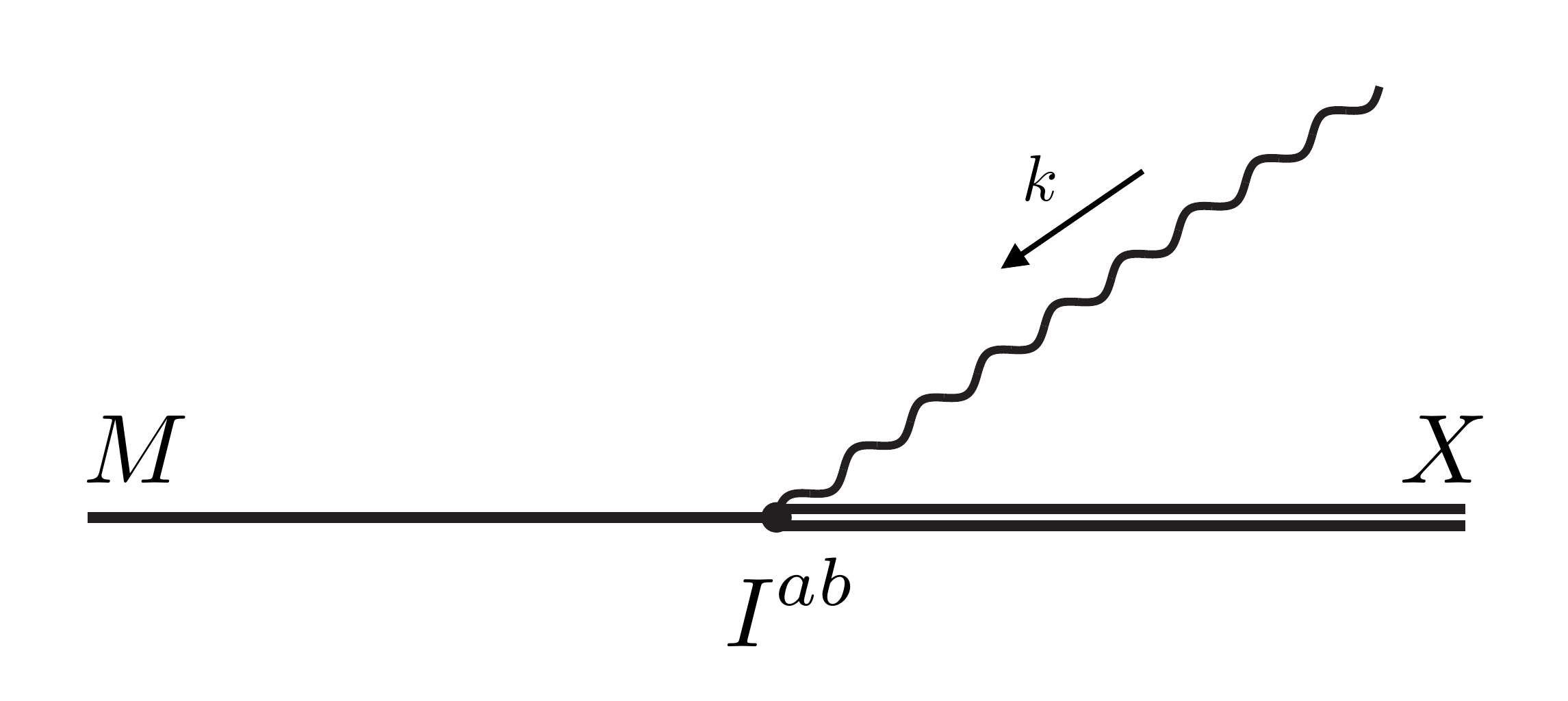}+\mbox{mag.}\\
&=& {i\over 2 m_{Pl}}\int dt e^{-i\omega t} \langle X|I_{ab}(t) |M\rangle\times \langle 0|E_{ab}(t,0)|k,h\rangle +\mbox{mag.},
\end{eqnarray}
in the rest frame, to leading order in $1/m_{Pl}$.  Here, the matrix element $\langle 0|E_{ab}(t,0)|k,h\rangle$ between the one-graviton state of four-momentum $k^\mu$ ($k^0=\omega>0)$, helicity $h=\pm 2$, and the vacuum is readily computed by standard canonical quantization of the Einstein-Hilbert Lagrangian~\cite{Einstein:1916vd}, as established in~\cite{Gupta:1952zz,Feynman:1963ax,DeWitt:1967yk,DeWitt:1967ub,DeWitt:1967uc}.   The transition matrix element $ \langle X|I_{ab}(t) |M\rangle$  from the initial black hole of mass $M$ to some unknown final state $|X\rangle$ is not calculable in the EFT, but assuming unitarity
\begin{equation}
\sum_X |X\rangle \langle X| = \mathbf{1}_{\cal H},
\end{equation}
we can express the inclusive absorption cross section for a graviton incident on the horizon,
\begin{eqnarray}
\sigma_{abs}(\omega) &=& \lim_{T\rightarrow\infty} {1\over 2\omega}\sum_X {|{\cal A}(M\rightarrow X)|^2 \over T}\\
\nonumber
&=&G_N\pi \omega^3 \int dt e^{i\omega t} \epsilon^{*}_{cd,h}(k) \langle I_{cd}(t) I_{ab}(0)\rangle  \epsilon_{ab,h}(k) + \mbox{mag.},
\end{eqnarray}
in terms of the two-point correlators of the $\ell=2$ multipole operators in the initial state of a black hole of mass $M$ and spin $J$.    

The cross section $\sigma_{abs}(\omega)$ is a physical quantity that can be compared against the predictions of classical general relativity in the limit $r_s\omega\ll 1$, where the EFT description is useful.  Using the classical absorption probabilities calculated in refs.~\cite{staro,Page:1976df} one finds that $\sigma_{abs}(\omega)\approx 4\pi r_s^6\omega^4/45$, and exploiting the rotational invariance of the Schwarzschild black hole to write
\begin{equation}
\label{eq:wman}
\int dt e^{i\omega t} \langle I_{ab}(t) I_{cd}(0)\rangle ={1\over 2}\left[\eta^\perp_{ac}\eta^\perp_{bd}+\eta^\perp_{ad}\eta^\perp_{bc}-{2\over 3} \eta^\perp_{ab}\eta^\perp_{cd}\right] A_+^E(\omega) ,
\end{equation}
($\eta^\perp_{ab}=\eta_{ab} - p_a p_b/M^2$ is the spatial metric in the black hole's rest frame) one finds that the frequency space correlators are to lowest order 
\begin{equation}
A^E_+(\omega)=A^B_+(\omega)\approx 2 \theta(\omega) r_s^6\omega/45G_N,
\end{equation}
where the factor $\theta(\omega)$ ensures that for a classical black hole, the graviton emission matrix element $\mbox{BH}\rightarrow\mbox{BH}'+\mbox{graviton}$ is zero.  

 The point of this exercise is that the same correlators that one extracts from on-shell observables  in the one-body sector also control \emph{off-shell} graviton exchange processes in the two-body sector, where the binary dynamics is described by NRGR.    By including diagrams with insertions of the multipole operators $I_{ab}$, $J_{ab}$, it becomes possible to include the effects of horizon dynamics in the EFT while retaining a worldline description of the binary constituents.

 For example, single graviton exchange between two black holes generates a tidal friction ($T$-odd) term in the two-body equations of motion associated with the excitation of horizon modes, leading to a flux of energy across the event horizon.   This is encoded in the box diagram contribution to the in-in effective action of the binary
 \begin{equation}
 \label{eq:box}
i\Delta\Gamma[x_A,\eta;{\tilde x}_A,\eta]\supset \includegraphics[width=0.4\hsize,valign=c]{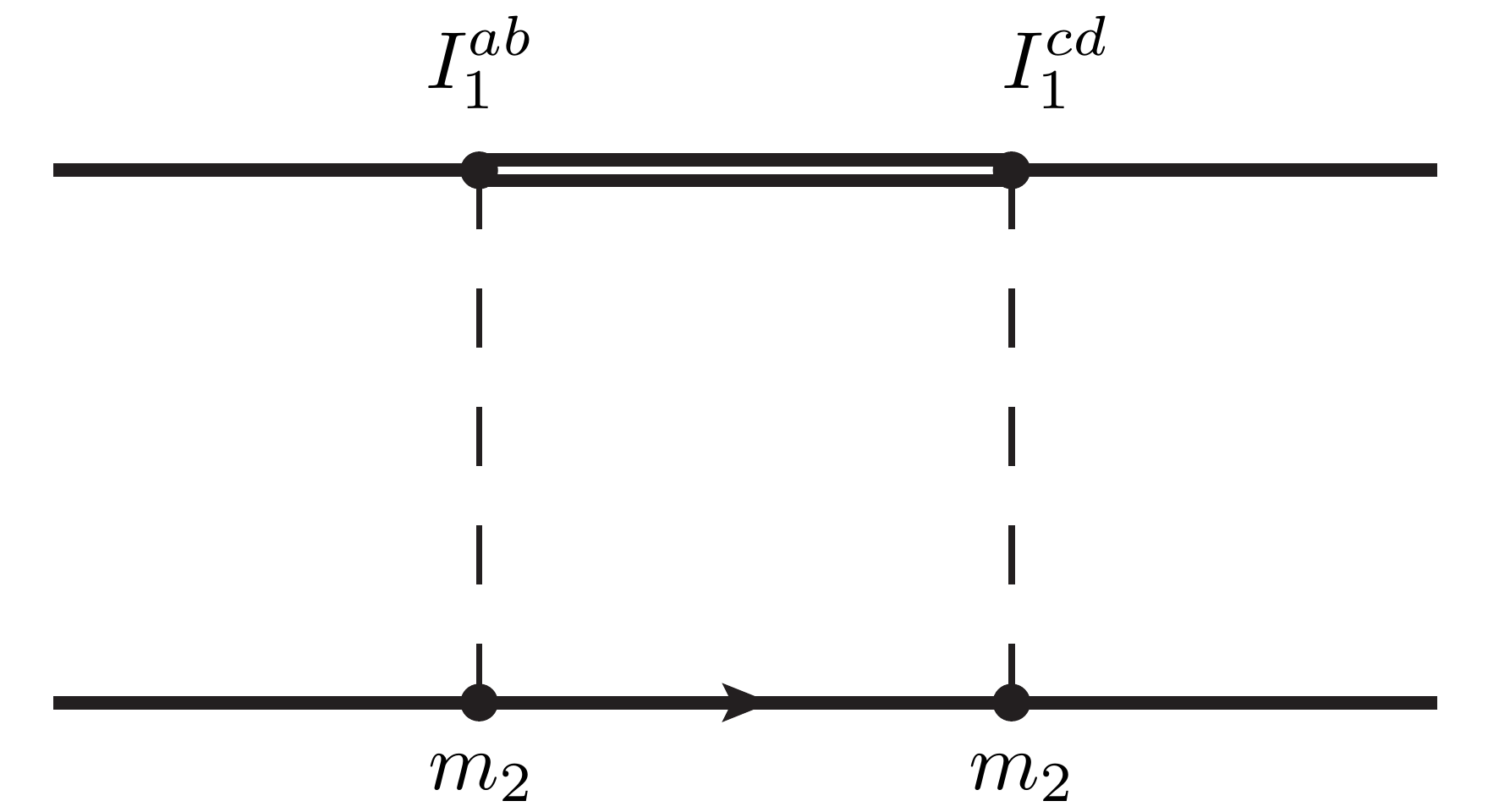}+(1\leftrightarrow 2),
 \end{equation}
plus a similar magnetic contribution. Given the form of the correlators $\langle I^{ab} I^{cd}\rangle$, $\langle J^{ab} J^{cd}\rangle$, this is a non-local in time quantity.     The frictional force on the particle trajectories in the black hole binary, associated with the absorption of gravitational energy by the horizons, follow from variation of the effective action, ${\vec F}_1=-\left.{\vec F}_2 =\delta\Delta\Gamma/\delta{\vec x}_1\right|_{{\tilde x}_A=x_A}$.

The variation of Eq.~(\ref{eq:box}), with respect to the particle trajectories is causal, in the sense that it depends only the retarded Green's function,
\begin{equation}
G_R^{ab,cd}(\tau)=-i\theta(\tau)\langle [I^{ab}(\tau),I^{cd}(0)]\rangle,
\end{equation}
 of the multipole operators on the black hole worldline.    This again is a consequence of the in-in Feynman rules, together with the operator identity Eq.~(\ref{eq:opid}).  Note that the retarded Green's function is not technically the same as the Wightman function in Eq.~(\ref{eq:wman}) .   They are, however, related by a dispersion relation, which in frequency space reads
 \begin{equation}
 G^R(\omega) =i\int_{-\infty}^{\infty} {d\omega'\over 2\pi} {A_+(\omega')-A_+(-\omega')\over \omega-\omega'-i0^+},
 \end{equation}
 where we have suppressed the tensor structure, which is identical to the one in Eq.~(\ref{eq:wman}).

 The dispersion relation implies that the imaginary part of $G^R(\omega)$, responsible for dissipation,  is controlled by the Wightman function $A_+(\omega)$ which we obtained by matching the microscopic theory.   On the other hand, the real part of the dispersion relation
 \begin{equation}
\mbox{Re}G^R(\omega) =\mbox{Pr} \int_0^\infty {d\omega'\omega'\over \pi}{A_+(\omega')-A_+(-\omega')\over \omega^2-\omega'^2},
 \end{equation}
involves arbitrarily large frequency scales.  Evaluation of the RHS becomes impossible while remaining in the regime of validity of the EFT.   In any case, in the EFT, the real part of the retarded response also receives a contribution from local counterterms on the black hole's worldline, \eg~from those in Eq.~(\ref{eq:s4}), so matching the physical response of the black hole requires knowledge of other observables beyond the graybody factor $\sigma_{abs}(\omega)$.   One would also need an analytic expression for the elastic graviton scattering amplitude, in the $\omega\rightarrow 0$ limit.
 
We can get around this by noting that, in light of Eq.~(\ref{eq:inq}), we can think of the static Love number as the $\omega\rightarrow 0$ limit of the AC response function $G^R(\omega)$.   Eq.~(\ref{eq:inq}) is actually a special case of a more general linear response relation~\cite{Goldberger:2005cd,Chakrabarti:2013lua,Goldberger:2020wbx}, between the induced moment $\langle I_{ab}\rangle$ on the surface of the black hole, (which, \eg~determines the long distance quadrupolar gravitational field as seen by asymptotic observers) and the input $E_{ab}(x(\tau))$,  a slowly varying background field in which the black hole propagates\footnote{Classical general relativity approaches to the motion of black holes in curved spacetime backgrounds are reviewed in~\cite{Poisson:2011nh}.}
\begin{eqnarray}
\nonumber
\langle I_{ab}(\omega)\rangle = \int_{-\infty}^{\infty} d\tau e^{i\omega\tau} \langle I_{ab}(\tau)\rangle &=& {2\over 3} {r_s^5\over G_N} k^E_{\ell=2}E_{ab}(\omega) + G_R^{ab,cd}(\omega) E_{cd}(\omega)\\
&\equiv& {2\over 3} {r_s^5\over G_N} \kappa^E_{\ell=2}(\omega) E_{ab}(\omega),
\end{eqnarray}
where $E_{ab}(\omega) =\int d\tau e^{i\omega\tau} E_{ab}(x(\tau))$ and we have defined, generalizing Eq.~({\ref{eq:inq}), the AC Love number of the $d=4$ Schwarzschild black hole, $\kappa^E_{\ell=2}(\omega)$ on the second line of this equation.  The static Love number is then the  $\omega\rightarrow 0$ limit of this quantity,
\begin{equation}
\label{eq:sr}
\kappa^E_{\ell=2}(0) = k^E_{\ell=2}+3 G_N r_s^{-5}\int_0^\infty {d\omega\over 2\pi} {A_+(\omega)\over \omega}.
\end{equation}
Calculations in full general relativity~\cite{Binnington:2009bb,Damour:2009va}, and in the EFT~\cite{Kol:2011vg}, indicate that for Schwarzschild black holes, the quadrupolar static response vanishes, \ie
\begin{equation}
\kappa^E_{\ell=2}(0) =0,
\end{equation}
at least in $d=4$ spacetime dimensions.     This pattern seems to extend to the static Love response at $\ell>2$~\cite{Pani:2015hfa}.   

If  $\kappa^E_{\ell=2}(0) =0$, we can intepret Eq.~(\ref{eq:sr}) as a sum rule~\cite{Rothstein:2014sra,GnR} enforcing a cancellation between the bare Love number $k^E_{\ell=2}$ and the radiative corrections from the  degrees of freedom localized on the worldline.   Given that the latter quantity receives contributions from UV modes of arbitrarily large frequency, this suggests that the vanishing of the static response requires a fine tuning of parameters that is unnatural (in the `t Hooft~\cite{tHooft:1979rat}} sense) from the point of view of the EFT~\cite{GnR,Rothstein:2014sra,Porto:2016zng}.   The question of wether this apparent tuning of parameters has a resolution in terms of some hidden symmetries of the EFT degrees of freedom has been a subject of recent active study, see~\cite{Charalambous:2021mea,Charalambous:2021kcz,Hui:2021vcv,Hui:2022vbh}.

Given that the static response is zero, we now have enough information to fix the low frequency AC Love number, 
\begin{equation}
\kappa^E_{\ell=2}(\omega)\approx -{i\pi\over 15} r_s|{\omega}|+\cdots.
\end{equation}
Inserting this into the box diagram, the instantaneous force on each black hole is local in time, but not derivable from the variation of a local action,
 \begin{equation}
{\vec F}_1=-{\vec F}_2 = -{32\over 5} {G_N^7 (m_1 m_2)^2 (m_1^4+m_2^4)\over |{\vec x}_1-{\vec x}_2|}\left[{\vec v}_{12} + {2 ({\vec v}_{12}\cdot {\vec x}_{12}) {\vec x}_{12}\over |{\vec x}_{12}|^2}\right],
 \end{equation}
 where ${\vec x}_{12}$ and ${\vec v}_{12}$ are the relative displacement and velocity of the binary.  This is a 6.5PN correction to the equations of motion, analogous to the Burke-Thorne radiation damping force from the previous section.     It gives a 4PN correction to the mechanical energy loss of the binary ${\dot E}=\sum_A {\vec v}_A\cdot {\vec F}_A$.
 
  In the case of spinning objects, the correlator is no longer determined by a single form factor $A^{E,B}_+(\omega)$ as more tensor structures, involving the spin vector of the object, can appear.   For non-zero spin, the EFT has been extended to slowly spinning black holes~\cite{Porto:2007qi} , to more general spinning sources in ref.~\cite{Endlich:2016jgc}, and generalized to rapidly spinning (close to extremal) Kerr black holes~\cite{Goldberger:2020fot}.     As in the non-spinning case, the full causal response is determined by matching to the black hole absorption rates in~\cite{staro,Page:1976df} as well as results of~\cite{LeTiec:2020spy,Chia:2020yla,LeTiec:2020bos}, to fix the local counterterms (the conservative part of the tidal response) on the worldline Lagrangian to zero.

For instance, the energy loss in the spinning case, for arbitrary spins
 \begin{equation}
 \label{eq:absEdot}
\left. {dE\over dt}\right|_{h}={8\over 5} {G_N^5 m_1^2 m_2\over |{\vec x}_1-{\vec x}_2|} (m_1+m_2)\left[1+3 \chi_1^2 -{15\over 4}\chi_1^2\left({{\vec s}_1\cdot ({\vec x_1}-{\vec x}_2)\over |{\vec x}_1-{\vec x}_2|}\right)^2\right] {\vec S}_1\cdot {\vec L}+(1\leftrightarrow 2),
 \end{equation}
 at leading PN order~\cite{Goldberger:2020fot} $({\vec s}_A={\vec S}_A/|{\vec S}_A|$).   For nearly maximally rotating black holes, with $\chi=|{\vec S}|/G_N M\lsim 1,$  Eq.~(\ref{eq:absEdot}) gets enhanced, by the superradiant effect~\cite{zeldovich,staro}, to 2.5PN order compared to 4PN in the Schwarzschild case, $\chi=0$.   Notice that the energy flux can have either sign depending on the relative orientation between the black hole spin and the orbital angular momentum ${\vec L}$.   In particular, it is possible to extract rotational energy from the black holes as in the Penrose process~\cite{Penrose:1971uk}. Eq.~(\ref{eq:absEdot}) generalizes to arbitrary orbits and spin orientations earlier results obtained by classical techniques in~\cite{Alvi:2001mx,Chatziioannou:2012gq,Chatziioannou:2016kem}.

\subsubsection{Quantum effects}
\label{sec:hawkrad}

Because the \zosouv~formalism is explicitly quantum mechanical, it is capable of describing processes involving quantum black holes~\cite{Hawking:1974sw,Hawking:1976ra} interacting with other particles or fields.   In particular, it can be used~\cite{Goldberger:2019sya,Goldberger:2020geb} to take into account the effects of Hawking radiation on scattering observables.   

The EFT description assumes that black holes behave according to the standard quantum mechanical rules, with unitary time evolution and a complete Hilbert space of microstates.    It is valid in the window of momentum transfers $q$ defined by
\begin{equation}
t^{-1}_{Page}\ll q\ll 1/G_N M_{BH}\ll m_{Pl},
\end{equation}
where, \eg~ in the Schwarzschild case, the black hole evaporation scale is of order the Page time~\cite{Page:1976df}, $t_{Page} \sim G_N^2 M_{BH}^3$.   The upper bound ensures that the worldline description of the black hole is reliable, while the lower bound allows us to treat black holes as approximately long-lived asymptotic states in the $S$-matrix.   Since $G_N M_{BH}\gg 1/m_{PL}$, the black holes can be regarded as being semiclassical.    Because the time scales are short compared to  $t_{Page}$, the processes we will consider are not sensitive to the apparent loss of unitarity in black hole evaporation~\cite{Hawking:1976ra}.

To construct the EFT, we have to determine how the structure of multipole operator correlation functions is modified by the presence of a Hawking thermal spectrum of radiation emission from the black hole.    As in the classical case, we match to the simplest possible observables that depend, on the EFT side, on the multipole correlators.    It is convenient in particular to calculate the on-shell transition probabilities $p(m\rightarrow n)$ for a Kerr black hole to emit $n$ gravitons out to future ${\cal I}^+$, given that $m$ particles of the same energy $\omega$ and angular momentum quantum number $\ell$  are incident on the black hole in the far past from ${\cal I}^-$ .    Explicit results\footnote{The results of refs.~\cite{Bekenstein:1977mv,Panangaden:1977pc} are for free scalar fields propagating in the black hole background, but they generalize naturally to higher spin $s>0$ fields by simply replacing the scalar transmission coefficients by their higher spin version found in refs.~\cite{staro,Page:1976df}.} are available for this observable, in the limit of free quantum field theory in the black hole background~\cite{Bekenstein:1977mv,Panangaden:1977pc} ,
\begin{eqnarray}
p_\ell(m\rightarrow n) &=& {(1-x) x^n (1-|R_{\ell}|^2)^{n+m}\over (1- x|R_{\ell}|^2)^{n+m+1}}\\
\nonumber
& & \times \sum_{k=0}^{\min(n,m)} {(n+m-k)!\over k! (n-k)! (m-k)!}\left[{(x^{-1}|R_\ell|^2-1) (1-x|R_{\ell}|^2)\over (1-|R_{\ell}|^2)^2}\right]^k,
\end{eqnarray}
where $x=\exp[-\beta_H\omega]$, $|R_{\ell}(\omega)|^2$ is the Boltzmann factor for the Kerr black hole and $|R_{\ell}(\omega)|^2$ are the partial wave reflection coefficients obtained in~\cite{staro,Page:1976df}.

In \zosouv, one finds that to leading order in $\omega/m_{Pl}\ll 1$, the $m\rightarrow n$-particle probabilities are controlled by $n+m$-point Wightman correlation functions of the multipole operators in Eq.~(\ref{eq:zoso}).  Given the structure of the full theory result~\cite{Bekenstein:1977mv,Panangaden:1977pc}, these higher points correlators factorize into suitable products of two-point functions, up to non-Gaussianities suppressed by $\omega^2/m^2_{Pl}\ll 1$.    Somewhat surprisingly, one finds that at $r_s\omega \ll 1$ the effects of Hawking radiation are not Planck suppressed at the level of the Wightman functions.   Instead, they become \emph{enhanced}  in the limit $\hbar\omega/T_H= 4\pi r_s\omega \ll 1$ where the EFT is valid, a consequence of the high temperature behavior of the Planck distribution.

Despite this enhancement at the level of the Wightman functions, in the full theory the effects of Hawking radiation cancel in the free field \emph{retarded} correlators.  Up to corrections suppressed by $1/m_{Pl}$, these take the same form in the  Unruh state~\cite{Unruh:1976db} that describes an evaporating black hole or the Boulware state~\cite{Boulware:1974dm} where the black hole does not emit radiation.    Consequently,  finite size effects associated with the quantized nature of the black hole horizon, \eg~ in the observables discussed in the previous section, are suppressed by at least one power of $\omega^2/m_{PL}^2\ll 1$.  Thus, if one assumes that black holes evolve according to the usual rules of quantum mechanics, the results of~\cite{Goldberger:2019sya} imply a no-go theorem on the possibility of detecting black microstate ``hair'' or other possible exotic signatures of quantum behavior in  binary black hole mergers at LIGO/VIRGO or any other foreseeable experiment.

On the other hand, the emission of Hawking radiation does modify observables that depend on the Wightman functions directly.   While not of phenomenological important, an example~\cite{Goldberger:2020geb} of formal interest is the inelastic scattering of elementary particles incident on a semiclassical black hole, mediated by the exchange of \emph{virtual} (off-shell) Hawking gravitons.

For illustration, consider a scalar particle $\phi$ with mass in the range $k_B T_H\ll m_\phi \ll M_{BH}$.   In this window, direct $s$-channel production of $\phi$-particle Hawking pairs is exponentially (Boltzmann) suppressed, so that the scattering process
\begin{equation}
\phi(p)+\mbox{BH}\rightarrow \phi(p')+{\mbox{BH}}^\prime
\end{equation}
proceeds instead through off-shell graviton exchange.   One finds the result~\cite{Goldberger:2020geb}
\begin{eqnarray}
\nonumber
{d^3\sigma \over dq^2 d(q\cdot v)} &\approx& {7 G_N  r_s^5\over 270\pi [(v\cdot p)^2 - m_\phi^2]} \left[(v\cdot p)^4  -m_\phi^2 (v\cdot p)^2\left(1 - {12\over 7} {(v\cdot q)^2\over q^2}\right) \right.\\
\label{eq:result}
& & \left.+{1\over 7}m^4 \left(1 -3 {(v\cdot q)^2\over q^2}+  6{(v\cdot q)^4\over q^4}\right)\right],
\end{eqnarray}
for the differential cross section in the black hole rest frame $v^\mu=(1,{\vec 0})$, as a function of the momentum transfer $q^\mu=p^\mu-p'^\mu$.   The point of this result is that, in generic regions of phase space, the integrated cross section scales as $\sim q^2/m^2_{Pl}$ relative to the leading order classical gravitational scattering between point masses $M$, $m_\phi$.   It is  parametrically of the same size as the sort of ${\cal O}(\hbar)$ corrections from graviton vacuum polarization loops that appear in scattering processes involving elementary particles, \eg~\cite{Donoghue:1993eb,Donoghue:1994dn,Bjerrum-Bohr:2002gqz}.   Eq.~(\ref{eq:result}) can be therefore be regarded as a specific realization of a qualitatively new  phenomenon in low energy quantum gravity, associated with the exchange of virtual Hawking gravitons and tractable (calculable) by the methods of effective field theory applied to gravity.

\section{Acknowledgments}
This chapter is partially based on lectures delivered at the Pauli Center workshop \emph{From Scattering Amplitudes to Gravitational-Wave Predictions for Compact Binaries} held at the University of Zurich and the ETH Zurich on July 4-15, 2022.  I thank the organizers V. Del Duca, H. Ita, and L. Mayer for the invitation to present this subject, and the students for many stimulating questions.    The research presented here is partially supported by the US Department of Energy under grant DE-SC00-17660.

\end{document}